\documentclass{article}
\usepackage[T1]{fontenc}
\usepackage[latin1]{inputenc} 
\usepackage{graphics}
\usepackage{graphicx}
\usepackage{longtable}
\usepackage{color}%\usepackage{pslatex}
\usepackage{hyperref}
\usepackage{sverb}
\usepackage{subfigure}
\usepackage{mathrsfs}

\usepackage{graphicx}
\usepackage{epstopdf}
\usepackage{a4p}
\usepackage{rotating}

\begin{document}
\begin{titlepage}
 
\begin{flushright} 
{ Preprint no IFJ-PAN-IV-2018-14 %\\  
} 
\end{flushright}
 
\vskip 15 mm
\begin{center}
{\bf\huge The {\tt TauSpinner} approach for electroweak corrections in LHC  $Z \to ll$ observables}
\end{center}
\vskip 15 mm

\begin{center}
%  {\bf T. Przedzinski$^{a}$, Elzbieta Richter-Was$^{a}$ and Zbigniew Was$^{b}$  }\\
  {\bf E. Richter-Was$^{a}$ and Z. Was$^{b}$  }\\
  {\em $^a$ Institute of Physics, Jagellonian University, 30-348 Krakow, Lojasiewicza 11, Poland} \\
       {\em $^b$  Institute of Nuclear Physics, Polish Academy of Sciences,
        31-342 Krakow, ul. Radzikowskiego 152, Poland}\\ 
\end{center}
\vspace{1.3 cm}
\begin{center}
{\bf   ABSTRACT  }
\end{center}

The LHC   Standard Model $Z$-boson couplings  measurements 
approach  the LEP legacy precision. The
calculations of electroweak (EW) corrections available for the Monte Carlo generators
become of relevance. Predictions of $Z$-boson production and decay require
classes of QED/EW/QCD corrections and  
separatly from the production process QCD dynamics. \\
At the LEP time  electroweak form-factors
and {\it Improved Born Approximation}  were introduced for non QED genuine weak and 
line-shape
corrections.
%The multi-loop vacuum polarization contribution and multi-photon bremsstrahlung were necessary.
This formalism was well suited for  observables,
so-called doubly-deconvoluted the $Z$-pole region where
initial- and final-state QED real and virtual emissions were treated separately
or were integrated over. The approach was convenient for implementation into Monte Carlo programs for LEP,
Belle-BaBar and other future $e^+e^-$ colliders and for  invariant mass of  outgoing lepton pair from a few GeV to well above $WW$ and even $t\bar t$ threshold.
We attempt now to profit from that, for the LHC $pp$  and   70 to 150 GeV window for the
outgoing lepton pair invariant mass.\\
Our technical focus is on  the EW corrections for LHC
 $Z \to \ell \ell$ observables.
For this purpose the {\tt TauSpinner} package, for the reweighting
of previously generated events, is  enriched with the
genuine EW corrections (QED effects subtracted) of
the {\tt Dizet} electroweak library, taken from the LEP era {\tt KKMC} Monte Carlo.
Complete genuine EW $O(\alpha)$ weak loop corrections 
and dominant
higher-order terms are taken into account.
For the efficiency and numerical stability
 look-up tables are used.
For LHC observables: the  $Z$-boson line-shape, the outgoing leptons forward-backward
asymmetry, the effective leptonic weak mixing angles and finally for
the spherical harmonic expansion coefficients  of the lepton distributions,
corrections are evaluated.
Simplified calculations of 
{\it Effective Born} of modified EW couplings are  compared with 
of {\it Improved Born Approximation} of complete set of  EW form-factors. \\
Approach  uses  LEP precision tests definitions and thus offers consistency checks. 
The package can be useful to evaluate of observables precision limits and
to determine which
corrections are then important for LHC and FCC projects phenomenology.
 
\vskip 1 cm

%Draft as of: {\bf \today}. 

%\vspace{0.2 cm}
%\centerline{ \bf DRAFT \today }
 
\vspace{0.1 cm}
\vfill
{\small
\begin{flushleft}
{   Preprint no IFJ-PAN-IV-2018-14
\\ August 2018, June 2019
}
\end{flushleft}
}
%%%%%%%%%%%%%%%%%%%%%%%%%%%%%%%%%%%%%%%%%%%%%%%%%%%%%%
\vspace*{1mm}
\footnoterule
\noindent
{\footnotesize \noindent  $^{\dag}$

This project was supported in part from funds of Polish National Science
Centre under decision UMO-2014/15/B/ST2/00049.

Majority of the numerical calculations were performed at the PLGrid Infrastructure of 
the Academic Computer Centre CYFRONET AGH in Krakow, Poland.

}
\end{titlepage}

%\tableofcontents
\clearpage

\section{Introduction}

A theoretically sound separation of QED/EW effects between the QED emissions and genuine weak effects
was essential for the phenomenology of LEP precision physics~\cite{ALEPH:2005ab}. It was motivated by the structure of the amplitudes
for single $Z$ or (to a lesser degree) $WW$ pairs production in $e^+e^-$ collisions, and by the
fact that QED bremsstrahlung occurs at a different energy scale than the electroweak processes.
Even more importantly, with this approach multi-loop calculations for complete electroweak sector could be avoided.
The QED terms could be resumed in an exclusive exponentiation scheme implemented in Monte Carlo~\cite{Jadach:2013aha}. 
Note that QED corrections modify the cross-section at the peak by as much as 40\%.
The details of this paradigm are explained in~\cite{Altarelli:1989hv}.
It was obtained as a consequence of massive efforts, we will not recall them here.
For the  present study, the  observation that spin amplitudes  semi-factori\-ze into a Born-like terms and
functional factors responsible for bremsstrahlung~\cite{Berends:1983mi} was very important.

A similar separation can be also achieved for dynamics of production process in $pp$ collisions,
which can be isolated from QED/EW corrections.
It was explored recently in the case of configurations with
high-$p_T$ jets associated with the Drell-Yan production of $Z$~\cite{Richter-Was:2016mal} or
$W$ bosons~\cite{Richter-Was:2016avq} at LHC.
The potentially large electroweak Sudakov logarithmic
corrections discussed in~\cite{Kuhn:2005az} (absent in our work) represent
yet another class of weak effects,
separable from those discussed throughout this paper.
They are very small for lepton pairs with a virtuality close to the $Z$-boson pole mass and, if accompanied by the jet
when virtuality of $\ell \ell j$ system is not much larger than 2 $M_W$.
Otherwise the Sudakov corrections have to be revisited and calculation of electroweak corrections  extended, even if
invariant mass of the lepton pair is close to the $Z$ mass.

To assess precisely the size and impact of genuine weak corrections to the Born-like cross
section for lepton pair production with a virtuality below threshold for $WW$ pair production,
the precision calculations and programs prepared for the LEP era: {\tt KKMC} Monte Carlo~\cite{Jadach:1999vf}
and {\tt Dizet} electroweak (EW) library, were adapted to provide pre-tabulated EW corrections
to be used by LHC specific event reweighting programs like {\tt TauSpinner} package~\cite{Czyczula:2012ny}. Even at present {\tt KKMC} Monte Carlo
use {\tt Dizet} version {\tt 6.21}~\cite{Bardin:1989tq,Bardin:1999yd}. We restrict ourselves
to that reference version. 
The {\tt TauSpinner} package was initially created as a tool to correct with per-event weight
longitudinal spin effects in the generated event samples including $\tau$ decays.  Algorithms implemented there
turned out to be of more general usage.
The possibility to introduce one-loop electroweak corrections from {\tt SANC}
library~\cite{Andonov:2008ga} in case of Drell-Yan production of the $Z$-boson became 
available in {\tt TauSpinner} since~\cite{Przedzinski:2014pla}.
Pre-tabulation prepared for
EW corrections of {\tt SANC} library, was  useful 
to introduce weights for complete spin effects at each
individual event  level. 
% for each individual spin configurations of outgoing leptons.
However no higher loop contributions  were available.

{\tt TauSpinner}  provides a reweighting technique to modify hard process matrix elements (also matrix elements for $\tau$ decays)
which were used for Monte Carlo generation. For each event no changes of any details
for event kinematic configurations are  introduced.
The reweighting algorithm can be used for events where final state
QED bremsstrahlung photons and/or high $p_T$ jets are present.
For  matrix element calculation 
used for re-weigh\-ting, some contributions such as of QED bremsstrahlung
or of jet emissions have to be removed. For that purpose factorization
and  detailed inspection of fixed order  perturbation expansion amplitudes
is necessary.
The most recent summary on  algorithms and their applications is given in~\cite{Przedzinski:2018ett}.
%Detailed technical description of the {\tt TauSpinner} reweighting algorithm was given recently
%in Ref.~\cite{Przedzinski:2018ett}.
The reference explains in detail how kinematical configurations are reduced to Born-level configurations
used for the correcting weights, also for  electroweak corrections%
\footnote{In Ref.~\cite{Davidson:2010rw} (on {\tt Tauola Universal Interface})
other than {\tt TauSpinner}
solution was prepared. Then parton level history entries for generated event record were
used. For  {\tt TauSpinner} use of history event record entries was abandoned, because of too many variants how corresponding information was required to be interpreted.
  Instead, contributions from all possible parton level processes, weighted with parton distribution functions are averaged.
  This could also be used for configurations generated with multi jet matrix elements, when Born level marix element configurations
  can not be identified. %We return to this aspect of the algorithm briefly, later in the paper.
}.

Used for both {\tt Tauola Univesal Interface} and {\tt TauSpinner }, 
SANC library \cite{Andonov:2008ga} of year 2008 calculates one loop i.e.
NLO electroweak corrections 
 in two  $\alpha(0)$ and $G\mu$ ($G_F$)  
 schemes. It was  found numerically insufficient for practical applications.
 For example, it was missing sizable $\alpha_s$ corrections to the
 calulated $Z$ boson width.
  Two aspects of EW corrections implementation~\cite{Przedzinski:2014pla} had  to be
  enhanced.

  First,
in~\cite{Richter-Was:2016mal,Richter-Was:2016avq} we have studied separation
 of QCD higher order corrections and the Born-level spin amplitudes  calculated in the
adapted {\it Mustraal} lepton pair rest frame\footnote{ Over the paper we use several variants of coordinate
  system orientation for the lepton pair rest-frame. The {\it Mustraal} frame resulted from careful
  analysis of the cross section for  the initial and final state bremsstrahlung
  that is $e^+e^- \to \mu^+\mu^- \gamma$. It was found that it can be represented,
  without any approximation as sum of four incoherently added distributions with well defined
  probabilities (two for initial and two for final state emission),
  each factorized into Born cross section calculated 
  in reference frame oriented as required by the form of matrix element and the factor dependent
  on kinematical variables for the  $\gamma$. One should keep in mind that the spin carried
  by the photon cancels out with its orbital momentum. That property of the matrix element
  originates from the properties of the Lorentz group representations, their combinations
  for the ultra-relativistic states. That is why it generalizes unchanged to the 
  $q \bar q \to l^+l^- g$ and approximately also to other processes of single or even double
  jet emissions in a bulk of parton emissions in $pp$ collisions. It was checked numerically in
  Refs.~\cite{Richter-Was:2016mal,Richter-Was:2016avq}.
}. It is  defined like for QED brems\-strahlung of Ref. ~\cite{Berends:1983mi}.
The separation holds to a good approximation
for the  Drell-Yan processes where one or even two high $p_T$ jets are present.
This frame is now used as option for EW weight calculation.

Second, the {\tt TauSpinner} package and algorithms are now adapted to  EW corrections
from the {\tt Dizet} library\footnote{This legacy library of EW corrections,
  features numerically important,   corrections
  beyond NLO, n particular  to $Z$ and $\gamma^*$ propagators.
  Contributions corresponding to QED are carefully removed 
  and left for the independent treatment.}, more accurate than { SANC}. 
%   directly into spin amplitudes and weight calculations for the Drell-Yan $Z$-boson
%   production process.
%Beyond NLO EW corrections can be thus  implemented as well.
The EW corrections are introduced 
with  form-factor corrections of Standard Model couplings and propagators which
enter 
spin amplitudes of the {\it Improved Born Approximation}, used for EW weights calculation. They represent complete $\cal O (\alpha) $ electroweak
corrections with 
QED contributions removed but augmented with carefully selected dominant higher order terms.
This  was very successful in analyses of LEP I  precision physics. We attempt
a similar
strategy for the  $Z$-boson pole LHC precision physics; the  approach to EW corrections   already attracted attention.
It was used in  the  preliminary measurement of effective leptonic weak mixing angle
recently published by ATLAS Collaboration~\cite{ATLAS:2018gqq}. 

This paper is organized as follows. In Section \ref{sec:IBA} 
we collect the main formulae of the formalism, in particular we recall the definition
of the {\it Improved Born Approximation}.
In Section~\ref{sec:formfactors} we present numerical results for the electroweak form-factors.
Some details on commonly used EW schemes are discussed in Section~\ref{sec:EWschemes},
which also recall the definition of the {\it Effective Born}.
In Section~\ref{sec:IBAaveraging} we comment on the issues of using the Born approximation in $pp$ collisions
and in Section~\ref{sec:QCDcorr} we give more explanation why the Born approximation
of the EW sector is still valid in the presence of NLO QCD matrix elements.
In Section~\ref{sec:EWweight} we define the concept of EW weight which can be  applied
to introduce EW corrections into already existing samples, generated with Monte Carlo
programs with EW LO  hard process matrix elements only.
In Section~\ref{sec::EWobserv} we discuss, in numerical detail, EW corrections
to different observables of interest for precision measurements: $Z$-boson line-shape,
lepton forward-backward asymmetry and for coefficients of  lepton  spherical harmonic expansion.
In this Section we include also a discussion of the effective weak mixing
angle in case of $pp$ collision.  
For results presented in Section~\ref{sec::EWobserv} we use QCD NLO {\tt Powheg+MiNLO}~\cite{Alioli:2010xd} $Z+j$
Monte Carlo sample, generated for $pp$ collision with $\sqrt{s}$ = 8 TeV and EW LO implementation
in matrix elements. Section~\ref{sec:sumary} summarizes the paper.

In Appendix~\ref{app:TauSpinnerInit} details on the technical implementation of EW weight
and how it can be calculated with help of the {\tt TauSpinner} framework are given.
In Appendix~\ref{app:SW2scan} formulae which have been implemented to allow variation
of the weak mixing angle parameter of the Born spin amplitudes are discussed.
%Illustrative numerical results are provided, but detailed discussion is left for the forthcoming work.
In Appendix~\ref{app:DizetInit} initialization details, and  options valuable for future discussions,
for the {\tt Dizet} library  are collected.

\section{Improved Born Approximation} 
\label{sec:IBA}

At LEP times, to match higher order QED effects with the loop corrections of electroweak
sector, the concept of electroweak  form-factors was introduced~\cite{Altarelli:1989hv}.  This arrangement was very beneficial and enabled
common treatment of one loop electroweak effects with not only higher order QED corrections including 
bremsstrahlung, but also to incorporate higher order loops into $Z$ and photon propagators, see e.g. documentation
of {\tt KKMC} Mon\-te Carlo~\cite{Jadach:2013aha} or {\tt Dizet}~\cite{Bardin:1999yd}.
Such description has its limitations for the LHC applications, but for the processes of the Drell-Yan type 
with a moderate  virtuality of produced lepton pairs is expected to be useful, even in the case when high $p_T$ jets are present. 
For the LEP applications~\cite{ALEPH:2005ab}, the EW form-factors were used together with multi-photon bremsstrahlung amplitudes,
but for the purpose of this paper we discuss their use with parton level Born processes only (no QED ISR/FSR%
\footnote{Presence in reweighted events of  QED initial and final state bremsstrahlung,
  does not lead to complications of
  principle, but would obscure presentation. Necessary extensions \cite{Przedzinski:2018ett}
  are technically simple, thanks to properties of QED matrix elements, presented for the first time in \cite{Berends:1983mi}.}).

The terminology  {\it double-deconvoluted  observable} was widely used since LEP time and is explained e.g. in
\cite{Bardin:1999gt}.
The so called  {\it Improved Born Approximation} (IBA)~\cite{Bardin:1999yd}
is employed. It absorbs  some
of the higher order EW corrections into a redefinition of couplings and propagators of the Born spin
amplitude. This allows for straightforward 
calculation of  {\it doubly-deconvoluted  observables} like various cross-sections and asymmetries. QED effects are then removed or integrated over.

It is possible, because
the excluded initial/final QCD and QED corrections form separately gauge invariant subsets of diagrams~\cite{Bardin:1999yd}. The QED subset consists
of QED-vertices, $\gamma \gamma$ and $\gamma Z$ boxes and  bremsstrahlung diagrams. %Fermionic self-energies have to be also taken into account.
 The subset corresponding to the initial/final QCD corrections can be constructed as well.
All the remaining corrections contribute to the IBA: purely EW loops, boxes and {\it internal} QCD corrections for loops (line-shape corrections).
They can be split into two more gauge-invariant subsets, giving rise to two {\it improved (or dressed)} amplitudes: (i)
improved $\gamma$ exchange amplitude with running QED coupling where  fermion loops of low $Q^2$  contribute dominantly and (ii) improved
$Z$-boson exchange amplitude with four complex
{\it EW form-factors}: $\rho_{\ell f}$, ${\mathscr K}_{\ell}$, ${\mathscr K}_{f}$, ${\mathscr K}_{\ell f}$.
%The complete formulae for improved spin amplitudes was shown in (\ref{Eq:BornEW}).
Components of those corrections are as follows:
\begin{itemize}
\item
  Corrections to photon propagator, where  fermion lo\-ops contribute dominantly
  the so called vacuum-pola\-ri\-zation corrections.
\item
Corrections to $Z$-boson propagator and couplings, called EW form-factors.
\item
Contribution from the purely weak $WW$ and $ZZ$ box  diagrams. They are negligible at the
$Z$-peak (suppressed by the factor $(s-M^2_Z)/s$), but very important at higher energies.
They enter as corrections to form-factors and introduce non-polynomial dependence on the $\cos$ of the scattering angle.
\item
Mixed $O(\alpha \alpha_s, \alpha \alpha_s^2, ...)$ corrections which originate from gluon insertions to the fermionic components of bosonic self-energies.
They  enter  as corrections to all form-factors.
\end{itemize}

Below, to define notation we present the formula of the Born spin amplitude ${\mathscr A}^{Born}$.
We recall  conventions from~\cite{Bardin:1999yd}. 
Let us start with defining the lowest order coupling constants (without EW corrections) of the $Z$ boson to fermions:
$ s^2_W = 1- M_W^2/M_Z^2=\sin \theta^2_W $ defines weak Weinberg angle  in the on-mass-shell scheme and $T_3^{\ell, f}$
third component of the isospin.
The vector $v_{\ell}, v_f$ and axial $a_{\ell}, a_f$ couplings for leptons and quarks are defined
with the formulae below\footnote{We will use ``$\ell$'' for lepton, and ``$f$'' for quarks.}
\begin{eqnarray}
   \label{Eq:avLO}
  v_{\ell} && = (2 \cdot T_3^{\ell} - 4 \cdot q_{\ell} \cdot s^2_W)/\Delta , \nonumber \\
  v_f && = (2 \cdot T_3^f - 4 \cdot q_f \cdot s^2_W)/\Delta ,  \\
  a_{\ell} && = (2 \cdot T_3^{\ell} )/\Delta , \nonumber \\
  a_f && = (2 \cdot T_3^f )/\Delta . \nonumber 
\end{eqnarray}
where
\begin{equation}
  \label{Eq:Delta}
  \Delta = \sqrt{ 16 \cdot s^2_W \cdot (1 - s^2_W)} ,
\end{equation}
and $ q_f$, $q_l$ denote charge of incoming fermion (quark) and outgoing lepton.
With this notation, the  ${\mathscr A}^{Born}$ spin amplitude for the $q \bar q \to Z/\gamma^* \to \ell^+ \ell^- $
can be written as:
\begin{eqnarray}
  \label{Eq:Born}
 && {\mathscr A}^{Born} = \frac{\alpha}{s}\ \  \{ \nonumber  \\ 
            && [\bar u \gamma^{\mu} v   g_{\mu \nu}  \bar v \gamma^{\nu} u] \cdot ( q_{\ell} \cdot q_f)  \cdot \chi_{\gamma}(s)
               + [\bar u \gamma^{\mu} v g_{\mu \nu} \bar \nu \gamma^{\nu} u  \cdot  ( v_{\ell} \cdot v_f ) \nonumber\\
            &&   +  \bar u \gamma^{\mu} v g_{\mu \nu} \bar \nu \gamma^{\nu} \gamma^5 u  \cdot  (v_{\ell} \cdot a_f) % \nonumber \\
                 +  \bar u \gamma^{\mu} \gamma^5 v g_{\mu \nu} \bar \nu \gamma^{\nu}  u  \cdot  (a_{\ell} \cdot v_f)\nonumber \\
            &&
               + \bar u \gamma^{\mu} \gamma^5  v  g_{\mu \nu}\bar \nu \gamma^{\nu} \gamma^5  u  \cdot  (a_{\ell} \cdot a_f) ] \cdot \chi_Z (s)\ \ \  \} , 
\end{eqnarray}
where $u, v$ denote fermion spinors and, $\alpha$ stands for QED coupling constant. The  $Z$-boson and photon propagators are defined respectively as:
\begin{equation}
  \chi_{\gamma}(s) = 1,   \\
\end{equation}
\begin{equation}
  \label{Eq:Zprob}
%  \chi_Z(s) =   \frac{G_{\mu} \cdot M_{z}^2  \cdot \Delta^2 }{\sqrt{2} \cdot 8 \pi \cdot \alpha}\cdot \frac{s}{s - M_Z^2 + i \cdot \Gamma_Z \cdot M_Z}. \\
  \chi_Z(s) =   \frac{G_{\mu} \cdot M_{z}^2  \cdot \Delta^2 }{\sqrt{2} \cdot 8 \pi \cdot \alpha}\cdot \frac{s}{s - M_Z^2 + i \cdot \Gamma_Z \cdot s/M_Z}. \\
\end{equation}

For the IBA,  we redefine vector and axial couplings and introduce EW form-factors
{\small $\rho_{\ell f}(s,t), {\mathscr K}_{\ell}(s,t)$, ${\mathscr K}_f(s,t)$, ${\mathscr K}_{\ell f} (s,t)$}
as follows:
\begin{eqnarray}
  v_{\ell} && = (2 \cdot T_3^{\ell} - 4 \cdot q_{\ell} \cdot s^2_W \cdot {\mathscr K}_{\ell}(s,t))/\Delta , \nonumber \\
  v_f && = (2 \cdot T_3^f - 4 \cdot q_f \cdot s^2_W \cdot {\mathscr K}_f(s,t))/\Delta ,  \\
  a_{\ell} && = (2 \cdot T_3^{\ell} )/\Delta , \nonumber \\
  a_f && = (2 \cdot T_3^f )/\Delta . \nonumber 
  \label{Eq:avNLO}
\end{eqnarray}
Normalization correction $ Z_{V_{\Pi}} $  to the $Z$-boson propagator is defined as
\begin{equation}
  Z_{V_{\Pi}} = \rho_{\ell f}(s,t) \ .
  \label{Eq:ChiZ}
\end{equation}
Re-summed vacuum polarization corrections $ \Gamma_{V_{\Pi}}$  to the $\gamma^*$ propagator are expressed as 
\begin{equation}
  \Gamma_{V_{\Pi}} = \frac{1}{ 2 - (1 + \Pi_{\gamma \gamma}(s))},
   \label{Eq:ChiGamma}
\end{equation}
where $\Pi_{\gamma \gamma}(s)$ denotes vacuum polarization loop corrections of virtual  photon exchange.
Both $\Gamma_{V_{\Pi}}$ and $Z_{V_{\Pi}}$ are multiplicative correction factors. The $\rho_{\ell f}(s,t)$ could be also absorbed
as multiplicative factor into the definition of vector and axial couplings.

The EW form-factors  {\small $\rho_{\ell f}(s,t), {\mathscr K}_{\ell}(s,t)$, ${\mathscr K}_f(s,t)$, ${\mathscr K}_{\ell f}(s,t) $}
depend on two Mandelstam invariants $(s,t)$
due to contributions of the $WW$ and $ZZ$ boxes. The Mandelstam variables  satisfy the identity
\begin{equation}
  \label{Mandelstam}
  s+t+u = 0 \ \ \ where \ \ \ \ t = -\frac{s}{2}(1 - \cos \theta)
\end{equation}
and $ \cos \theta$ is the cosine of the scattering angle, i.e. the angle between incoming and outgoing fermion directions.

Note, that in this approach the mixed EW and QCD loop corrections,
originating from gluon insertions to  fermionic components of bosonic self-energies, are included in  $\Gamma_{V_{\Pi}}$ and  $Z_{V_{\Pi}}$. 

One has to pay special attention to the angle dependent product of the vector couplings.
The corrections  break factorization, formula (\ref{Eq:Born}),
of the couplings into ones associated with either $Z$ boson production or decay.
The mixed term has to be added: 
\begin{eqnarray}
  vv_{\ell f} =&& \frac{1}{v_{\ell} \cdot v_f} [
    ( 2 \cdot T_3^{\ell}) (2 \cdot T_3^f) - 4 \cdot q_{\ell} \cdot s^2_W \cdot {\mathscr K}_f(s,t)( 2 \cdot T_3^{\ell})\nonumber
    \\&&
    - 4 \cdot q_f \cdot s^2_W \cdot {\mathscr K}_{\ell}(s,t) (2 \cdot T_3^f) \\
    && + (4 \cdot q_{\ell} \cdot s^2_W) (4 \cdot q_f \cdot s^2_W) {\mathscr K}_{\ell f}(s,t)] \frac{1}{\Delta^2}.  \nonumber
   \label{Eq:vvNLO}
\end{eqnarray}

Finally, we can write the spin amplitude for Born with EW corrections, ${\mathscr A}^{Born+EW} $, as:
\begin{eqnarray}
  \label{Eq:BornEW}
  {\Huge \mathscr A}^{Born+EW} &=& \frac{\alpha}{s} \{
              [\bar u \gamma^{\mu} v   g_{\mu \nu}  \bar v \gamma^{\nu} u] \cdot ( q_{\ell} \cdot q_f)]  \cdot  \Gamma_{V_{\Pi}} \cdot \chi_{\gamma}(s) \nonumber \\ &&
               + [\bar u \gamma^{\mu} v g_{\mu \nu} \bar \nu \gamma^{\nu} u  \cdot  ( v_{\ell} \cdot v_f \cdot  vv_{\ell f})\\
            &&+  \bar u \gamma^{\mu} v g_{\mu \nu} \bar \nu \gamma^{\nu} \gamma^5 u  \cdot  (v_{\ell} \cdot a_f)  \nonumber \\ &&
                 +  \bar u \gamma^{\mu} \gamma^5 v g_{\mu \nu} \bar \nu \gamma^{\nu}  u  \cdot  (a_{\ell} \cdot v_f)
                 \nonumber \\ &&
  + \bar u \gamma^{\mu} \gamma^5  v  g_{\mu \nu}\bar \nu \gamma^{\nu} \gamma^5  u  \cdot  (a_{\ell} \cdot a_f) ] \cdot Z_{V_{\Pi}}\cdot \chi_Z (s)\}. \nonumber
\end{eqnarray}

The EW form-factor corrections: $\rho_{\ell f}, {\mathscr K}_{\ell}, {\mathscr K}_f, {\mathscr K}_{\ell f}$
can be calculated using the {\tt Dizet}  library.
This library invokes also  calculation of vacuum polarization corrections to the photon propagator $\Pi_{\gamma \gamma}$.
For the case of $pp$ collisions we do not introduce QCD corrections to vector and axial couplings
of incoming  fermions. They are assumed to
 be included elsewhere as a part of the QCD NLO calculations for the initial parton state, including   convolution with proton structure
functions.

The {\it Improved Born Approximation} uses the spin amplitude ${\mathscr A}^{Born+EW}$ of Eq.~(\ref{Eq:BornEW}) and $2 \to 2$ body kinematics
to define the differential cross-section with EW corrections for $q \bar q \to Z/\gamma^* \to l l$ process.  The formulae presented above very closely
follow the approach taken for implementation\footnote{Compatibility with this program is also part of the motivation
  why we leave updates for the {\tt Dizet} library to the forthcoming work. {\tt Dizet 6.21} is also well documented.}
of EW corrections to {\tt KKMC} Monte Carlo~\cite{Jadach:2013aha}.

%For completeness let us note that the above discussion was presented for scattering process, however one may be interested in the decay process only.
%For this {\it effective couplings} of $Z$-decay are often introduced; these are complex-valued constants as well.
%
%The ratio of effective vector and axial couplings defines $g_Z^f$ %(here we use ``f'' for quark or lepton)
%\begin{equation}
%  g_Z^f = \frac{v_Z^f}{a_Z^f} = 1 - 4 |q_f| (K^f_Z s^2_W + I_f^2)
%\end{equation}
%with
%\begin{equation}
%  \label{Eq:I2f}  
%  I_f^2 = \alpha^2(s) \frac{35}{18} [ 1 - \frac{8}{3} Re(K^f_Z) s^2_W].
%\end{equation}
%and the flavour dependent {\it effective weak mixing angles} as
%\begin{equation}
%  \label{Eq:sweff}
%  \sin^2\theta^f_{eff} = Re({\mathscr K}_Z^f) s^2_W + I^2_f
%\end{equation} 

\section{Electroweak form-factors} \label{sec:formfactors}

For the calculation of EW corrections, we use the {\tt Dizet} library,
as of the 2010 {\tt KKMC}
Monte Carlo~\cite{Jadach:2013aha} version. For this and related projects, massive theoretical effort 
was necessary. Simultaneous study of several processes, like
of $\mu^+\mu^-$, $u \bar u$, $d \bar d$, $\nu \bar \nu$ production in $e^+e^-$ collisions and
also in $p \bar p$ initiated parton processes, like at Tevatron, was performed. Groups of diagrams
for  the $Z/\gamma^*$ propagators, production and  decay  vertices
could be identified and incorporated into form-factors. The core of the Dizet library relies
on such separation. It also opened the possibility that for one group of diagrams, such as
vacuum polarizations, higher order contributions  could be included while for others were not.
That was particularly important for
quark contributions to vacuum polarizations. Otherwise, the required precision would not be achieved.
%The vertex contributions corresponding to QED were separated and
%could be switched off in Dizet; left to be treated by other programs such as  {\tt KKMC} and
%to higher than the first orders.
%Also QCD corrections which correspond to initial state interactions for $p p (\bar p)$ collisions
%(or quark pair production at LEP) are separated from the one
%of vacuum polarizations.
%
The above short explanation only indicates fundamental  importance of the topic, we delegate
the reader to Refs.~\cite{Jadach:2013aha,Bardin:1980fe,Bardin:1981sv} and experimental
papers of LEP and Tevatron experiments quoting these papers.

The interface in {\tt KKMC} prepares look-up tables with EW form-factors
and vacuum polarization corrections. The tabulation grid granularity and ranges of the centre-of-mass energy of outgoing
leptons and lepton scattering angle are adapted to variation of the tabulated functions.
Theoretical uncertainties on the predictions for EW form-factors have been estimated in times of LEP precision measurements,
in the context of either benchmark results like \cite{Bardin:1999gt} or specific analyses ~\cite{Altarelli:1989hv}.
The predictions are now updated with the known Higgs boson and  top-quark masses.
In the existing code of the {\tt Dizet} library, certain types
of the corrections or options of the calculations of different corrections can be switched off/on. In Appendix~\ref{app:DizetInit},
we show in Table~\ref{TabApp:GammaZCoupl_2} an almost  complete list of
 options useful for discussions.
We do not  attempt to estimate the size of theoretical uncertainties, delegating it to the
follow up work in the context of LHC EW Precision WG studies. The other versions of electroweak calculations, like
of~\cite{Andonov:2008ga,Akhundov:2013ons}, can and should be studied then as well.
%This will be necessary once precision requirements will become higher than now.
Already now the precision requirements of LHC experiments \cite{ATLAS:2018gqq} are comparable to those of
individual LEP measurements, but phenomenology aspects are more involved.

\subsection{Input parameters to Dizet}
\label{inputDizet}
The {\tt Dizet} package relies on the so called {\it on-mass-shell} (OMS) normalization
scheme~\cite{Bardin:1980fe,Bardin:1981sv} but modifications are present. The OMS uses the masses of all fundamental particles, both fermions and bosons, the
electromagnetic coupling constant $\alpha(0)$ and the strong coupling $\alpha_s(M_Z^2)$.
The dependence on the ill-defined masses of the light quarks {\it u, d,c, s} and {\it b} is solved
by dispersion relations, for details see \cite{Bardin:1999yd}. Another exception is the $W$-boson mass $M_W$, which
still can be predicted with better theoretical accuracy than experimentally measured.
The Fermi constant $G_{\mu}$ is precisely known from $\mu$-decay.
For this reason,  $M_W$ was usually, in time of LEP analyses, replaced by $G_{\mu}$ as an input.

The knowledge about the hadronic vacuum polarization is contained in  $\Delta \alpha_h^{(5)}(s)$,
which is used as external, easy to change, parametrization. It can be either computed from quark masses or, preferably, fitted to
experimental low energy $e^+ e^- \to hadrons$ data.

%The two important constants used are therefore: $\alpha(0)$  - electromagnetic coupling  $\alpha$ in Thomson limit and 
%$G_{\mu}$- Fermi constant in $\mu$-decay. The following parameters are also passed to the main {\tt Dizet} subroutine:
%\begin{equation}
%  M_W, \ M_Z, \ m_t, \ \Delta \alpha_h^{(5)}(M_Z), \ \alpha_s(M_Z).
%\end{equation}

%Note that the above list is over-complete, only two out of the three parameters
%\begin{equation}
%  G_{\mu}, \ M_W, \ M_Z
%\end{equation}
%are independent. They can be selected with appropriate flags setting.
%The basic choice implemented in the {\tt Dizet} library, for calculating EW corrections at the Z-resonance, is to use
%$ G_{\mu}$ and $\ M_Z$ as input parameters and then calculate $M_W$.

The $M_W$ is calculated iteratively from the  equation

\begin{equation}
  M_W = \frac{M_Z}{\sqrt{2}} \sqrt{ 1 + \sqrt{ 1 - \frac{ 4 A^2_0}{M^2_Z ( 1 - \Delta r)}}},
  \label{Eq:MWiterative}
\end {equation}
where
\begin{equation}
  A_0 = \sqrt{ \frac{\pi \alpha(0)}{\sqrt{2}G_{\mu}}}.
\end {equation}

The Sirlin's parameter $\Delta r$ \cite{Sirlin:1980nh}
\begin{equation}
  \label{Eq:Dr}
  \Delta r = \Delta \alpha(M_Z^2) + \Delta r_{EW}
\end {equation}
is also calculated iteratively, and the definition of $ \Delta r_{EW}$ involves re-summation and higher order corrections.
This term implicitly depends on $M_W$ and $M_Z$, and the iterative procedure is needed.
The re-summation term in formula (\ref{Eq:Dr}) is not formally justified by renormalisation group arguments, the
correct generalization is to compute higher order corrections, see discussion in ~\cite{Bardin:1999yd}. 

Note that once the $M_W$ is recalculated from formula (\ref{Eq:MWiterative}), the lowest order Standard Model
relationship between the weak and electromagnetic couplings
\begin{equation}
  G_{\mu} = \frac{\pi \alpha}{\sqrt{2} M^2_W \sin^2\theta_W}
  \label{Eq:SMrelation}
\end{equation}
is not fulfilled anymore, unless the $G_{\mu}$ is redefined away from the measured value.
This is an approach of some EW LO schemes, but not the one used by {\tt Dizet}.
It requires therefore the
complete expression for $\chi_Z(s)$ propagator in
spin amplitude of Eq.~ (\ref{Eq:BornEW}), as defined by formula (\ref{Eq:Zprob}).

In the OMS renormalisation scheme the weak mixing angle is defined uniquely through the gauge-boson masses:
\begin{equation}
  \label{Eq:sw2onshell}
  \sin^2\theta_W = s^2_W = 1 - \frac{M^2_W}{M^2_Z}.
\end{equation}
With this scheme, measuring $\sin^2\theta_W$ will be equivalent to indirect measurement of $M^2_W$ through the
relation (\ref{Eq:sw2onshell}).

%Let us return to the {\tt Dizet} scheme.
%After $M_W$ is computed, the list of input parameters of the main subroutine is ful%ly specified.
%We will consider this configuration as a starting point for discussion on EW corrections to double deconvoluted
%observables.
In Table~\ref{Tab:GammaZCoupl_1} we collect numerical values for all parameters used in the
presented below evaluations. Note that formally they are not representing EW LO scheme,
as the relation~(\ref{Eq:SMrelation}) is not obeyed. The $M_W$ in (\ref{Eq:sw2onshell}) is recalculated
with
(\ref{Eq:MWiterative}) but $G_{\mu}$, $M_Z$ remain unchanged.

%\begin{sidewaystable}
\begin{table}
 \vspace{2mm}
 \caption{The {\tt Dizet} initialization: masses and couplings. The
   calculated $M_W$ and $s^2_W$ 
   are shown also.} 
 \label{Tab:GammaZCoupl_1}
  \begin{center}
    \begin{tabular}{|l|c|c|}
        \hline\hline
         Parameter  & Value & Description \\ 
         \hline\hline
      $M_Z$    &  91.1876 GeV  &  mass of $Z$ boson \\
      $M_H$    &  125.0  GeV   &  mass of Higgs boson \\
      $m_t$    &  173.0  GeV   &  mass of top quark \\
      $m_b$    &  4.7    GeV   &  mass of b quark \\
         $1/\alpha(0)$    &  137.0359895(61)   & QED coupling  \\
 %                  &     & in Thomson limit \\
         $G_{\mu}$ &  $1.166389(22) \cdot 10^{-5}$   & Fermi constant  \\
               &  GeV$^{-2}$  & in $\mu$-decay \\
  \hline
      $M_W$    &  80.353 GeV  & formula (\ref{Eq:MWiterative}) \\
      $s^2_W$  &  0.22351946   & formula (\ref{Eq:sw2onshell})\\
  \hline
    \end{tabular}
  \end{center}
  %\end{sidewaystable}
\end{table}

\subsection{The EW form-factors}

\begin{figure*}
  \begin{center}                               
{
  \includegraphics[width=7.5cm,angle=0]{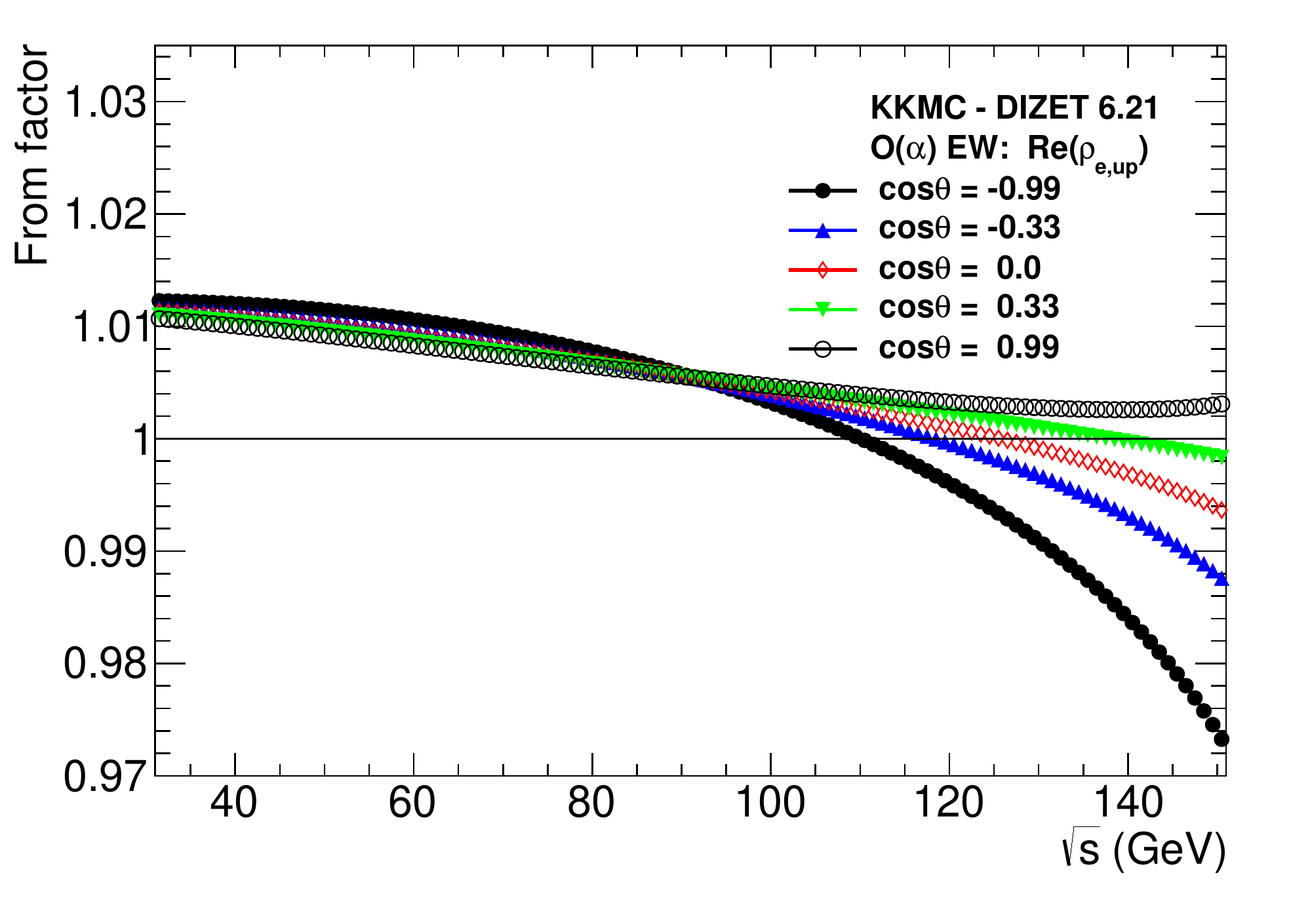}
  \includegraphics[width=7.5cm,angle=0]{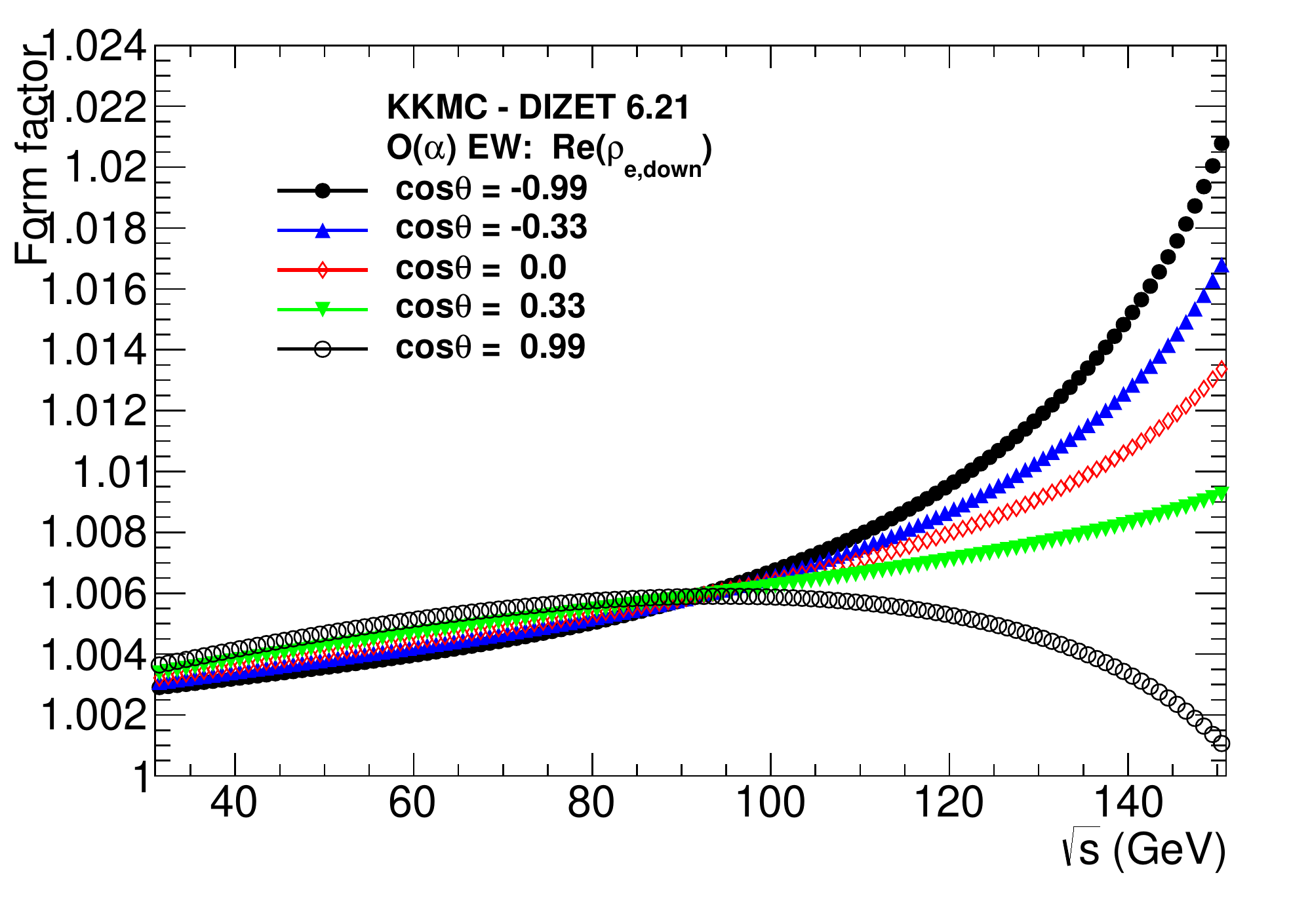}
  \includegraphics[width=7.5cm,angle=0]{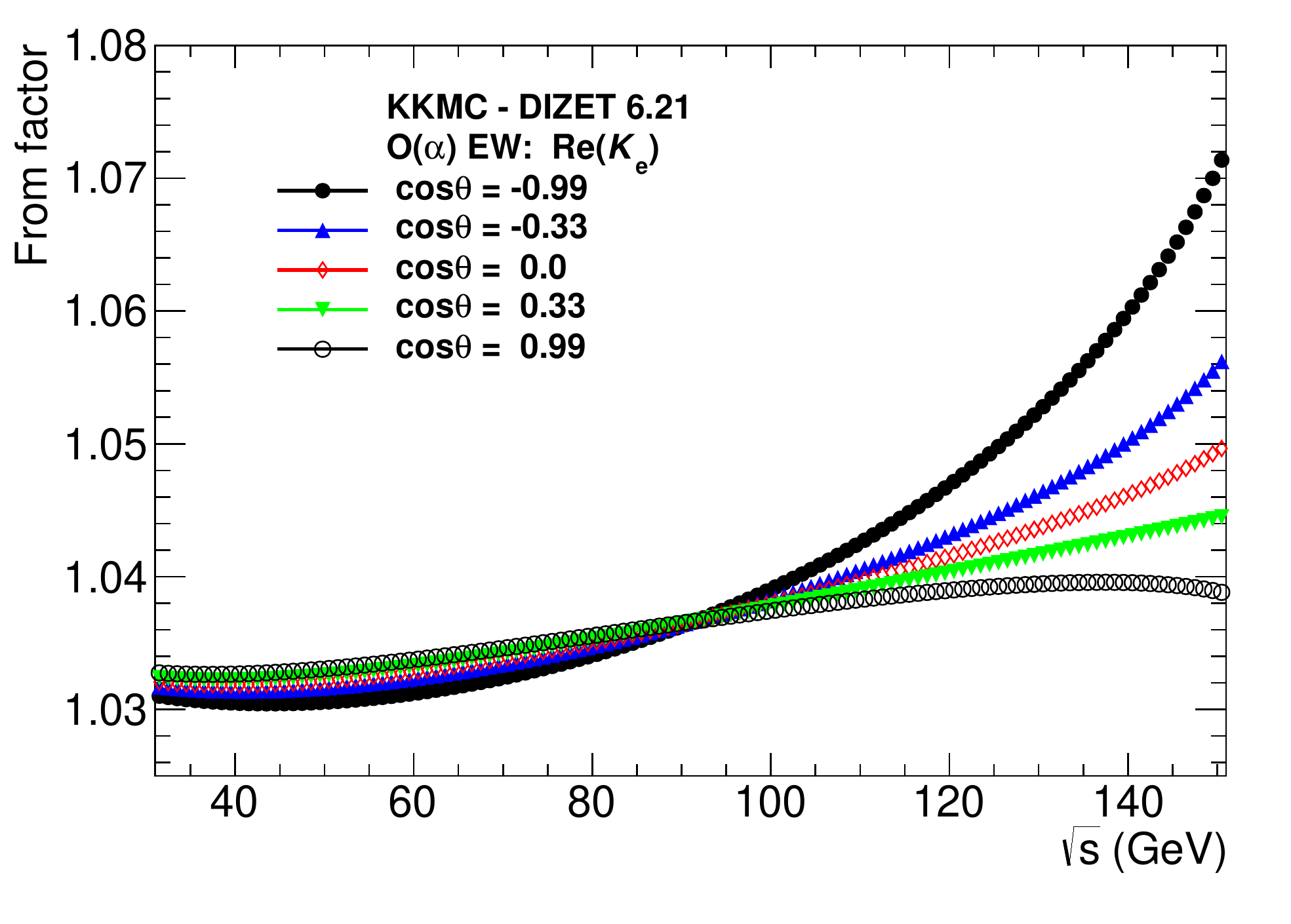}
  \includegraphics[width=7.5cm,angle=0]{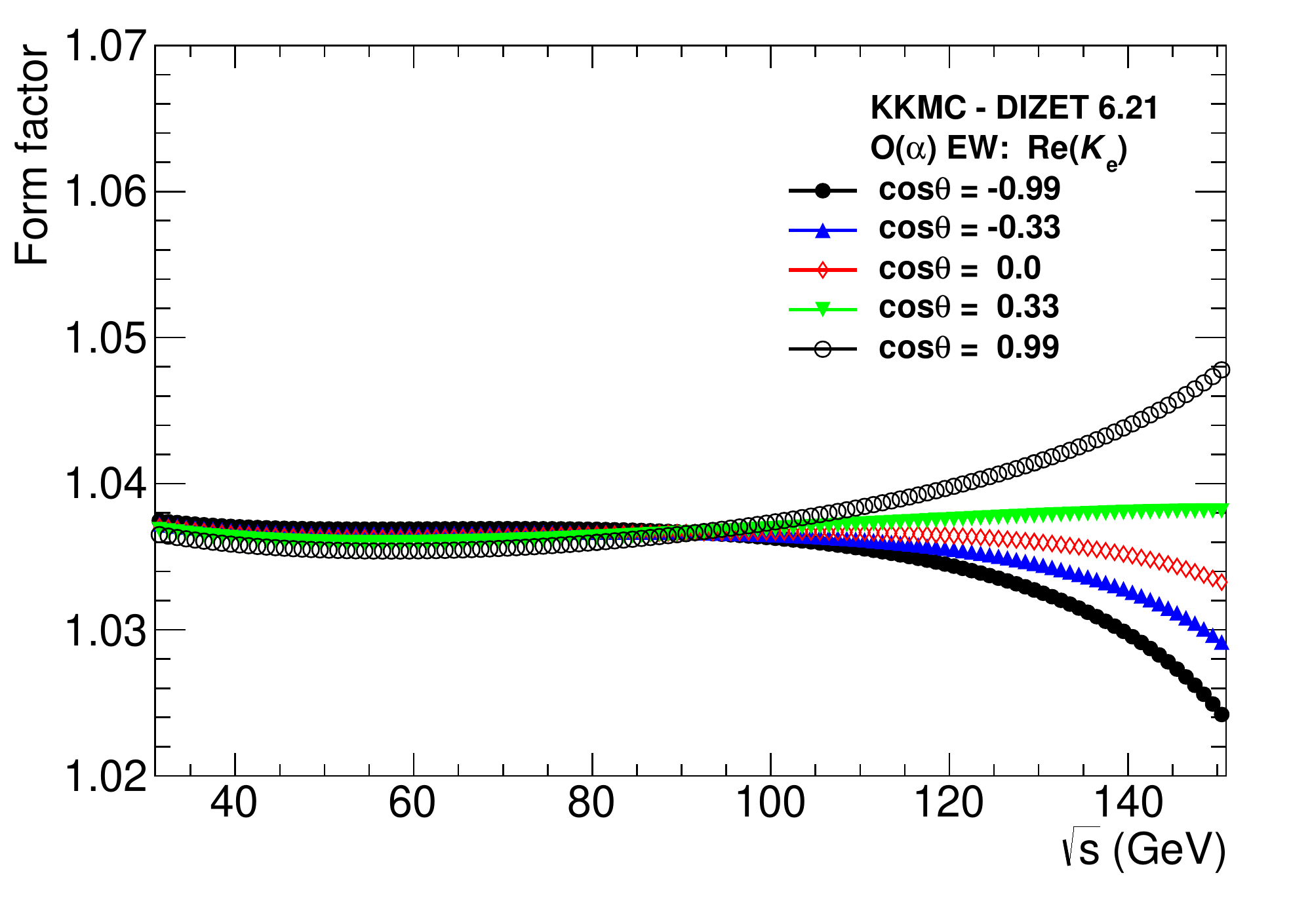}
  \includegraphics[width=7.5cm,angle=0]{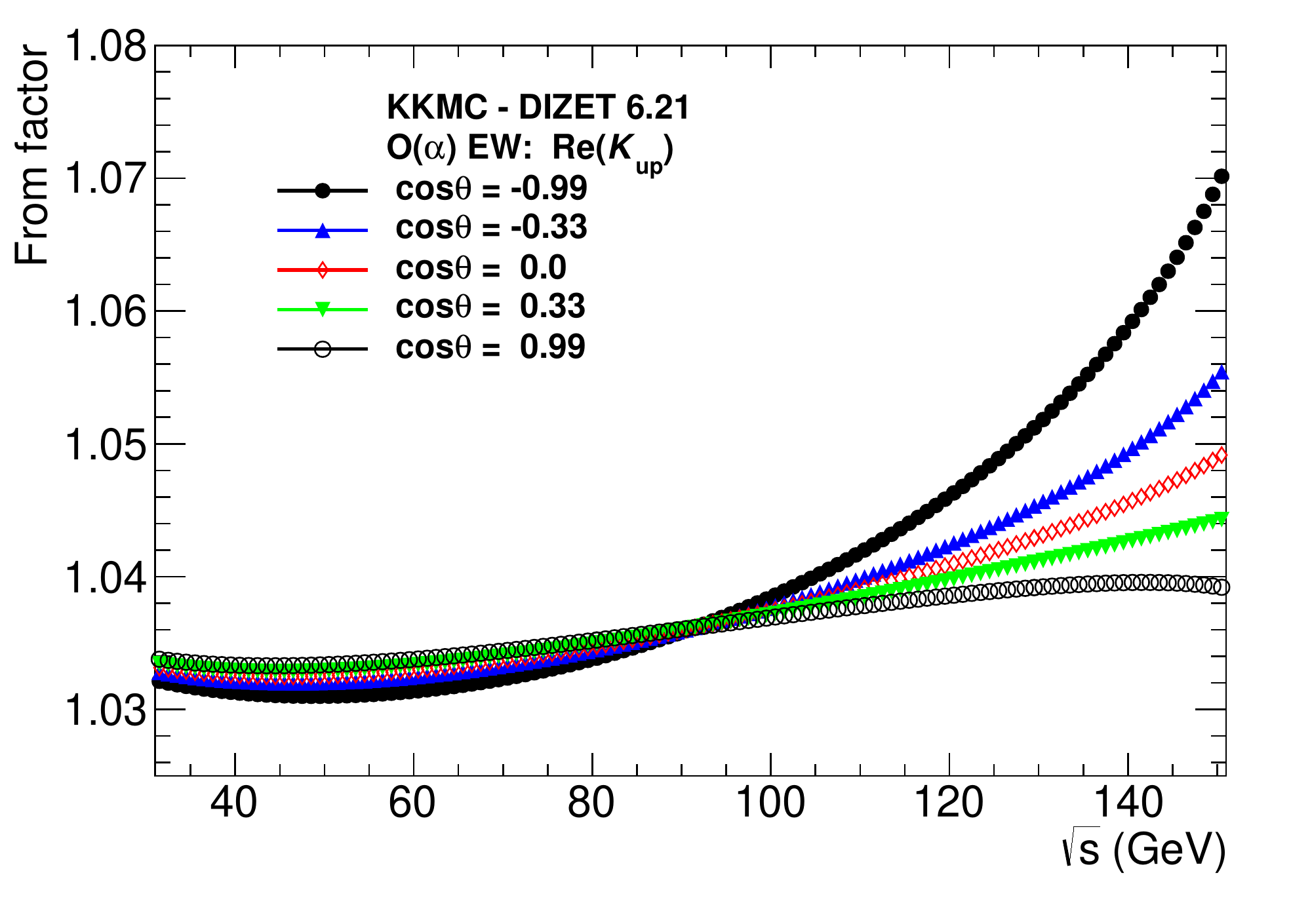}
  \includegraphics[width=7.5cm,angle=0]{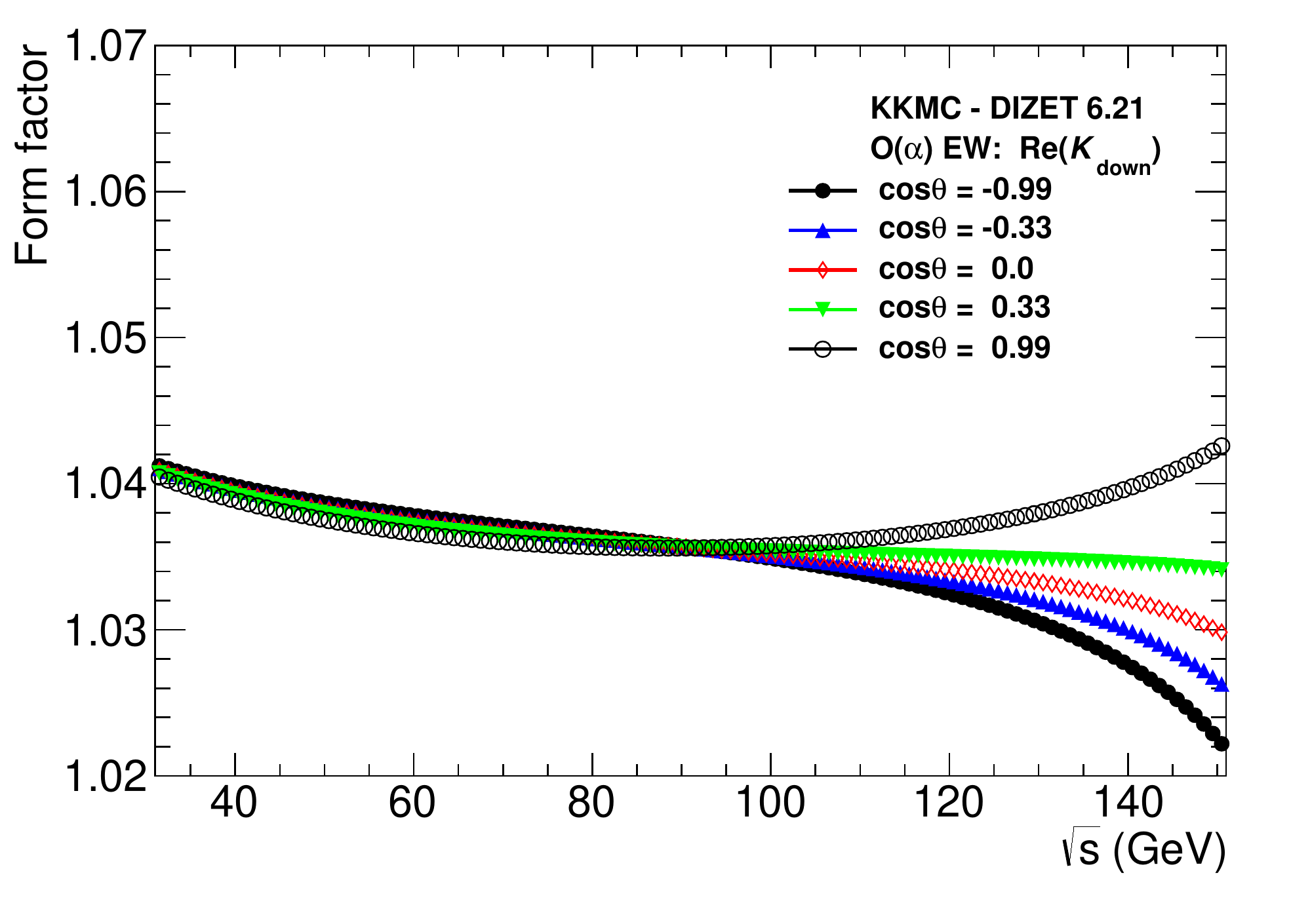}
  \includegraphics[width=7.5cm,angle=0]{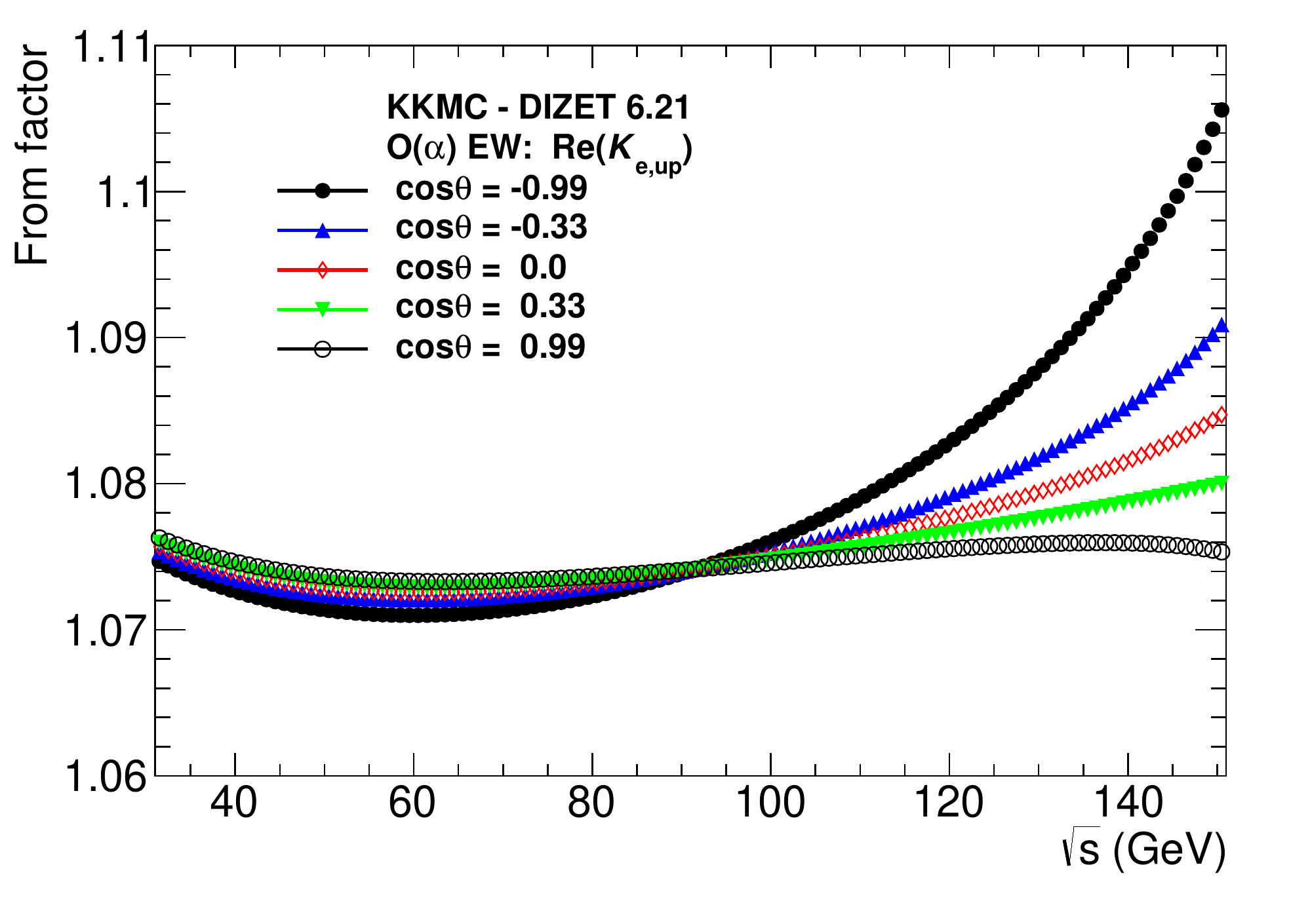}
  \includegraphics[width=7.5cm,angle=0]{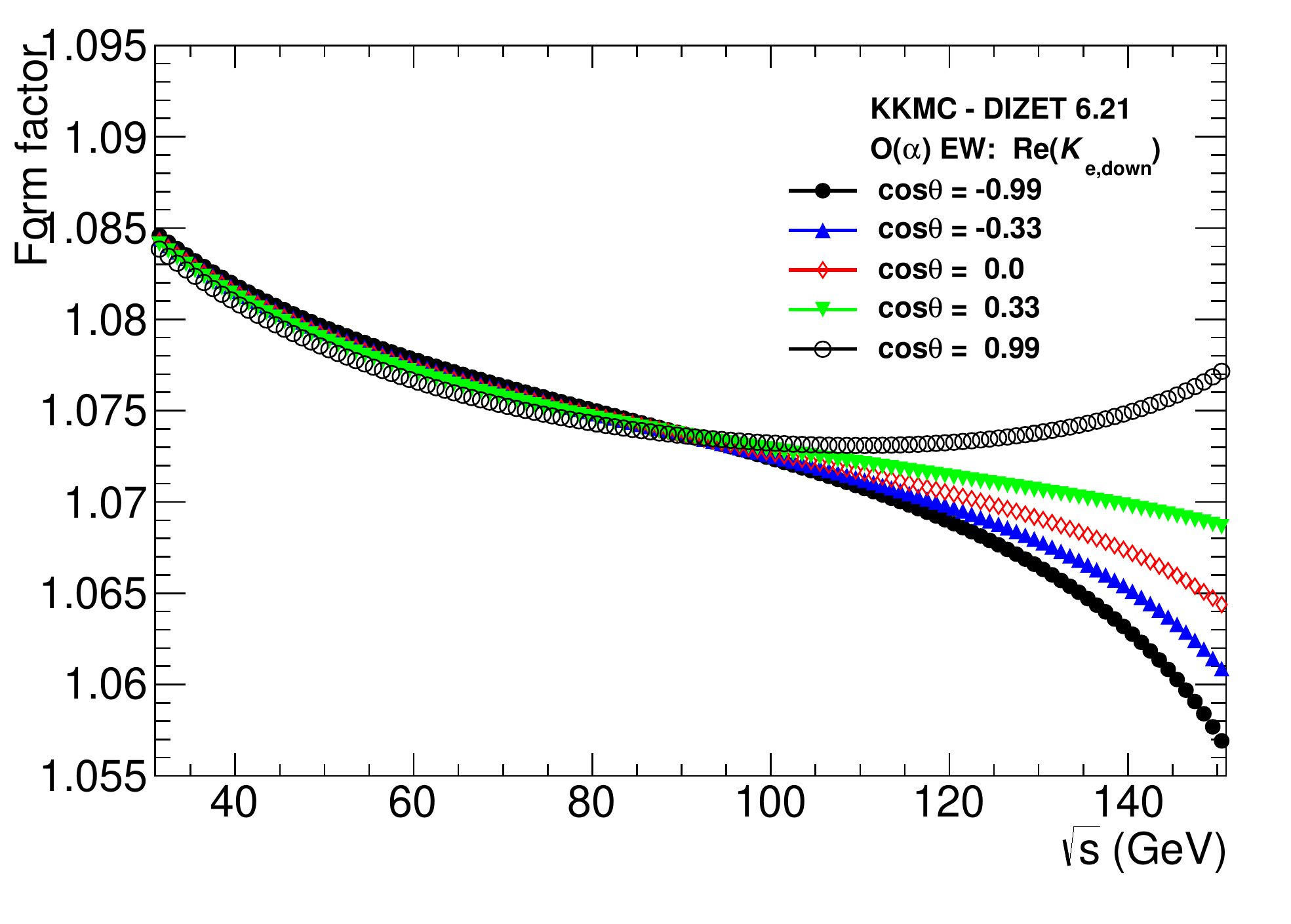}
}
\end{center}
  \caption{Real parts of the
 $\rho_{e, up}$, ${\mathscr K}_{e}$, ${\mathscr K}_{up}$ and ${\mathscr K}_{e, up}$
    EW form-factors of $q \bar q \to Z \to ee$ process,   
    as a function of $\sqrt{s}$ and for the few values of $\cos \theta$.
    For the up-type quark flavour, left side
    plots are collected and for the down-type the right side plots.
    Note, that ${\mathscr K_e}$ depends on the flavour of incoming quarks.
\label{Fig:Re_Rho_K_b_wide} }
\end{figure*}

Real parts of the
$\rho_{\ell f}(s,t)$, ${\mathscr K}_f(s,t)$, ${\mathscr K}_{\ell}(s,t)$, ${\mathscr K}_{\ell f}(s,t)$
EW form-factors are shown in Figure~\ref{Fig:Re_Rho_K_b_wide}
for a few values of $\cos \theta$, the
angle between directions of the incoming quark and the outgoing lepton, calculated  in the outgoing
lepton pair centre-of-mass frame.
Eq.~(\ref{Mandelstam}) relates Mandelstam variables $(s,t)$ to the invariant
mass and  $\cos \theta$. The  $\cos \theta$ dependence of the box correction
is more
sizable for the up-quarks. 

Note, that at the $Z$-boson peak, Born-like couplings are only weakly modified;
form-factors are close to 1 and of no numerically
significant angular dependence.
At lower virtualities corrections are relatively larger because the $Z$-boson contributions are non resonant
and thus smaller. In this phase-space region the $Z$-boson is itself dominated 
by the virtual photon contribution. 
Above the peak, the  $WW$  and later also $ZZ$ boxes contributions become sizable,
the dependence on the $\cos \theta$  appears; contributions become gradually doubly resonant and sizable.

\subsection{Running $\alpha(s)$}

Fermionic loop insertion of the photon propagator, i.e. vacuum polarization corrections,  are summed together as a multiplicative factor $\Gamma_{V_{\Pi}}$,
Eq.~(\ref{Eq:ChiGamma}), for the photon exchange in Eq.~(\ref{Eq:BornEW}). But
it can be interpreted as  the {\it running QED coupling}:
\begin{equation}
  \label{Eq:runningalpha}
  \alpha(s) = \frac{\alpha(0)}{1 - \Delta \alpha_h^{(5)}(s) - \Delta \alpha_{\ell}(s) - \Delta \alpha_t(s) - \Delta \alpha^{\alpha \alpha_s}(s)}.
\end{equation}

The hadronic contribution at $M_Z$  is a significant~\cite{Bardin:1999yd} correction: $ \Delta \alpha_h^{(5)}(M_Z^2)$ = 0.0280398.
It is calculated in the five flavour scheme with use of dispersion relation
and  input from low energy experiments. We will continue to use 
LEP times parametrization, while the most recent measured  $ \Delta \alpha_h^{(5)}(M_Z^2)$ = 0.02753 $\pm$ 0.00009 ~\cite{Davier:2017zfy}.
The changed value modifies predicted form-factors, in particular the effective leptonic mixing angle \\
$\sin^2\theta_{eff}^{lep}(M_Z^2)=Re{({\mathscr K}_l(M_Z^2))} s^2_W$ is shifted
by almost \\ $20 \cdot 10^{-5}$ closer to the measured  LEP value.
This is not included in the numerical results presented  as we consistently
remain with the defaults used  in {\tt KKMC}.

The leptonic loop contribution $\Delta \alpha_{\ell}(s)$ is calculated analytically up to the 3-loops,
and is a
comparably significant correction, $ \Delta \alpha_{\ell}(M_Z^2)$ = 0.0314976.
The other contributions are very small.
%The top contribution depends on the mass of the top quark, and for $m_t = 173.8$ GeV
% $ \Delta \alpha_t(s) = -0.585844 \cdot 10^{-4}$. The mixed two-loop $O(\alpha \alpha_s)$ corrections arising from $t \bar t$ loops
%with gluon, for the same top-quark mass and $\alpha_s = 0.119$ gives  $ \Delta \alpha^{\alpha \alpha_s}(M_Z^2) = -0.103962 \cdot 10^{-4}$.

\begin{figure}
  \begin{center}                               
{
   \includegraphics[width=7.5cm,angle=0]{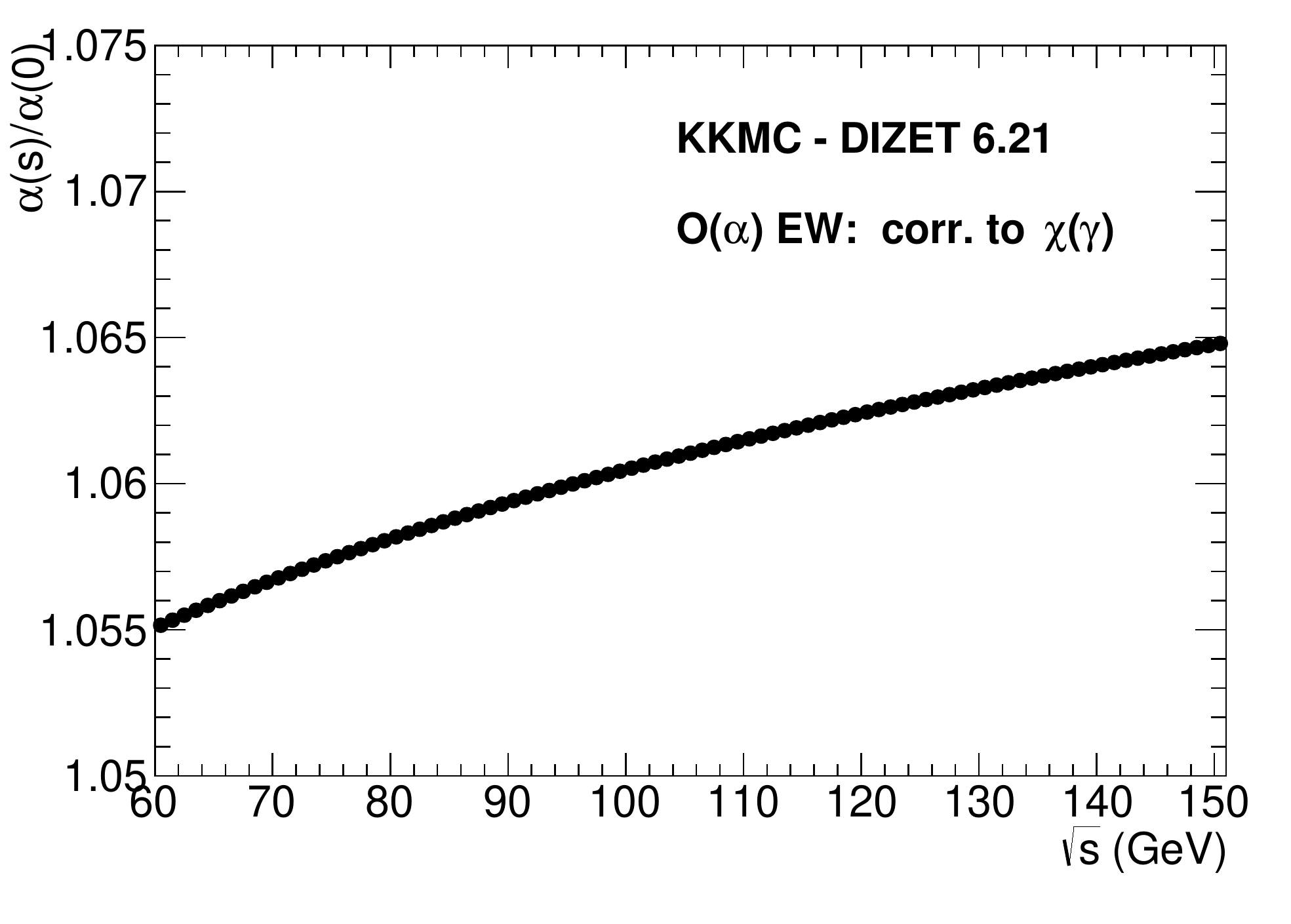}
}
\end{center}
  \caption{The  vacuum polarization ($\alpha(s)/\alpha(0)$) correction of $\gamma$ propagator,
    Eq.~(\ref{Eq:runningalpha}).
%    as a function of~$\sqrt{s}$. 
\label{Fig:ChiGamma} }
\end{figure}

Fig.~\ref{Fig:ChiGamma} shows the vacuum polarization corrections to the $\chi_{\gamma}(s)$ propagator, directly representing  the ratio $\alpha(s)/\alpha(0)$ of Eq.~(\ref{Eq:runningalpha}).

 \section{EW input schemes and Effective Born}
 \label{sec:EWschemes}

 Formally, at the lowest EW order, only three independent parameters can be set,
 other are calculated following the structure of $ SU(2) \times U(1)$ group
 from Standard Model constraints. Formula~(\ref{Eq:SMrelation})
 represents one of such constraints. Following report ~\cite{Alioli:2016fum},
the most common choices at hadron colliders
are:  {\it $G_{\mu}$ scheme}  $(G_{\mu}, M_Z, M_W)$
and {\it $\alpha(0)$ scheme} $(\alpha(0)$, $M_Z, M_W)$. There exists by now a family of different modifications of the
{\it $G_{\mu}$ scheme}, see discussion in~\cite{Alioli:2016fum}, and they are considered as preferred schemes
for  hadron collider physics\footnote{%
The Monte Carlo generators usually allow user to define set of input parameters $(\alpha, M_Z, M_W)$,
$(\alpha, M_Z, G_{\mu})$ or $(\alpha, M_Z, s_W^2)$. However,  within this flexibility, formally multiplicative factor  $\chi_Z(s)$
in the $Z$-boson propagator, see formula (\ref{Eq:Zprob}), is always kept to be equal to 1: 
\begin{equation}
  \label{Eq:StandardModelRel}
  \frac{G_{\mu} \cdot M_{z}^2  \cdot \Delta^2 }{\sqrt{2} \cdot 8 \pi \cdot \alpha} = 1,%\ \   \ \  \Delta^2 = 16 \cdot c_W^2 \cdot s_W^2.
\end{equation}
where $\Delta$ is given by Eq.~(\ref{Eq:Delta}).
The multiplicative factor of (\ref{Eq:StandardModelRel}) in the definition of
$\chi_Z(s)$ is quite often absent in the programs code. With the choice of
primary parameters, the others are adjusted to match the constraint Eq.~(\ref{Eq:StandardModelRel}),
regardless if they fall outside their measurement uncertainty window or not.}.

Let us recall, that the calculations of EW corrections available in {\tt Dizet}  work with
a variant
of the {\it $\alpha(0)$ scheme}. It is defined by the  input parameters $(\alpha(0), G_{\mu}, M_Z)$.
Then $M_W$ is calculated iteratively from formula~(\ref{Eq:MWiterative}) and $s^2_W$ of Eq.~(\ref{Eq:sw2onshell}) uses that value of $M_W$.
This formally brings it beyond EW LO scheme.
The numerical value of $s_W^2$  calculated from (\ref{Eq:sw2onshell}) does not fulfill the EW LO relation (\ref{Eq:SMrelation})
 anymore. 

 At this point we introduce two options for the {\it Effective Born} spin amplitudes
 parametrization,
  which works well for parametrizing  EW corrections near the $Z$-pole and
 denote them respectively as {\it LEP} and {\it LEP with improved norm.}:

\begin{itemize}
\item
The {\it LEP} parametrization  uses formula (\ref{Eq:BornEW}) for spin amplitude but with
$\alpha (s) = \alpha (M_Z^2) = 1./128.8667$,  $s^2_W = \sin^2\theta_W^{eff} (M_Z^2) = 0.23152$,
i.e. as measured at the $Z$-pole and reported in~\cite{Olive:2016xmw}. 
All form-factors are set to 1.0.
\item
  The {\it LEP with improved norm.} parametrization also uses formula (\ref{Eq:BornEW})
  for spin amplitude with parameters
   set as for {\it LEP} parametrization. All form-factors are set to  1,
  but  $\rho_{\ell f} = 1.005$.
  This  corresponds to the measured  $\rho(M_Z^2)$ = 1.005, as reported in~\cite{Olive:2016xmw}. 
\end{itemize}

Table~\ref{Tab:BornEff} collects initialization constants
 of  EW schemes relevant for our discussion.
We specify parameters which enter formula (\ref{Eq:BornEW})  for
Born spin amplitudes used for: (i) actual MC events generation~\footnote{The EW LO initialization is  consistent
with PDG  $\sin^2\theta_{eff}^{lep}$    = 0.23113, but  commonly used $G_{\mu}$ scheme,
($G_{\mu}=1.1663787 \cdot 10^{-5}$ GeV$^{-2}$, $M_Z$= 91.1876 GeV, $M_W$=80.385~GeV)
correspond to $s^2_W$ = 0.2228972.}, (ii) the  EW LO $\alpha(0)$ scheme,
(iii) effective Born ({\it LEP}) parametrization and (iv) effective Born  ({\it LEP with improved norm.}).
In each case parameters are chosen such that the SM relation, formula~(\ref{Eq:StandardModelRel}), is obeyed.

In the {\it Improved Born Approximation} complete $O(\alpha)$ EW corrections,
supplemented by selected higher order terms, are handled thanks to s-, t-dependent form-factors,
which multiply couplings
and propagators of the usual Born expressions.  Instead, the {\it Effective Born} absorbs the bulk of 
EW corrections into a redefinition of a few fixed parameters (i.e. couplings).

In the following, we will systematically compare predictions obtained with the  EW  corrections 
and those calculated with  {\it LEP} or  {\it LEP with improved norm.} approximations.
As we will see,
effective Born with {\it LEP with improved norm.} works very well around $Z$-pole both for
the line-shape and forward-backward asymmetry. 

\begin{table*}
 \vspace{2mm}
 \caption{The EW parameters used for: (i) MC events generation, (ii) the  EW LO $\alpha(0)$ scheme,
   (iii) effective Born spin amplitude around the $Z$-pole and (iv) effective Born with improved normalization. 
   In each case parameters are chosen such that the SM relation, formula~(\ref{Eq:StandardModelRel}), is obeyed.
   The $G_{\mu}$ = $1.166389 \cdot 10^{-5}$ GeV$^{-2}$, $M_Z$ = 91.1876 GeV and ${\mathscr K}_f, {\mathscr K}_e, {\mathscr K}_{\ell f}$ = 1 are taken.}
 \label{Tab:BornEff}
  \begin{center}
    \begin{tabular}{|l|l||l|l|}
      \hline\hline
        EW LO            &       EW LO                &     Effective Born       &   Effective Born                    \\
        MC generator     &  $\alpha(0)$ scheme            &     {\it LEP}          &   {\it LEP with improved norm.}   \\
        \hline\hline
      $\alpha$  = 1/128.8886     &  $\alpha$  = 1/137.3599     &      $\alpha$   = 1/128.8667       &   $\alpha$   = 1/128.8667     \\
      $s^2_W$    = 0.23113       &   $s^2_W$    = 0.21215       &      $s^2_W$    = 0.23152          &   $s^2_W$    = 0.23152        \\
      $\rho_{\ell f}$  = 1.0      &  $\rho_{\ell f}$  = 1.0        &      $\rho_{\ell f}$  = 1.0         &   $\rho_{\ell f}$  = 1.005     \\
      \hline 
    \end{tabular}
  \end{center}
  %\end{sidewaystable}
\end{table*}

%\clearpage
\section{Born kinematic approximation and  $p p$ scattering}
\label{sec:IBAaveraging}

The solution to define Born-like parton level kinematics for $p p$ scattering
process is encoded  in the 
{\tt TauSpinner} package~\cite{Przedzinski:2018ett}. It does not
exploit
hard-process, so-called history entries
which only sometimes are stored for
the generated events.
In particular, the flavour and momenta  of the incoming partons
have to be  emulated  from the kinematics of 
final states and incoming protons momenta. 
Probabilities calculated from parton level cross-sections and PDFs
weight all possible contributions.
Let us now recall briefly principles and choices for optimization.

 \subsection{Average over incoming partons flavour}

The parton level Born cross-section
$\sigma^{q \bar q}_{Born}(\hat s, \cos\theta)$ has to be convoluted with the structure functions, and sum\-med over all possible
flavours of incoming partons and all possible helicity states of outgoing leptons. The lowest order formula\footnote{Valid for the ultra-relativistic leptons.} is given below

\begin{eqnarray}
  \label{Eq:Bornpp}
  d\sigma_{Born}&&( x_1, x_2, \hat s, \cos \theta) =\sum_{q_f, \bar q_f} \nonumber\\ 
   &[& f^{q_f}(x_1,...)f^{\bar q_f}(x_2,...)
  d\sigma^{q_f \bar q_f}_{Born}( \hat s, \cos \theta) \\
  &+& \ f^{\bar q_f}(x_1,...)f^{ q_f}(x_2,...)
  d\sigma^{ q_f \bar q_f}_{Born}( \hat s, -\cos \theta) ] \nonumber,
\end{eqnarray}
where $x_1$, $x_2$ denote fractions of incoming protons momenta
carried by the corresponding parton,
$\hat s = x_1\  x_2\ s $ and $f/\bar f$ denotes parton (quark-/anti-quark) density functions. We assume that kinematics is reconstructed
from four-momenta of the outgoing leptons.
The  incoming quark and anti-quark  may come respectively either from the first and second proton or
reversely from the second and first. Both possibilities
are taken into account\footnote{ One should mention photon induced  contributions. They are of the same coupling order 
   as electroweak corrections.
   For production of the lepton pairs in $pp$ collisions, contributions 
   were  evaluated e.g. in \cite{Manohar:2016nzj}.
  
  In general, for the calculation of {\tt TauSpinner} weights, sum over partons is not restricted as  in eq. (\ref{Eq:Bornpp})
  to the quarks and anti-quarks only.  Gluon PDF's are used   when weight calculation with matrix
  elements for lepton pair with two jets in final state is used \cite{Bahmani:2017wbm}. The  $\gamma \gamma \to l^+ l^-$
  contributions can be then
  taken into account as  a part of the $2 \to 4$ matrix elements.
  
  Photon induced processes are however usually generated and stored separately.
  That is why our reweighting algorithm for EW corrections 
  does not need to take such (rather small) contributions into account
   in eq. (\ref{Eq:Bornpp}).
  }
by the two terms of~(\ref{Eq:Bornpp}).
The sign in front of $\cos\theta$,
the cosine of the scattering angle, is negative for the second term.
Then the parton of the first incoming proton which carries
$x_1$
and follows the direction of the $z$-axis
 is an anti-quark, not a quark.
%orientation being the one of the parton carrying $x_1$.
The formula is used  for calculating the differential cross-section $d\sigma_{Born}( x_1, x_2, \hat s, \cos \theta)$ 
of each analyzed event, regardless if its  kinematics and flavours of incoming partons  may be
available from the event history entries or not. The formula can be used to a good approximation in case of NLO QCD spin amplitudes. 
The momenta of outgoing leptons are used to construct {\it effective} kinematics of the Drell-Yan production
process and decay, without the  need  of information on parton-level
hard-process itself.
Born-like kinematics can be constructed, as we will see later, even 
for events of quark-gluon or gluon-gluon parton level collisions
(as inspected for test in the event history  entries) too.

 \subsection{Effective beams kinematics}

 The $x_1,  x_2$ are calculated from the kinematics of outgoing leptons, following formulae of~\cite{Davidson:2010rw}
 \begin{equation}
   x_{1,2} =  \frac{1}{2}\ {\Big (}\  \pm \frac{p_z^{ll}}{ E} + \sqrt{ (\frac{p_z^{ll}}{ E})^2 +  \frac{m^2_{ll}}{ E^2}} \; \; {\Big )} , 
 \end{equation}
 where $E$ denotes energy of the proton beam and $p_z^{\ell \ell}$ denotes $z$-axis momentum of outgoing lepton pair
 in the laboratory frame and $m_{ll}$ lepton pair virtuality. Note that this formula can be used, as approximation, for the events with hard jets too.

 \subsection{Definition of the polar angle}
 \label{Sec:polarAngle}

 For the polar angle $\cos \theta$, of factorized Born level $q \bar q \to Z \to \ell \ell$ process,
 weighted average 
 of the outgoing leptons angles with respect to the beams' directions, denoted as  $\cos \theta^*$, was
 used. In ~\cite{Was:1989ce} it was found helpful to compensate the effect of initial
 state hard bremsstrahlung photons  of $e^+e^- \to Z n\gamma$, $Z  \to \ell \ell m\gamma $, where $m,\; n$ denote the number of accompanying photons.
 Extension to $pp$ collisions required to take both options in Eq.~(\ref{Eq:Bornpp})  into account;
 when the $z$-axis is parallel- and anti-parallel
 to the incoming quark.

 For the
further calculation, boost of all four-momenta (also of incoming beams)
into the rest frame of the lepton pair need to be  performed.
The  $\cos \theta^{*}$  is then calculated from
 \begin{eqnarray}   
   \cos \theta_1 = \frac{\tau_x^{(1)} b_x^{(1)} + \tau_y^{(1)} b_y^{(1)} + \tau_z^{(1)} b_z^{(1)}}
                         { | \vec \tau^{(1)}| |\vec b^{(1)}|},\nonumber\\     
    \cos \theta_2 = \frac{\tau_x^{(2)} b_x^{(2)} + \tau_y^{(2)} b_y^{(2)} + \tau_z^{(2)} b_z^{(2)}}
         { | \vec \tau^{(2)}| |\vec b^{(2)}|},
 \end{eqnarray}
  as follows:
 \begin{equation}   
   \cos \theta^*   =  \frac{\cos\theta_1 \sin\theta_2 + \cos\theta_2 \sin\theta_1}{\sin\theta_1 + \sin\theta_2}
   \label{Eq:costhetaStar}
 \end{equation}  
 where $\vec \tau^{(1)}, \vec \tau^{(2)}$ denote 3-vectors of outgoing leptons and  $\vec b^{(1)}, \vec b^{(2)}$ denote
 3-vectors of incoming beams' four-mo\-menta.
 %The sign is reversed  for the  second  term of~(\ref{Eq:Bornpp}) is calculated.
% All these 3-vectors are taken from lepton pair centre-of-mass system.

 The polar angle definition,  Eq.~(\ref{Eq:costhetaStar}), is at present
 the {\tt TauSpinner} default. For tests 
 we have  used variants;  {\it Mustraal}~\cite{Berends:1983mi} and  {\it Collins-Soper}~\cite{Collins:1977iv} frames, which differ  when high $p_T$ jets are present.
 We will return later to the frame choice, best suitable
  when NLO QCD corrections are included in the
 production process of generated events.

\section{QCD corrections and angular coefficients} \label{sec:QCDcorr}

For the Drell-Yan production~\cite{DrellYan70} one can separate QCD and EW components of the fully differential cross-section
and
describe the  $Z/\gamma^* \to \ell \ell$ sub-process with 
lepton angular ($\theta,  \phi$) dependence 
\begin{equation}
 \label{Eq:master1}
  \frac{ d\sigma}{dp_T^2 dY d\Omega} = 
     \Sigma_{ \alpha=1}^{9} g_{ \alpha}( \theta,  \phi)
     \frac{3}{ 16 \pi} \frac{d \sigma ^{\alpha }}{ dp_T^2 dY}, % 
\end{equation}
where the $ g_{ \alpha }( \theta,  \phi)$ denotes second order spherical harmonics,
multiplied by normalization constants and $d \sigma ^{\alpha }$ denotes helicity cross-sections, for
each of nine helicity configurations of $q \bar q \to Z/\gamma^* \to \ell \ell$.
The  polar and azimuthal ($\theta$ and $\phi$)  angles of
$d\Omega = d \cos\theta d\phi$ are defined
in  the $Z$-boson rest-frame. The $p_T$, $Y$ denote laboratory frame transverse momenta and rapidity 
of the intermediate  $Z/\gamma^*$-boson. 
Thanks  to th effort~\cite{Mirkes:1992hu,Mirkes:1994dp,Mirkes:1994eb}
from the  early 90's one expects such factorization to break with non-logarithmic ${\cal O}(\alpha_s^2) \sim 0.01$  QCD corrections\footnote{
  Also the impact of final state QED bremsstrahlung can be overcome with a proper definition of frames. The
  solution is available thanks to  Ref.~\cite{Berends:1983mi}.
  We use it with the 
 definition of frames $A$ and $A'$; Section 3.1 of \cite{Przedzinski:2018ett}. }
only.

There is some
flexibility for the $Z$-boson rest frame $z$-axis choice. 
The most common, so called {\it helicity frame},
is to take the  $Z$-boson  laboratory frame momentum.
For the {\it Collins-Soper} frame it is defined from directions of the two beams in the
$Z$-boson rest frame and is signed with the $Z$-boson $p_z$ laboratory frame sign. 

Eq.~(\ref{Eq:master1}) with  explicit spherical harmonics and coefficients
reads
\begin{eqnarray}
 \label{Eq:master2}
  \frac{ d\sigma}{dp_T^2 dY d \cos\theta d\phi} & = & \frac{3}{ 16 \pi} \frac{d \sigma ^{ U+ L }}{ dp_T^2 dY}  [  (1 + \cos^2\theta) \nonumber\\
  + 1/2\ A_0 (1 - 3 \cos^2\theta) &+&  A_1 \sin{2\theta}\cos\phi \\
  + 1/2\ A_2 \sin^2\theta \cos( 2 \phi)   &+&
  A_3 \sin \theta \cos \phi +   A_4 \cos \theta \nonumber\\
  + A_5 \sin^2 \theta \sin( 2 \phi) + &A_6& \sin{2\theta} \sin \phi 
  + A_7 \sin \theta\ \sin \phi ], \nonumber 
\end{eqnarray}
where $d \sigma ^{ U+ L }$ denotes the unpolarised differential cross-section 
(notation used in several papers of the 80's). The coefficients $A_i(p_T, Y)$ are related to ratios of definite intermediate state helicity   contributions
to the $d \sigma ^{ U+ L }$
cross-sections. 
The first  term   of the polynomial expansion
is $ (1 + \cos^2\theta)$ because  intermediate boson is of the spin 1.

The dynamics of the production process is  hidden in the angular 
coefficients $A_i (p_T, Y)$.  In particular, all the hadronic physics is described 
implicitly by the angular coefficients and it decouples from the well understood leptonic and intermediate boson physics. 

For the present paper, of particular interest are coupling constants
present in  coefficients $A_i$ of  Eq.~(\ref{Eq:master2})
representing ratios of the so-called helicity cross sections~\cite{Mirkes:1992hu,Mirkes:1994dp,Mirkes:1994eb}:
\begin{eqnarray}
  \label{Eq:Aicoupl}
\sigma ^{U+L }  & \sim & (v_{\ell}^2 + a_{\ell}^2)(v_{q}^2 + a_{q}^2), \nonumber \\ 
A_0, A_1, A_2  & \sim & 1 , \nonumber \\
A_3, A_4       & \sim & \frac{ v_{\ell} a_{\ell} v_q a_q}{(v_{\ell}^2 + a_{\ell}^2)(v_{q}^2 + a_{q}^2)} ,  \\ 
A_5, A_6       & \sim & \frac{(v_{\ell}^2 + a_{\ell}^2) ( v_q  a_q)}{(v_{\ell}^2 + a_{\ell}^2)(v_{q}^2 + a_{q}^2)}, \nonumber  \\ 
A_7            & \sim & \frac{ v_{\ell} a_{\ell} ( v_q^2 + a_q^2)}{(v_{\ell}^2 + a_{\ell}^2)(v_{q}^2 + a_{q}^2)}. \nonumber 
\end{eqnarray}

%Thanks to the orthonormality of the formula~(\ref{Eq:master2}) polynomials,
Integration\footnote{One can easily check that  
 $A_{FB}$ of Eq.~(\ref{Eq:AFB}) equals to $\frac{3}{8} A_4$. }
 over the azimuthal angle
$\phi$ redu\-ces Eq.~(\ref{Eq:master2}) to
%in the
%full phase-space of the outgoing leptons, Eq.~(\ref{Eq:master2}) reduces to  
\begin{eqnarray}
 \label{Eq:master3}
  \frac{ d\sigma}{dp_T^2 dY d \cos\theta}  =  \frac{3}{ 8 \pi} \frac{d \sigma ^{ U+ L }}{ dp_T^2 dY}  
                      [  (1 + \cos^2\theta)\nonumber \\+ 1/2\ A_0 (1 - 3 \cos^2\theta) +   A_4 \cos \theta ].
\end{eqnarray}

Both Eqs.~(\ref{Eq:master2}) and~(\ref{Eq:master3}) are valid in any rest frame of the outgoing lepton pairs,
however  the $A_i(p_T, Y)$ are frame dependent.
The {\it Collins-Soper} frame is the most convenient and usual choice  
for the analyses dedicated to QCD dynamics.
In this frame, in the  low $p_T$ limit,  $A_4$ is the only
non-zero coefficient. It carries
direct information on the EW couplings, as can be concluded from
formulae~(\ref{Eq:Aicoupl}). All other coefficients depart
from zero with increasing $p_T$ while at the same time $A_4$ gradually decreases.

Due to different transfer dependence of the  $Z$ and $\gamma^*$ propagators,
the $A_i$
vary with  $m_{ll}$.
%In case of sizable  $p_T$ and $Y$, orientation of the
%reference frame differ for the definition variants.
The  $A_i$ dependence on $(p_T,Y)$, expressing production dynamics, differ
with the frame definition  variants of distinct coordinate system orientations.
For the studies of EW couplings, it is convenient when the lepton-pair
rest-frame definition absorbs effects of
production dynamics
partly into the $z$-axis choice.
Then, those $A_i$ coefficients which are proportional to the product of
EW vector and axial couplings remain
non-zero over the full range of $p_T$.
 Promising for that purpose frame was developed at LEP times for the {\it Mustraal} Monte Carlo
program~\cite{Berends:1983mi}.
Recently, an extension of this  {\it Mustraal} frame,
for the case of hadron-hadron collisions, was introduced
and discussed in~\cite{Richter-Was:2016mal}. As shown in that paper, 
both {\it Collins-Soper} and  {\it Mustraal}  frames are equivalent in the $p_T = 0$ limit. Then  $A_4$ is the only non-zero   
coefficient for both frames and is  also numerically very close.
With increasing  $p_T$,  in the {\it Mustraal} frame 
$A_4$ remains as the only  sizably non-zero coefficient, while several 
$A_i$ coefficients depart from zero with the {\it Collins-Soper} frame.

In the collision of the same-charge protons the careful choice for the
$z$-axis orientation is necessary for the  $A_4$ coefficient to remain non-zero.
%is not straightforward even in the
%$p_T = 0$ limit. It requires to choose  $z$-axis orientation. 
%the frame prefers one of two possible orientations of the $z$-axis.
For the {\it Collins-Soper} frame, the $z$-axis follows the  direction of the
intermediate $Z$-boson in the laboratory frame. In case of the {\it Mustraal} frame
the choice of the sign is made stochastically using information of the system of leptons and outgoing accompanying visible jets. For
details see~\cite{Richter-Was:2016mal}, alternatively
the same sign choice  for the  $z$-axis  as in the {\it Collins-Soper} case, can be used.

\begin{figure*}
  \begin{center}                               
{
  \includegraphics[width=7.5cm,angle=0]{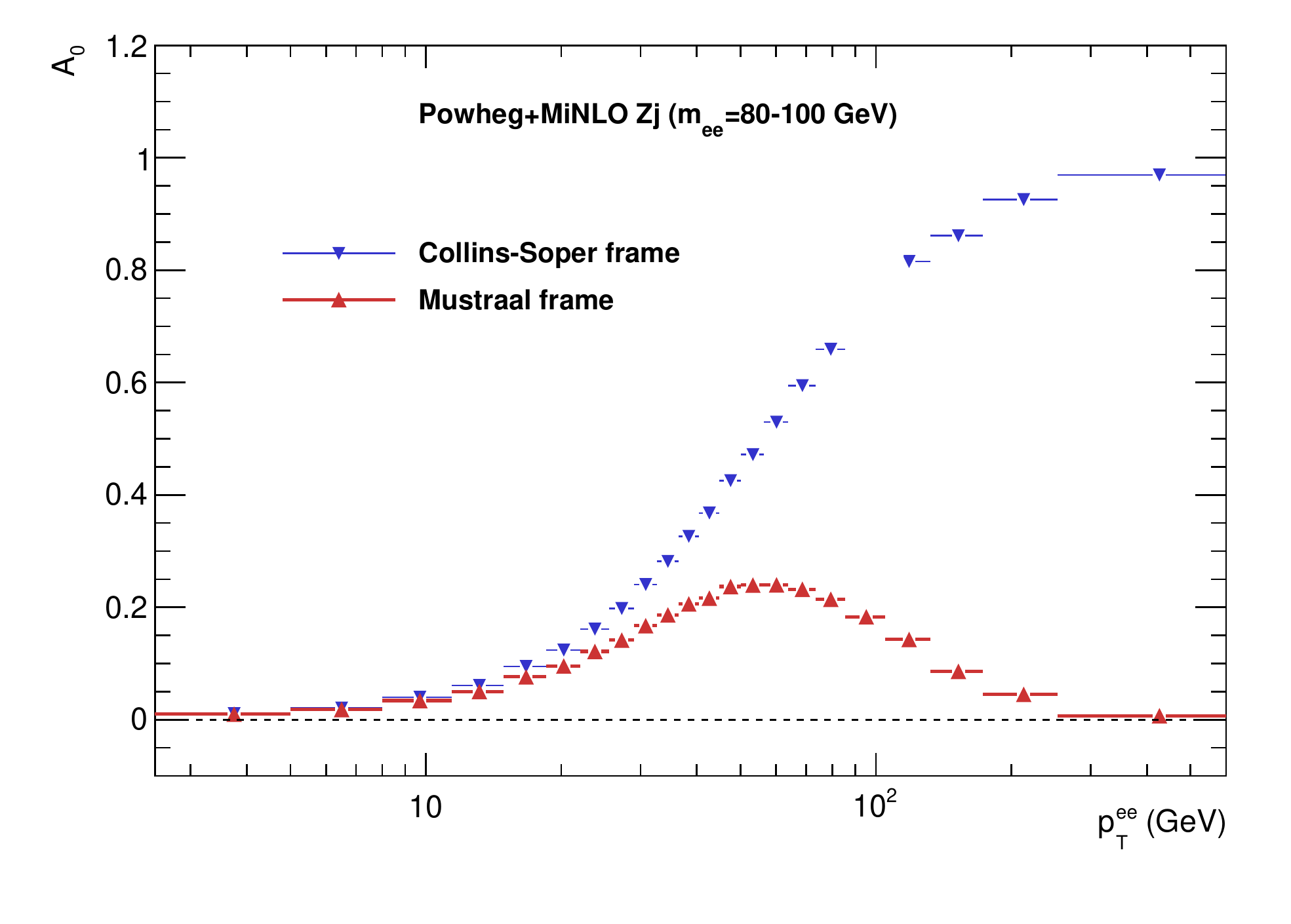}
  \includegraphics[width=7.5cm,angle=0]{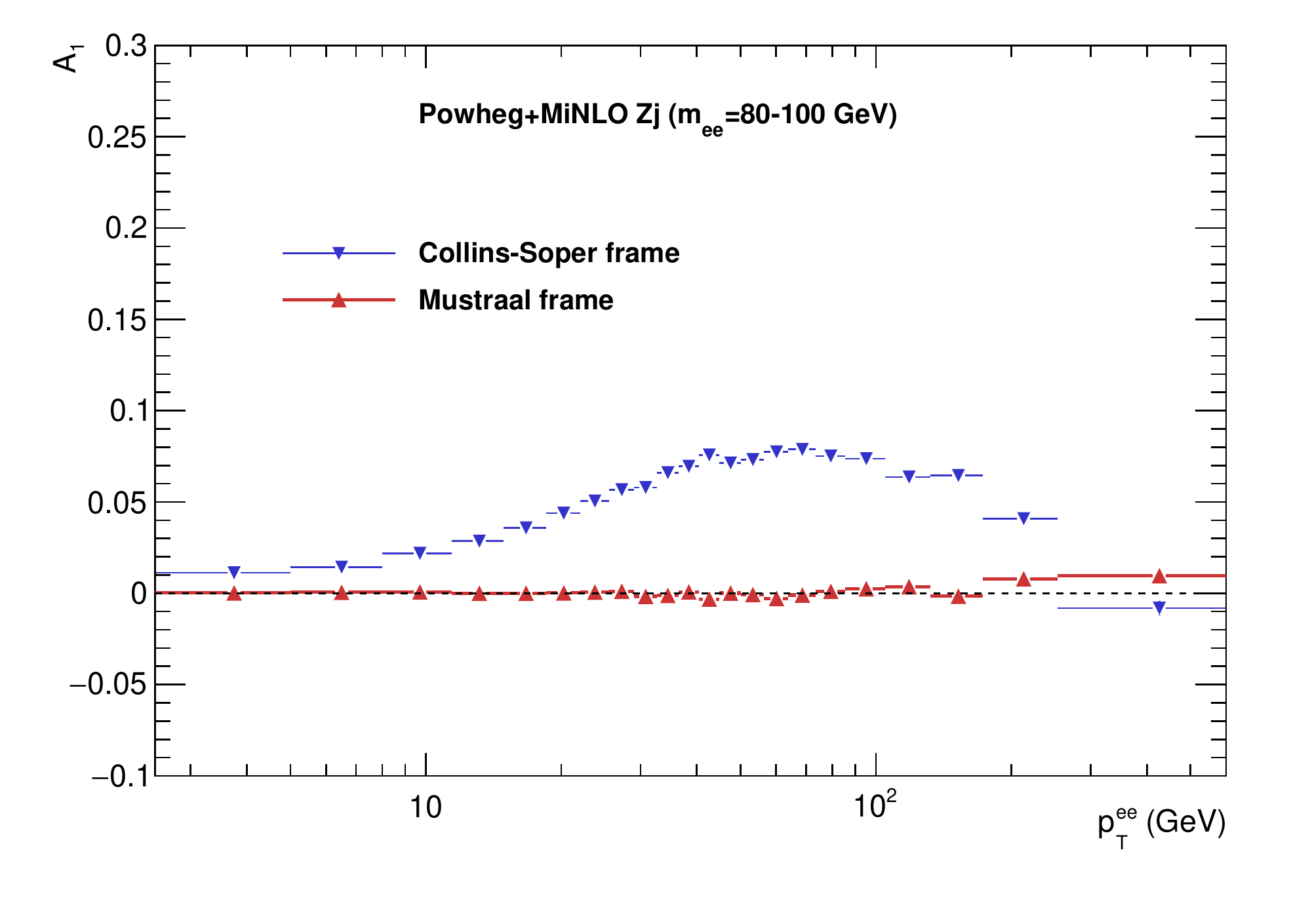}
  \includegraphics[width=7.5cm,angle=0]{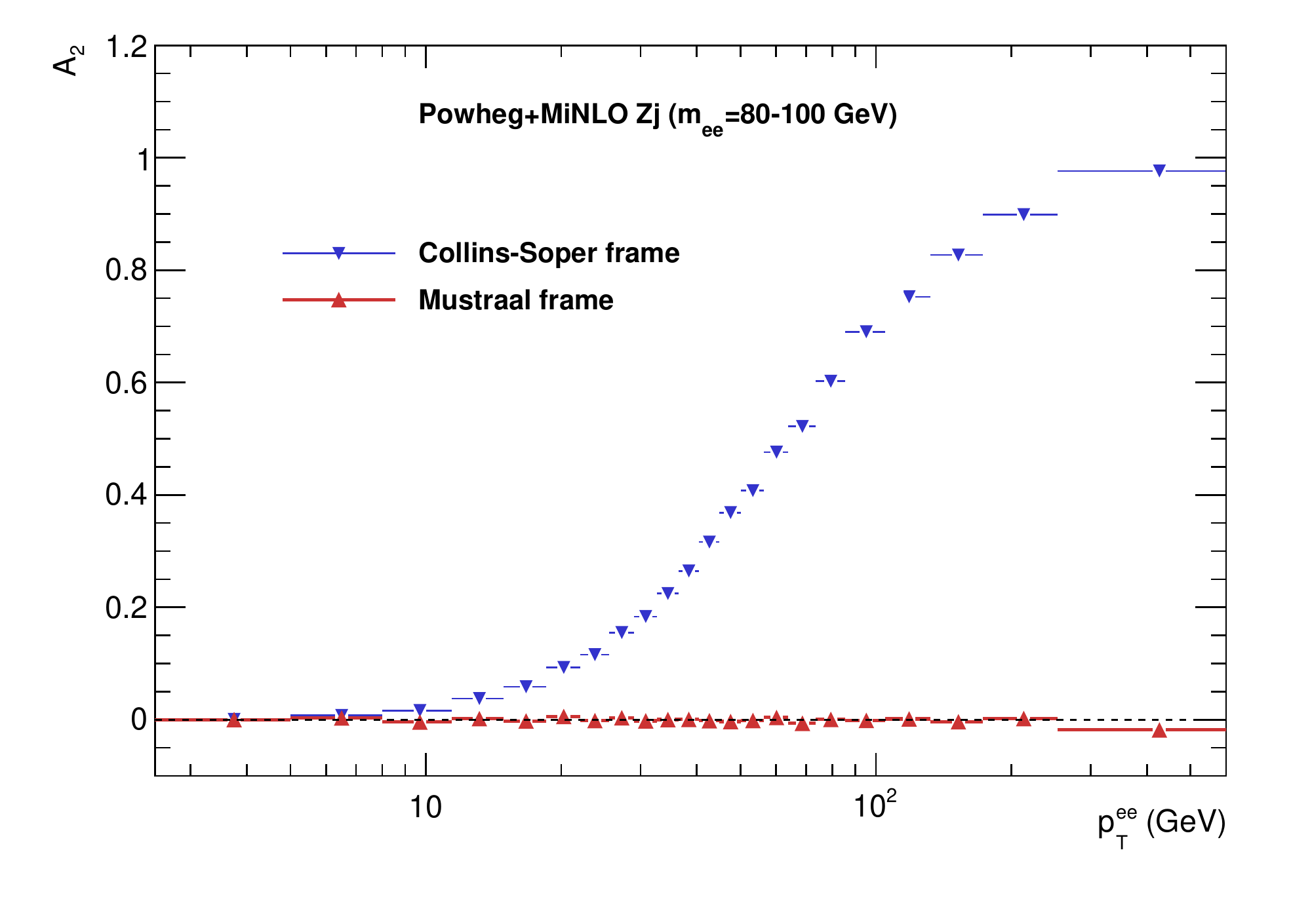}
  \includegraphics[width=7.5cm,angle=0]{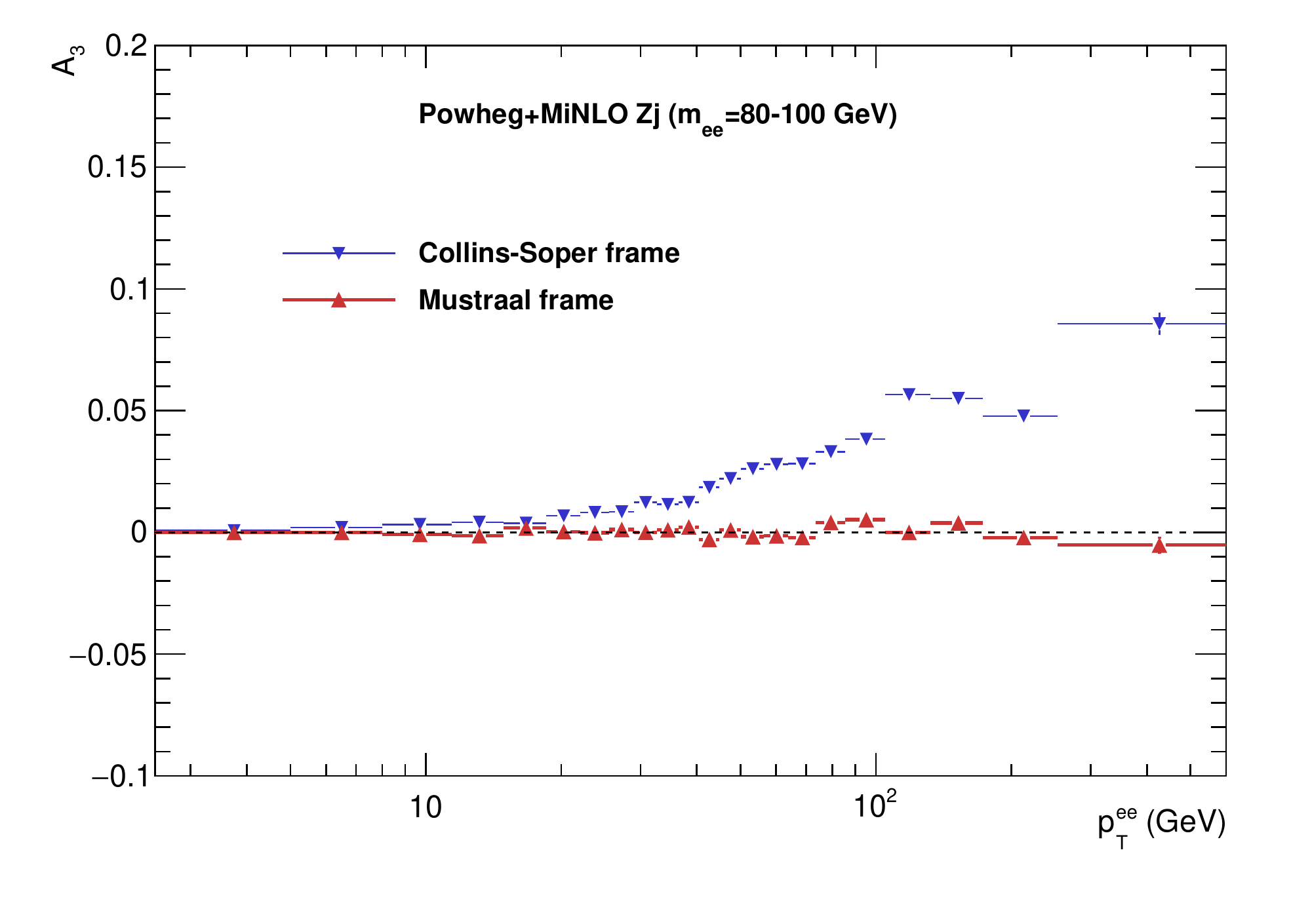}
  \includegraphics[width=7.5cm,angle=0]{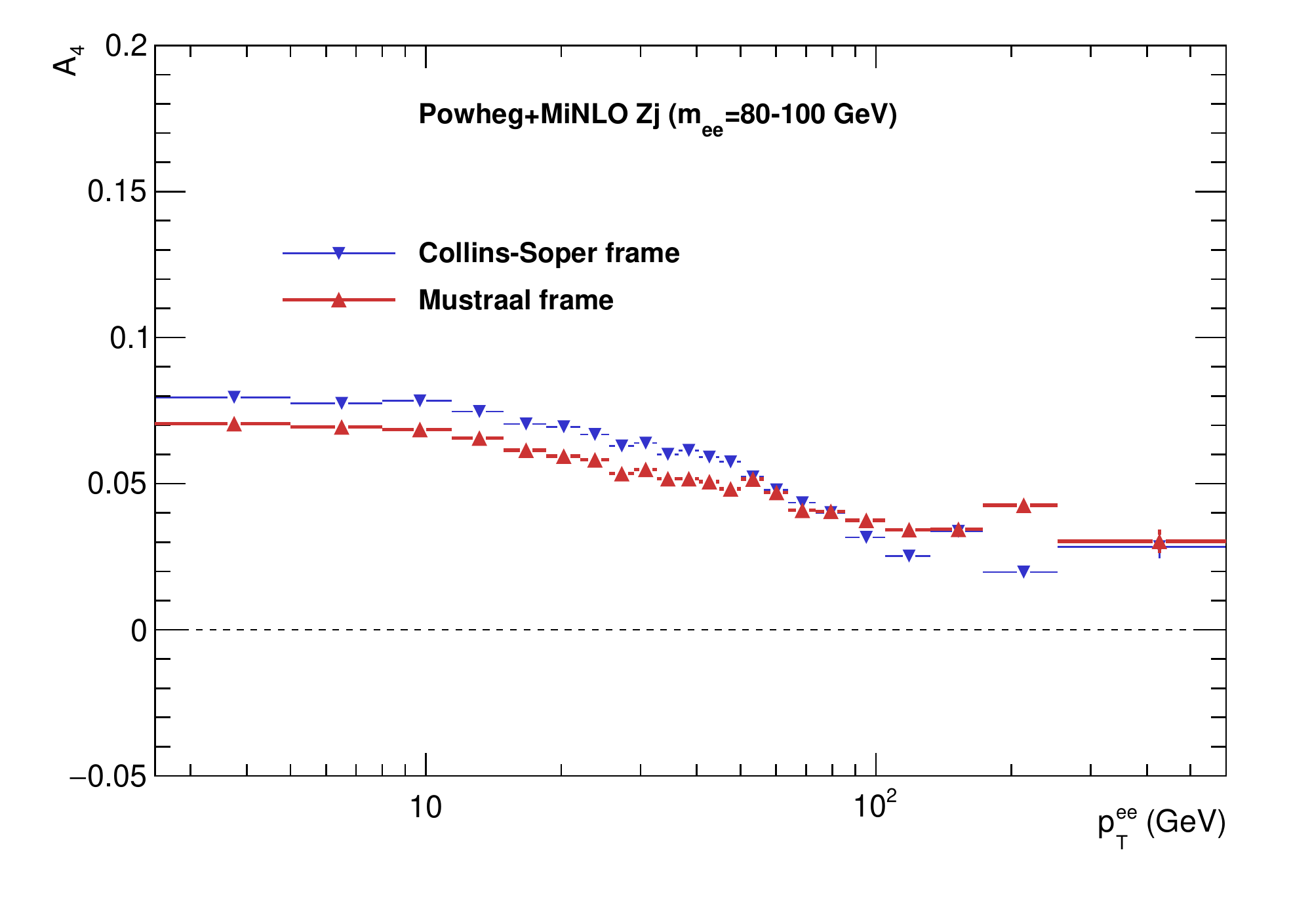}
}
  \end{center}
  \caption{ The $A_i$ coefficients  for  $Z\to e^+e^-$ in lepton pair invariant mass range
    $80 < m_{ee} < 100$ GeV.
    The $Z+j$ production process in $pp$ collisions
    at 8~TeV centre-of-mass energy, was used for the sample generation with {\tt Powheg+MiNLO} Monte Carlo.
    The $A_i$ coefficients are calculated in the {\it Collins-Soper} and {\it Mustraal} frames with moments
    method~\cite{Mirkes:1994dp}.
    \label{Fig:AisFrame} }
\end{figure*}

The shape of $A_i$ coefficients  as a function
of laboratory frame $Z$-boson transverse momenta $p_T$
depends on the choice of lepton pairs rest-frame. In Fig.~\ref{Fig:AisFrame},
$A_i$ coefficients of the {\it Collins-Soper} and {\it Mustraal} frames are shown.
%The {\it Mustraal} frame is designed specifically to preserve
% in the presence of multiple high $p_T$ jets,
% decomposition of the distribution into two Born-like terms.
As intended,
 even  for large $p_T$, with this frame, only $A_4$ coefficient is sizably non-zero.

\section{Concept of the EW weight}
\label{sec:EWweight}

The EW corrections enter the $\sigma_{Born}( \hat s, \cos \theta)$
through the definition of the vector and axial couplings, also photon and $Z$-boson  propagators.
They modify normalization of the cross-sections, the line-shape of the $Z$-boson peak,
polarization of the outgoing leptons and asymmetries. 

Given that, we were able to factorize QCD and EW components of the cross-section to a good approximation and define
per-event weights which specifically correct for EW effects. Such a weight may modify
events generated with EW LO to the ones including the EW corrections.
This is very much the same idea
as already implemented in {\tt TauSpinner} for introducing corrections for
other effects: spin correlations, production process, etc.

The per-event $ wt^{EW} $ is defined as ratio of the Born-level cross-sections with and without EW corrections 
\begin{equation}
  wt^{EW} = \frac{ d\sigma_{Born+EW}( s, \cos \theta)}{d\sigma_{Born}( s, \cos \theta)},
  \label{Eq:wt}
\end{equation}
where $\cos \theta$ can be taken according to  $\cos \theta^*$,  $\cos \theta^{Mustraal}$ ({\it Mustraal} frame) or $\cos \theta^{CS}$ ({\it Collins-Soper} frame) prescription.
For  most events, the three choices will lead to numerically very close
values for $\cos \theta$
and thus resulting $wt^{EW}$. The difference  originates from
distinct $\cos \theta$ dependence of
 $Z$ and $\gamma^*$ exchange amplitudes and not only from  electroweak boxes.  
The $wt^{EW}$   allows for flexible  implementation of the EW corrections 
using {\tt Tau\-Spinner} framework and form-factors calculated e.g. with {\tt Dizet}.

The formula for $wt^{EW}$ can be used to re-weight from one EW LO scheme
to another too. In that case, both the
numerator and denominator of Eq.~(\ref{Eq:wt}) will use lowest order $d\sigma_{Born}$, calculated in different EW schemes\footnote{
  In this way, in particular, the fixed width description for the $Z$-boson propagator can be replaced with the  $s$ dependent 
one.} though.

%\clearpage
\section{EW corrections to doubly-deconvoluted observables}
\label{sec::EWobserv}

Now that all components needed for  calculation of $wt^{EW}$ are explained, we can  
%From  all components available, we can now with  $wt^{EW}$ modify generated
%events and from them obtain and
present results for
selected examples of 
 doubly-deconvoluted observables around the $Z$-pole.

The {\tt Powheg+MiNLO} Monte Carlo, with NLO QCD and LO EW matrix elements, was used to generate
$Z+j$ events with $Z \to e^+ e^-$ decays in $pp$ collisions at 8~TeV. No selection was applied to generated events,
except for an outgoing electron pair invariant mass
range of $70 < m_{ee} < 150$~GeV.
For events generation, the EW parameters as 
shown in left-most column of Table~\ref{Tab:BornEff} were used.
It is often used
as a default for phenomenological studies at LHC.
The  $\alpha$ and $s^2_W$ close to the ones
of $\overline{\mathrm MS}$ scheme  discussed in \cite{Olive:2016xmw} were taken.
Note that they do not coincide accurately with the precise LEP experiments measurements  
at the $Z$-pole~\cite{ALEPH:2005ab}.
%The initialization of Table~\ref{Tab:BornEff}, left most column,
%and we will show later
%the estimated size of EW corrections for this setup.

To quantify the effect of the EW corrections, we re-weight events generated,
to EW LO with the scheme used by the
{\tt Dizet}: Table \ref{Tab:BornEff} second column. Only then we gradually introduce EW corrections and form-factors calculated with that library.
 For each step, the appropriate
 numerator of the $wt^{EW}$ is calculated, while for the denominator
 the EW LO  ${\mathscr A}^{Born}$ matrix element  Eq.~(\ref{Eq:Born}) is used;
parameters as in the left-most column of Table~\ref{Tab:BornEff}.
The sequential steps, in which we illustrate effects of EW corrections are given below:
  
\begin{enumerate}
\item
  Re-weight with $wt^{EW}$, from EW LO scheme used for MC events generation
  to EW LO scheme with $s^2_W$= 0.21215, Table~\ref{Tab:BornEff} second column.
  The ${\mathscr A}^{Born}$ matrix element, Eq.~(\ref{Eq:Born}),
  is used\footnote{The MC sample is generated with fixed width propagator.
  We remain with this  convention.
  This could also be changed with the help of $wt^{EW}$. 
%    This rewiegting is used also to change
%  fixed $Z$ width the $s$ dependent one, as of Eq.~\ref{Eq:Zprob}
  }
  for calculating numerator of $wt^{EW}$. 
\item
  As in step (1), but include EW corrections to $M_W$, effectively changing  to $s^2_W$= 0.22352 in calculation
  of $wt^{EW}$. Relation, formula~(\ref{Eq:SMrelation}), is not obeyed anymore.
\item
  As in step (2), but include EW loop corrections to the normalization of $Z$-boson and $\gamma^*$ propagators,
  i.e. QCD/EW corrections to $\alpha(0)$ and $\rho_{\ell f}(s)$ form-factor calculated without box corrections.
  The ${\mathscr A}^{Born+EW}$, Eq.~(\ref{Eq:BornEW}), is used for calculating numerator of  $wt^{EW}$.
\item
   As in step (3), but include EW corrections to $Z$-boson vector couplings: ${\mathscr K}_f, {\mathscr K}_l, {\mathscr K}_{\ell f}$,
  calculated without box corrections.
  The ${\mathscr A}^{Born+EW}$ is used for calculating numerator of  $wt^{EW}$.
\item
  As in step (4),
  but  $\rho_{\ell f}, {\mathscr K}_f, {\mathscr K}_l, {\mathscr K}_{\ell f}$ form-factors 
  include box corrections.
  The ${\mathscr A}^{Born+EW}$ is used for calculating numerator of  $wt^{EW}$.
  \end{enumerate}

After step (1) the sample is EW LO and QCD NLO, but with different EW scheme than used
originally for events generation. Then steps (2)-(5) introduce EW corrections. Step (3) effectively changes
 $\alpha$ back to be close to   $\alpha(M_Z^2)$, while steps (4)-(5) effectively shift back
$v_f, v_l$ close to the values used in generation. 
Parameters for EW LO scheme used for
event generation are   already close to measured at the $Z$-pole.
That is why we expect the total EW corrections
to the generated sample to be roughly at the percent level only.

In the following, we will  estimate how precise it would be to use effective Born approximation with {\it LEP}
or {\it LEP with improved norm.} parametrisations instead of complete EW corrections.
To obtain those predictions, re-weighting similar to step (1) listed above is needed, but in the numerator of $wt^{EW}$ the ${\mathscr A}^{Born}$
parametrisations as specified in the right two columns of Table~\ref{Tab:BornEff} are used. For {\it LEP with improved norm.}
the $\rho_{\ell,f} = 1.005$ has to be included as well.

The important flexibility of the proposed approach is that $wt^{EW}$ can be calculated using $d \sigma_{Born}$
in different frames: $\cos \theta^*$, {\it Mustraal} or {\it Collins-Soper}.
For some observables, frame choice used for $wt^{EW}$ calculation is not
numerically relevant at all; the simplest $\cos \theta^*$
frame can be used. We show later  an example, where only the {\it Mustraal} frame for the
$wt^{EW}$ calculation leads to correct results.

\subsection{The $Z$-boson line-shape}
\label{sec::lineshape}

In the EW LO, the $Z$-boson line-shape, assuming that the constraint (\ref{Eq:SMrelation}) holds,
depends predominantly  on  $M_Z$ and $\Gamma_z$.
The effects on the line-shape from EW loop corrections are due to corrections to the propagators:
vacuum polarization corrections (running $\alpha$) and $\rho$ form-factor, which  change 
relative contributions of the $Z$ to $\gamma^*$ and,  the  $Z$-boson vector to axial coupling ratio
($\sin^2\theta_{eff}$). The above affects not only shape but also normalization of the cross-section.
In the formulae (\ref{Eq:wt}) we do not use running $Z$-boson width, which remains fixed.% as for generated event.
%However, in the Appendix~\ref{app:Zwidth} we detail conversion formulae between
%fixed and running width schemes.
%If this convension was used to generated sample, it would lead to
%additional shift in the lineshape requiring modification of $wt^{EW}$ denominator.

\begin{figure*}
  \begin{center}                               
{
  \includegraphics[width=7.5cm,angle=0]{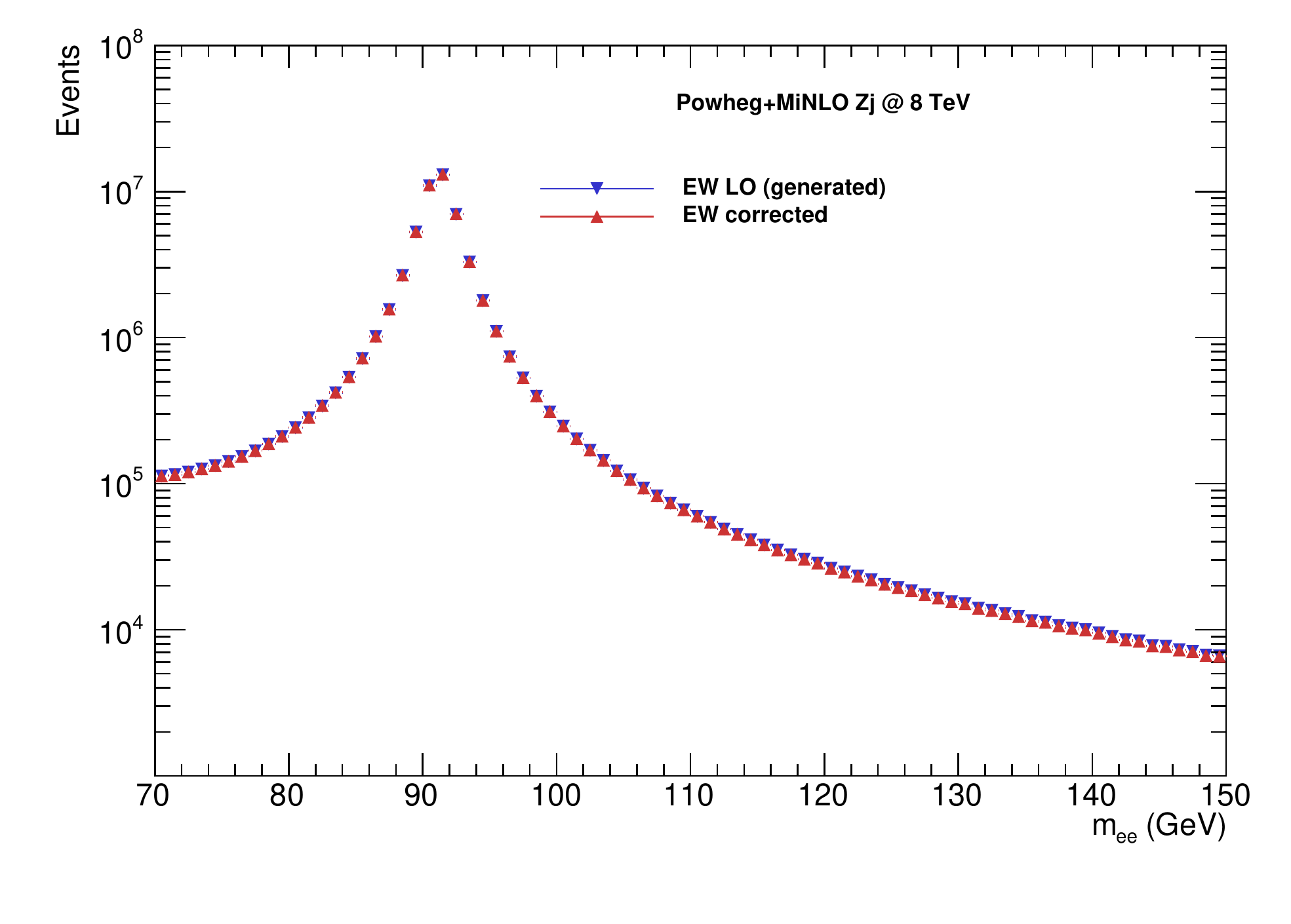}
  \includegraphics[width=7.5cm,angle=0]{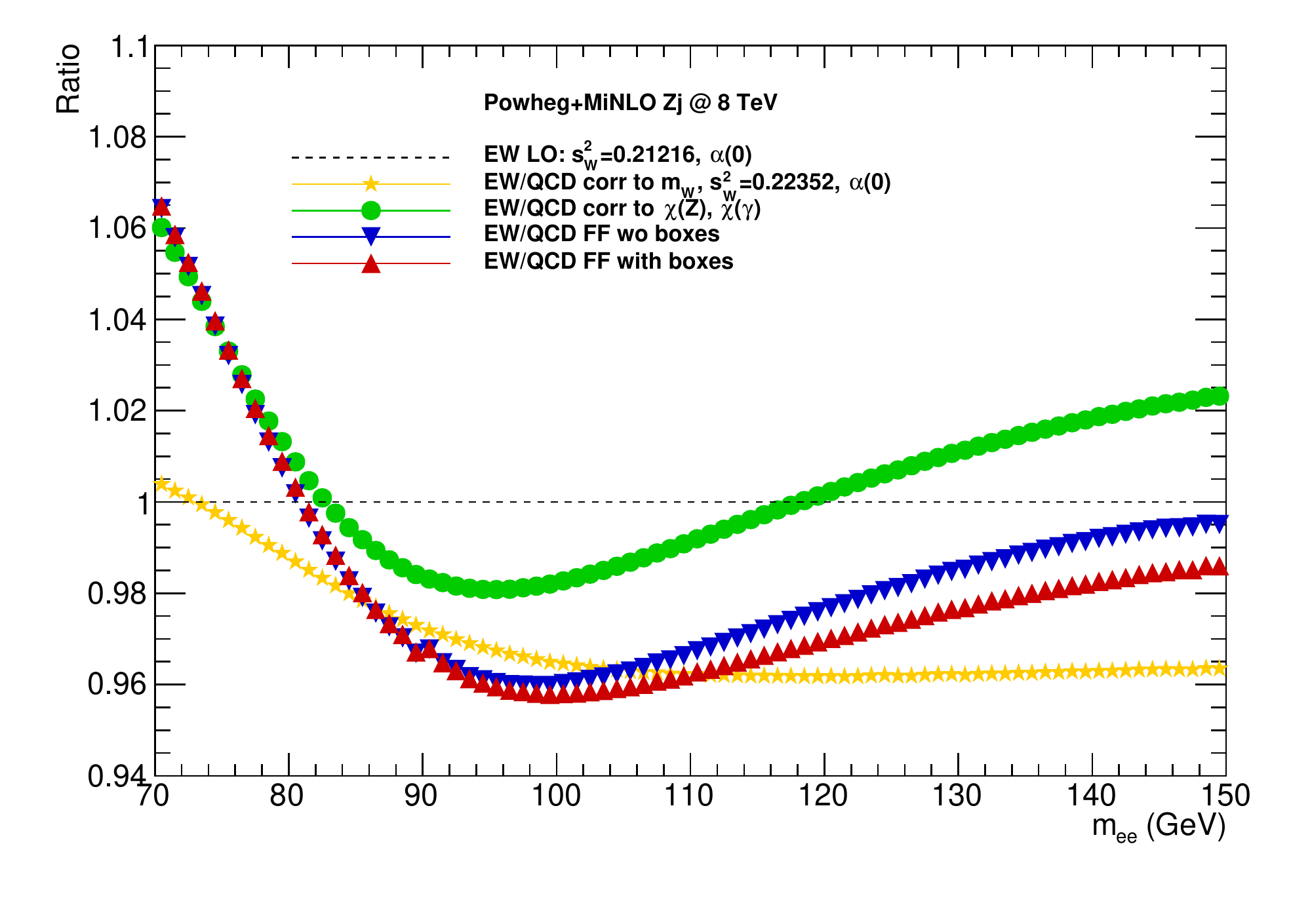}
  \includegraphics[width=7.5cm,angle=0]{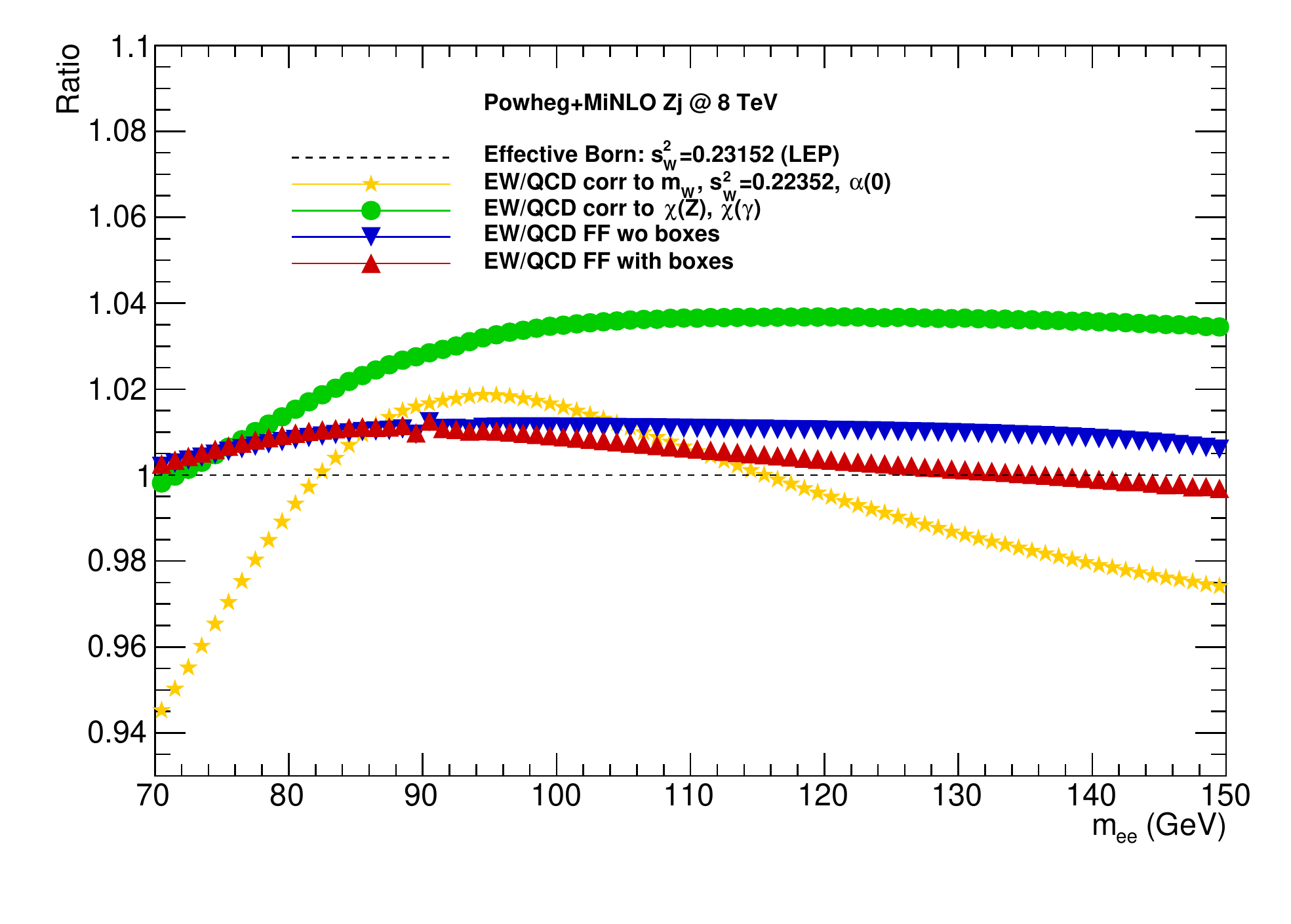}
  \includegraphics[width=7.5cm,angle=0]{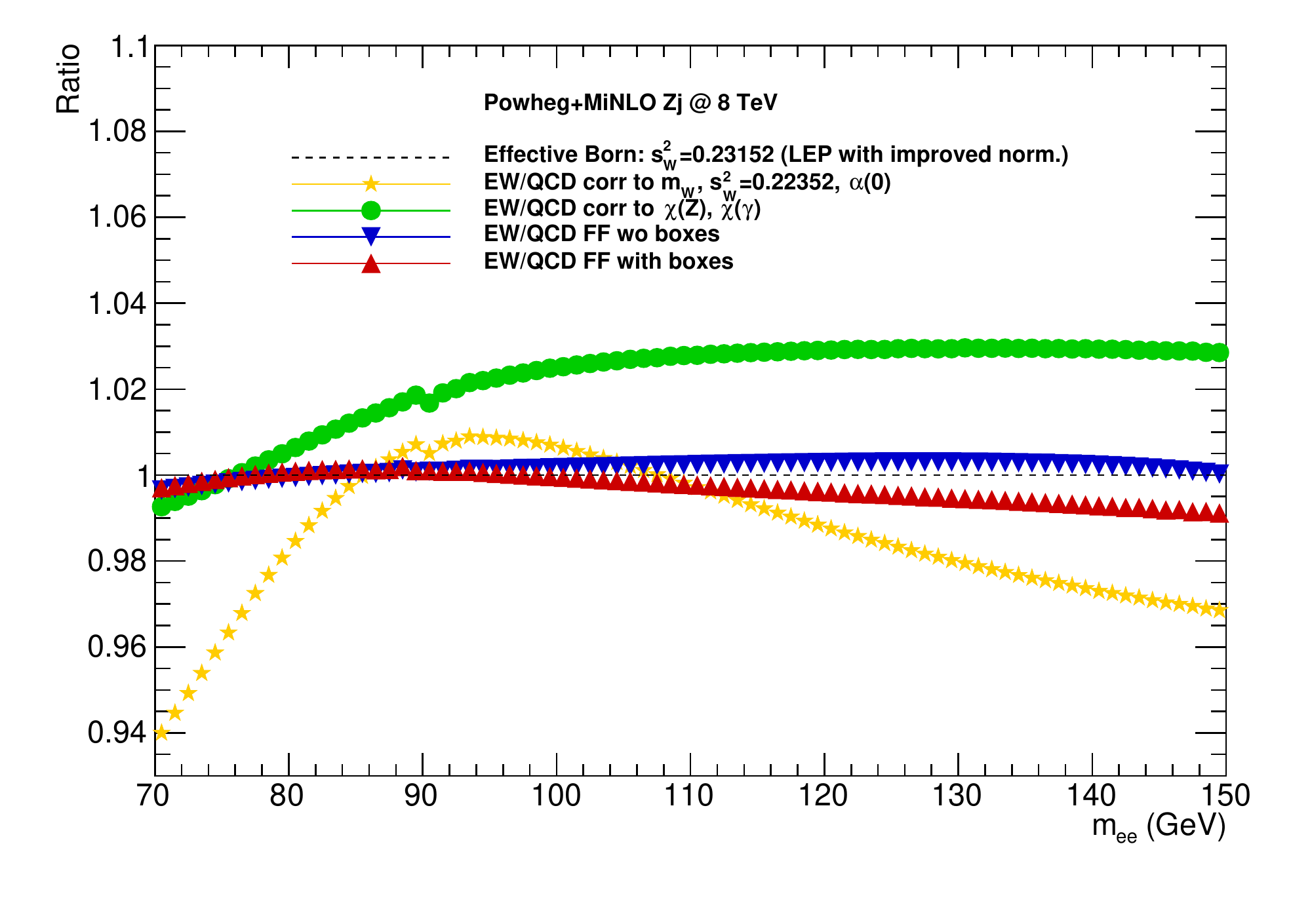}
}
\end{center}
  \caption{Top-left: line-shape distribution as generated with {\tt Powheg+MiNLO} (blue triangles)
    and after reweighting introducing all EW corrections (red triangles). The two choices are
    barely distinguishable. Ratios of the line-shapes with gradually introduced EW corrections
    are shown in consecutive plots, where  as a reference (black dashed line) respectively:
    (i)  EW LO  $\alpha(0)$ scheme (top-right),
    (ii)  effective Born {\it (LEP)} (bottom-left) and, (iii)
    effective Born {\it (LEP with improved norm.)} (bottom-right),  was used.
\label{Fig:Zlineshape} }
\end{figure*}

In Fig.~\ref{Fig:Zlineshape} (top-left) distributions of generated and EW corrected line-shapes are shown.
With the logarithmic scale, a difference is barely visible.
With the following plots of the same Figure we study  details.
The ratios of the line-shape distributions with gradually introduced
EW corrections are shown. For the reference
distributions (ratio-histograms  denominators) for the following
three plots:
(i) EW LO $\alpha(0)$ scheme,
(ii) effective Born {\it (LEP)} and (iii) effective Born {\it (LEP with improved norm.)} are used.
At the $Z$-pole, complete EW corrections contribute
about 0.1\%  with respect to the one of effective Born {\it (LEP with improved norm.)}. A use
of events generated with  EW LO matrix element but of different parametrisations  significantly
reduce the numerical size of missing EW corrections.

Table~\ref{Tab:EWnormcorr} details numerically EW corrections to the normalization
(ratio of the cross-sections) integrated
in the range $ 80 < m_{ee} <100$~GeV and $89 < m_{ee} < 93$~GeV. Results from   EW weight
with the $\cos \theta^*$ definition of the scattering angle are shown.
%Total EW correction factor for normalisation of EW LO $G_{\mu}$ cross-section is 1.010.
The total EW correction factor is about 0.965
for cross-section normalization and EW LO $\alpha(0)$ , while the total
correction for the effective Born {\it (LEP with improved norm.)} is of about 1.001.
In Table~\ref{Tab:EWnormcorr_optFR} results with $wt^{EW}$ calculated with different frames
 are compared. 
 If {\it Mustraal}  or {\it Collins-Soper} frames are used instead of $\cos \theta^*$
 for weight calculations,
the differences are at most at the 5-th significant digit.

\begin{table*}
 \vspace{2mm}
 \caption{EW corrections for cross-sections integrated over the specified mass windows.
   The EW weight is calculated with  $\cos \theta^*$.} 
 \label{Tab:EWnormcorr}
 \begin{center}
    \begin{tabular}{|l|c|c|}
        \hline\hline
         Corrections to cross-section                       & $ 89 < m_{ee} < 93$ GeV &  $80 < m_{ee} < 100$ GeV  \\ 
         \hline \hline
         $\sigma$(EW corr. to $m_W$)/$\sigma$(EW LO $\alpha(0)$)        &  0.97114 &  0.97162     \\
         \hline 
         $\sigma$(EW corr. to $\chi(Z),\chi(\gamma)$)/$\sigma$(EW LO $\alpha(0)$)   &  0.98246 &  0.98346     \\
         \hline 
         $\sigma$(EW/QCD FF no boxes)/$\sigma$(EW LO $\alpha(0)$)   &  0.96469 &  0.96602     \\
         \hline 
         $\sigma$(EW/QCD FF with boxes)/$\sigma$(EW LO $\alpha(0))$   &  0.96473 &  0.96607     \\
         \hline \hline
         $\sigma$({\it LEP})/$\sigma$(EW/QCD FF with boxes)   &  1.01102 &  1.01093    \\
         \hline
         $\sigma$({\it LEP with improved norm.})/$\sigma$(EW/QCD FF with boxes)   &  1.00100 &  1.00098     \\
    \hline
 \end{tabular}
  \end{center}
%\end{table}
%\begin{table}
 \vspace{2mm}
 \caption{EW corrections for cross-sections integrated over the  mass window around $Z$-pole;
   $89 < m_{ee} < $ 93 GeV. The EW weight is calculated 
   with $\cos \theta^*$, $\cos \theta^{Mustraal} $ or  $\cos \theta^{CS} $.} 
 \label{Tab:EWnormcorr_optFR}
 \begin{center}
    \begin{tabular}{|l|c|c|c|}
        \hline\hline
         Corrections to cross-section  ( $ 89 < m_{ee} < 93$ GeV)  & $wt^{EW}(\cos \theta^*)$ & $wt^{EW}(\cos \theta^{Mustraal})$  & $wt^{EW}(\cos \theta^{CS})$   \\ 
         \hline \hline
         $\sigma$(EW corr. to $m_W$)/$\sigma$(EW LO $\alpha(0)$)        &  0.97114 &  0.97115   &  0.97114     \\
         \hline 
         $\sigma$(EW corr. to $\chi(Z),\chi(\gamma)$)/$\sigma$(EW LO $\alpha(0)$)   &  0.98246 &  0.98247 &  0.98246     \\
         \hline 
         $\sigma$(EW/QCD FF no boxes)/$\sigma$(EW LO $\alpha(0)$)   &  0.96469 &  0.96471    &  0.96470     \\
         \hline 
         $\sigma$(EW/QCD FF with boxes)/$\sigma$(EW LO $\alpha(0))$   &  0.96473 &  0.96475   &  0.96474     \\
         \hline \hline
         $\sigma$({\it LEP})/$\sigma$(EW/QCD FF with boxes)   &  1.01102 &  1.01103&  1.01102     \\
         \hline
         $\sigma$({\it LEP with improved norm.})/$\sigma$(EW/QCD FF with boxes)   &  1.00100 &  1.00102 &  1.00100     \\
    \hline
 \end{tabular}
  \end{center}
\end{table*}

\subsection{The $A_{FB}$ distribution} \label{sec:Afb}

The forward-backward asymmetry  for $pp$ collisions reads

\begin{equation} \label{Eq:AFB}
  A_{FB} = \frac{\sigma(\cos \theta > 0) - \sigma(\cos \theta < 0)}{\sigma(\cos \theta > 0) + \sigma(\cos \theta < 0)},
\end{equation}
where $\cos \theta$ of the {\it Collins-Soper} frame is used.

\begin{figure*}
  \begin{center}                               
{
  \includegraphics[width=7.5cm,angle=0]{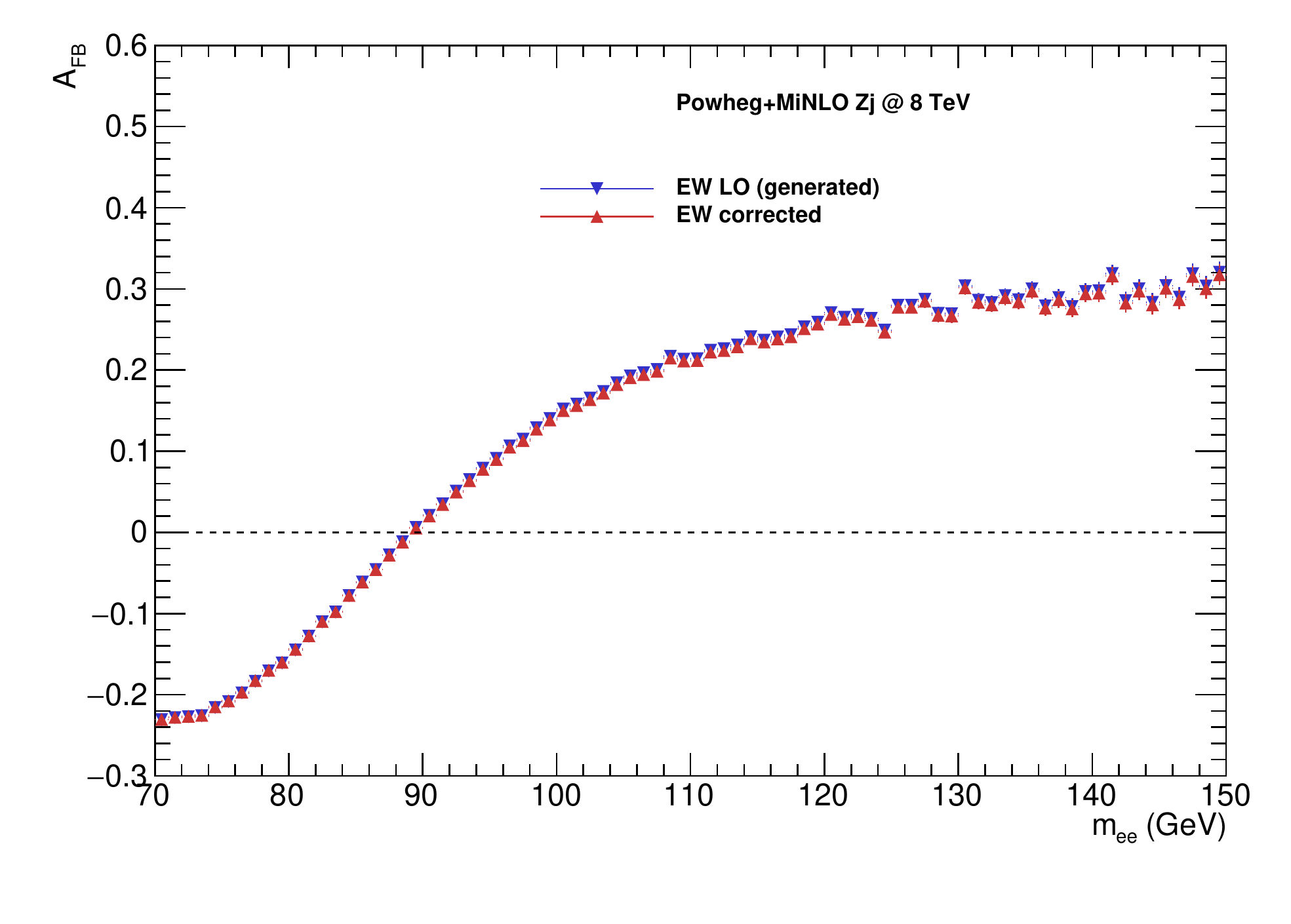}
  \includegraphics[width=7.5cm,angle=0]{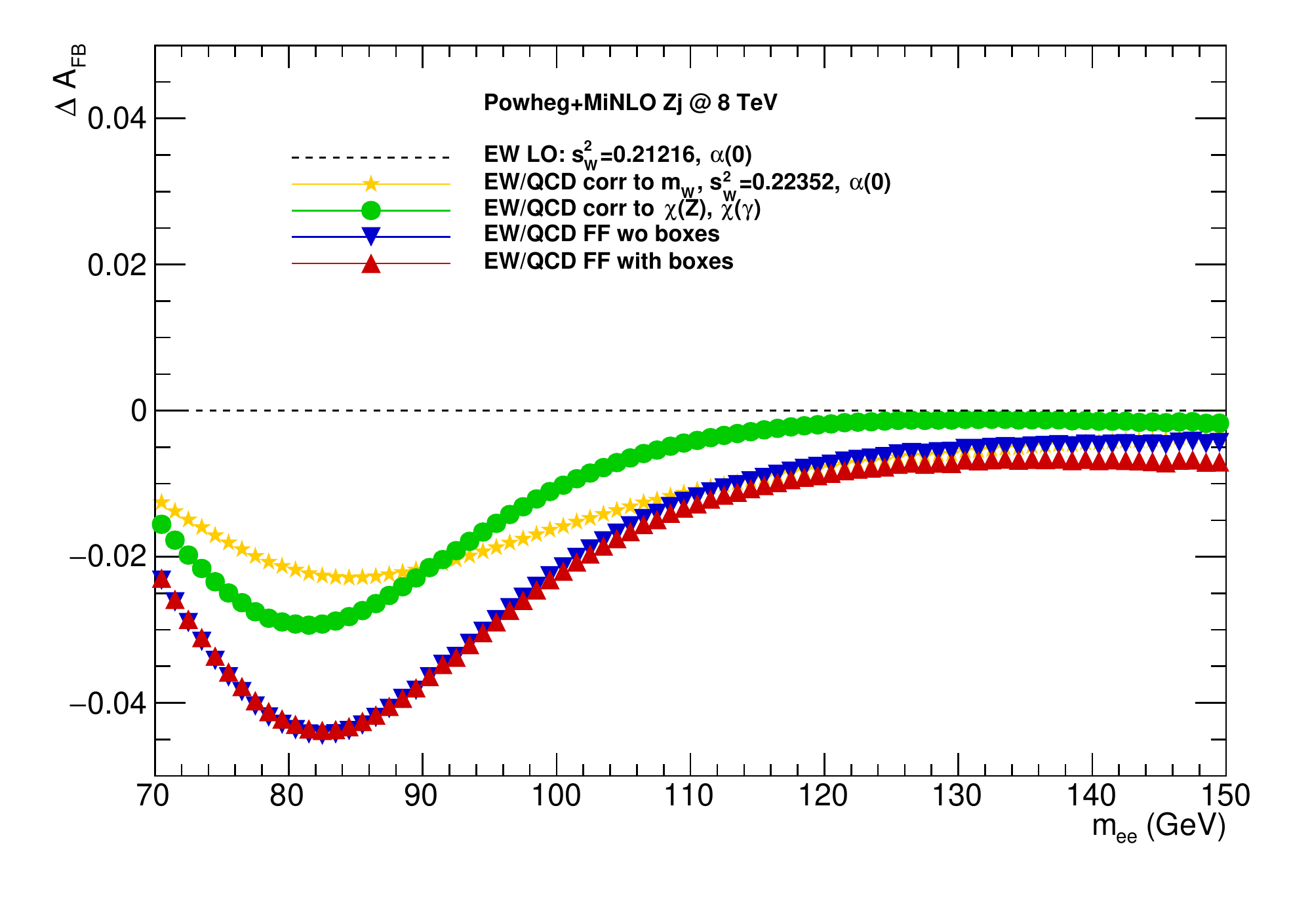}
  \includegraphics[width=7.5cm,angle=0]{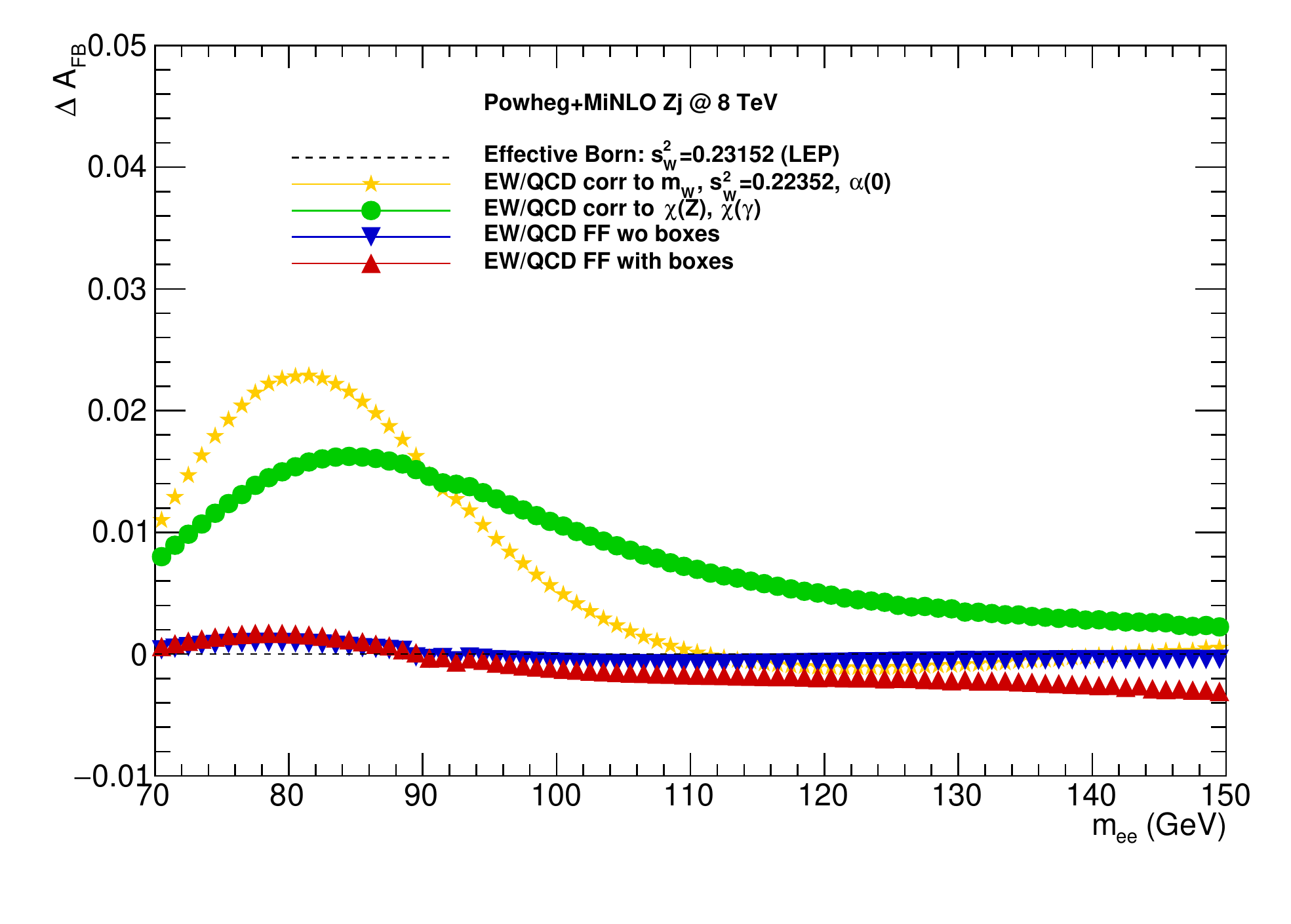}
  \includegraphics[width=7.5cm,angle=0]{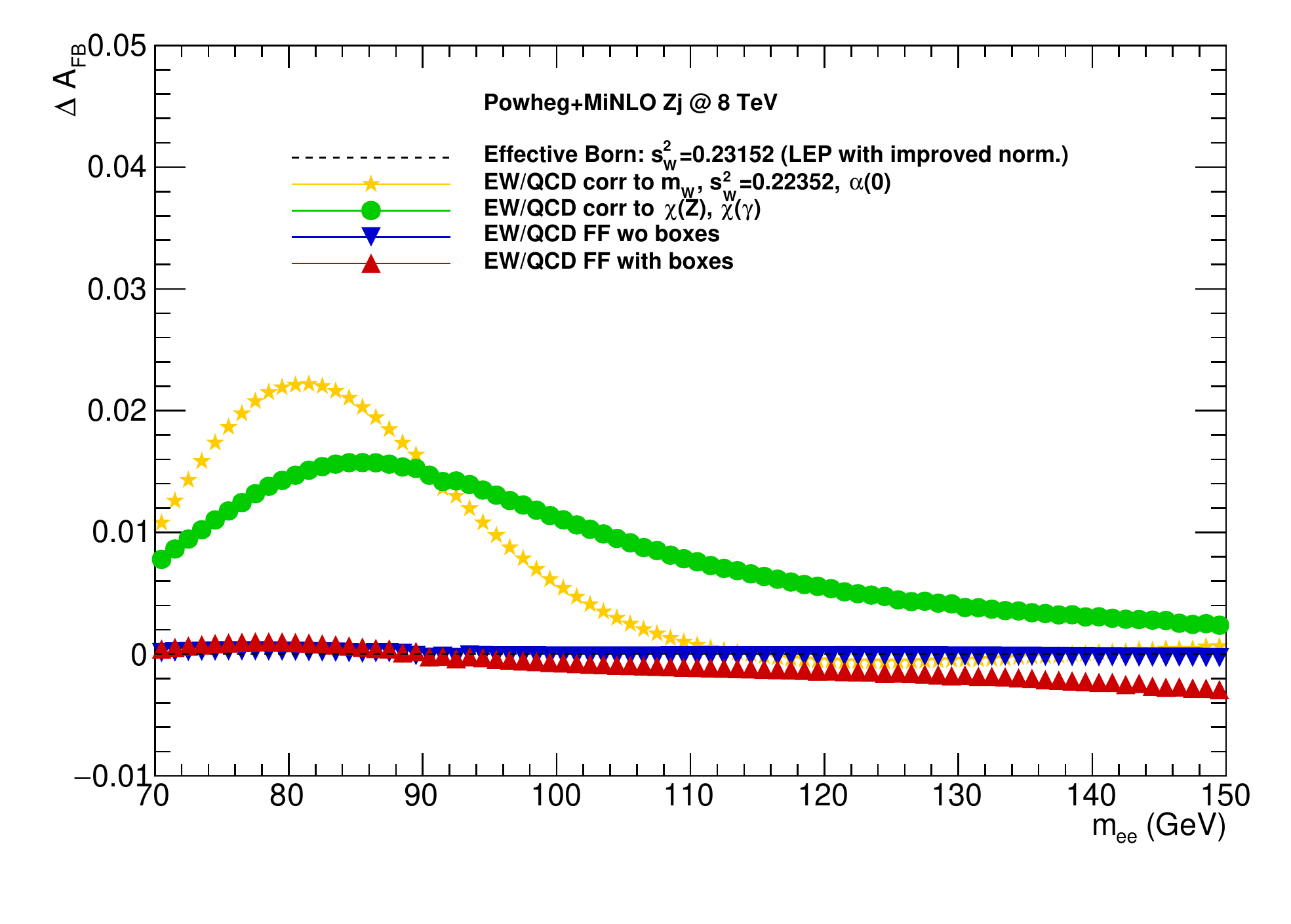}
}
\end{center}
  \caption{Top-left: the $A_{FB}$  as generated with {\tt Powheg+MiNLO}  (blue triangles)
    and after reweighting introducing all EW corrections (red triangles).
    The two choices are barely distinguishable. The differences
    $\Delta A_{FB} = A_{FB} - A_{FB}^{ref}$, due to gradually introduced EW corrections are shown
    in consecutive plots, where as a reference (black dashed line) respectively:
    (i)  EW LO  $\alpha(0)$ scheme (top-right),
    (ii)  effective Born {\it (LEP)} (bottom-left) and, (iii) 
    effective Born {\it (LEP with improved norm.)} (bottom-right),  was used.
\label{Fig:Afb} }
\end{figure*}

The EW corrections change $A_{FB}$, particularly around the $Z$-pole. 
In Fig.~\ref{Fig:Afb} (top-left), the $A_{FB}$  as generated (EW LO) and EW corrected is shown
as a function of $  m_{ee}$.
In the  following plots of this Figure, we study  details. The  $\Delta A_{FB} = A_{FB} - A_{FB}^{ref}$, with gradually introduced EW corrections to $A_{FB}$ is shown and
compared with the following  reference choices for $A_{FB}^{ref}$:
(i) EW LO $\alpha(0)$ scheme,
(ii) effective Born {\it (LEP)} and (iii) effective Born {\it (LEP with improved norm.)}.

Complete EW corrections  to predictions of EW LO $\alpha(0)$ scheme for   $A_{FB}$ integrated around $Z$-pole give
 $\Delta A_{FB}$ = -0.03534.
The EW correction  $\Delta A_{FB}$ to predicition of  effective Born ({\it LEP with improved norm.}),
is only  -0.00005. We observe that
effective Born {\it (LEP improved norm.)}  reproduces EW loop corrections  
precision better and  $\Delta A_{FB}$ = -0.0001 in the full presented mass range.
The remaining box corrections contribute around $ m_{ee}=150$ GeV about -0.002
to $\Delta A_{FB}$.

Table~\ref{Tab:AFBEWcorr} details numerically  EW corrections, for  $A_{FB}$ 
integrated
over  the $80 < m_{ee} < 100$~GeV and $ 89 < m_{ee} < 93$~GeV ranges. For calculating EW weight, the 
$\cos \theta^*$ definition of the scattering angle
%are shown, but for asymmetry definition $\cos \theta^{CS}$
was used. In Table~\ref{Tab:AFBEWcorr_optFR} 
results obtained with $wt^{EW}$ calculated in different frames are compared.  When  the {\it Mustraal}
or {\it Collins-Soper} frame is used instead of $\cos \theta^*$, the differences are at most at the 5-th
significant digit, similar as for the line-shape.

\begin{table*}
 \vspace{2mm}
 \caption{The difference $\Delta A_{FB}$ in forward-backward asymmetry calculated in the
   specified mass window. The  $\cos \theta^{CS}$ is used to define forward and backward hemispheres.
   The EW weight is calculated from  $\theta^*$ definition of the scattering angle.} 
 \label{Tab:AFBEWcorr}
 \begin{center}
    \begin{tabular}{|l|c|c|}
        \hline\hline
         Corrections to $A_{FB}$                       & $ 89 < m_{ee} < 93 $ GeV &  $80 < m_{ee} < 100 $ GeV  \\ 
         \hline \hline
         $A_{FB}$(EW corr. $m_W$) - $A_{FB}$(EW LO $\alpha(0)$)        &  -0.02097 &  -0.02103     \\
         \hline 
         $A_{FB}$(EW corr. prop. $\chi(Z),\chi(\gamma)$) - $A_{FB}$(EW LO $\alpha(0)$)   &  -0.02066 &  -0.02098     \\
         \hline 
         $A_{FB}$(EW/QCD FF no boxes) - $A_{FB}$(EW LO $\alpha(0)$)   &  -0.03535 &  -0.03569     \\
         \hline 
         $A_{FB}$(EW/QCD FF with boxes) - $A_{FB}$(EW LO $\alpha(0))$   &  -0.03534 &  -0.03567     \\
         \hline \hline
         $A_{FB}$({\it LEP}) - $A_{FB}$(EW/QCD FF with boxes)   &  -0.00006 &  -0.00001    \\
         \hline
         $A_{FB}$({\it LEP with improved norm.}) - $A_{FB}$(EW/QCD FF with boxes)   &  -0.00005 &  -0.00002     \\
    \hline
 \end{tabular}
  \end{center}
%\end{table}
%\begin{table}
 \vspace{2mm}
 \caption{The difference $\Delta A_{FB}$ in forward-backward asymmetry around $Z$-pole,
   $m_{ee}$ = 89 - 93 GeV.
   The  $\cos \theta^{CS}$ is used to define forward and backward hemispheres.
   The EW weight is calculated respectively from $\cos \theta^*$, $\cos \theta^{Mustraal}$ or $\cos^{CS}$.} 
 \label{Tab:AFBEWcorr_optFR}
 \begin{center}
    \begin{tabular}{|l|c|c|c|}
        \hline\hline
         Corrections to $A_{FB}$ ( $89 < m_{ee} < 93 $ GeV) & $wt^{EW}(\cos \theta^*)$ & $wt^{EW}(\cos \theta^{ML})$  & $wt^{EW}(\cos \theta^{CS})$   \\ 
         \hline \hline
        $A_{FB}$(EW/QCD corr. to $m_W$) - $A_{FB}$(EW LO $\alpha(0)$)        &  -0.02097 &  -0.02112   &  -0.02101     \\
         \hline 
         $A_{FB}$(EW/QCD corr. to $\chi(Z),\chi(\gamma)$) - $A_{FB}$(EW LO $\alpha(0)$)   &  -0.02066 &  -0.02081 &  -0.02070    \\
         \hline 
         $A_{FB}$(EW/QCD FF no boxes) - $A_{FB}$(EW LO $\alpha(0)$)   &  -0.03535 &  -0.03560  &  -0.03542     \\
         \hline 
         $A_{FB}$(EW/QCD FF with boxes) - $A_{FB}$(EW LO $\alpha(0))$   &  -0.03534 &  -0.03559  &  -0.03541    \\
         \hline \hline
         $A_{FB}$({\it LEP}) - $A_{FB}$(EW/QCD FF with boxes)   &  -0.00006 &  -0.00005  &  -0.00006    \\
         \hline
         $A_{FB}$({\it LEP with improved norm.}) - $A_{FB}$(EW/QCD FF with boxes)   &  -0.00005 &  -0.00005  &  -0.00005     \\
    \hline
 \end{tabular}
  \end{center}
\end{table*}

\subsection{Effective weak mixing angles}
\label{sec:weakmixing}

\begin{figure*}
  \begin{center}                               
{
   \includegraphics[width=7.5cm,angle=0]{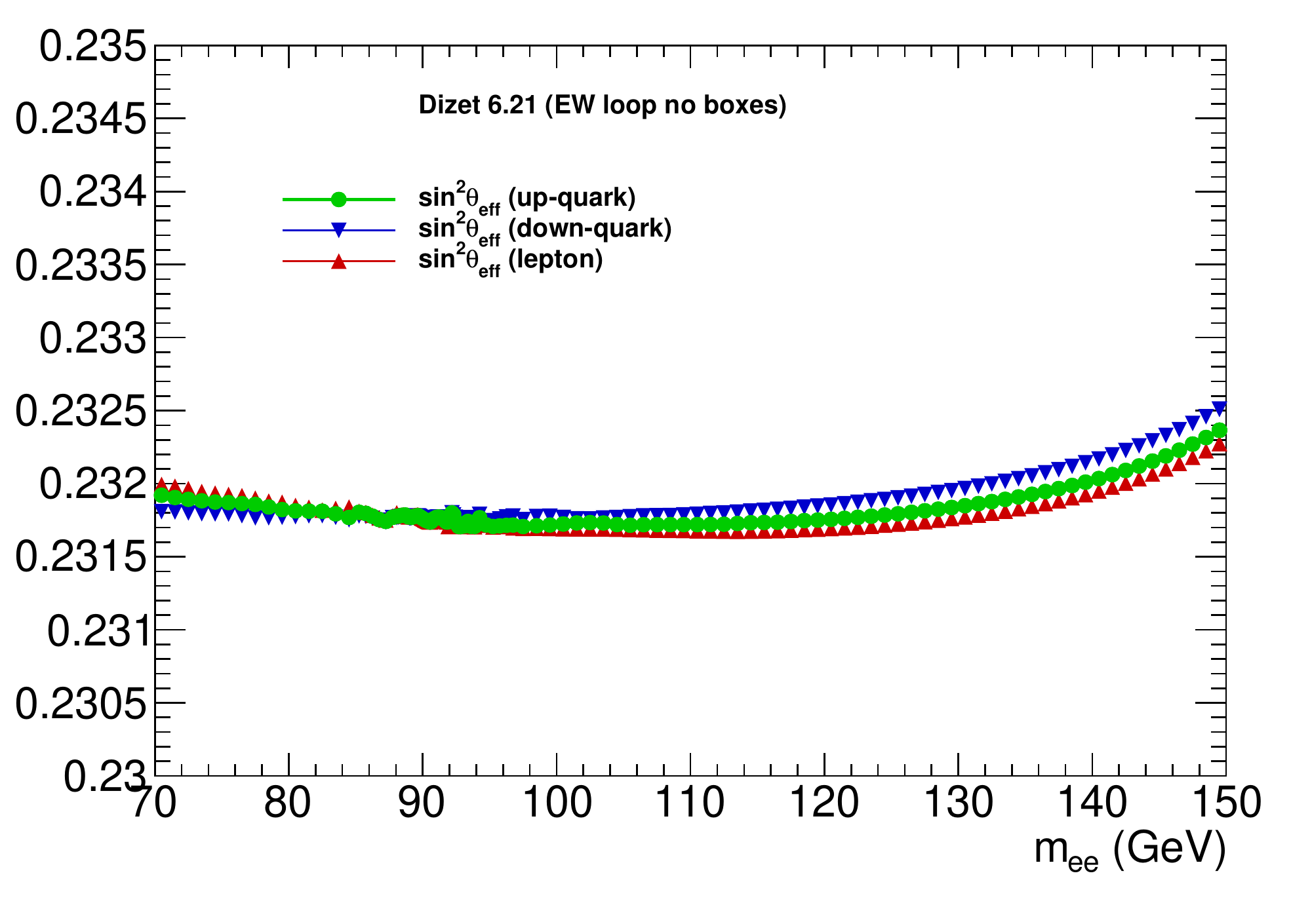}
   \includegraphics[width=7.5cm,angle=0]{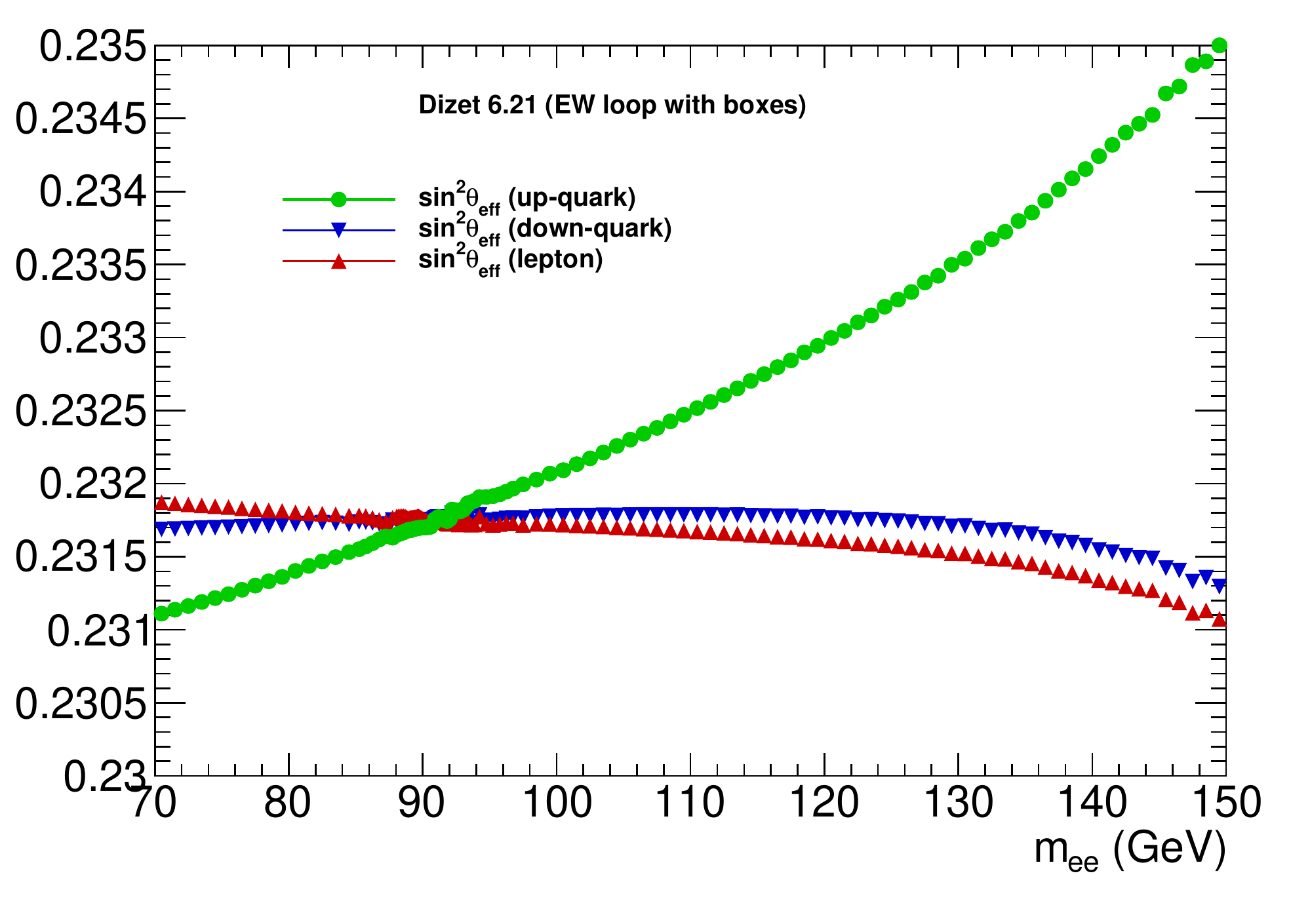}
}
\end{center}
  \caption{Effective weak mixing angles  $\sin^2\theta_{eff}^f(s,t)$
    as a function of $m_{ee}$ and $\cos \theta$ = 0,
    without (left-hand plot) and with (right-hand plot) box corrections.
    The ${\mathscr K}^f(s,t)$ form-factor calculated using {\tt Dizet}
    library and on-mass-shell  $s^2_W=0.22352$ were used.
    Only the real part is shown, imaginary part   of  $I^2_f(s,t)$ is only about $10^{-4}$.
\label{Fig:sw2effpp} }
\end{figure*}

The forward-backward asymmetry $A_{FB}$ at the $Z$-pole can be used as an observable for effective weak mixing Weinberg angles, 
 dependent on the invariant mass of lepton pairs.
 We extend standard LEP definition of effective weak mixing angles to 
\begin{equation}
  \label{Eq:sweffpp}
  \sin^2\theta^f_{eff}(s,t) = Re({\mathscr K}^f(s,t)) s^2_W + I^2_f(s,t),
\end{equation}
which is
 more suitable for  LHC and  for the off $Z$-pole regions.
The flavour dependent effective weak mixing angles, calculated using: Eq.~(\ref{Eq:sweffpp}), EW form-factors of {\tt Dizet} library,
and $s^2_W=0.22352$ are shown on Fig.~\ref{Fig:sw2effpp} as a function of the invariant
mass of outgoing lepton pair and for $\cos \theta = 0.5$.
The imaginary part   of  $I^2_f(s,t)$ is about $10^{-4}$ only.
In Table~\ref{Tab:MC_sw2eff} we display 
effective weak mixing angles averaged over specified mass windows.

The effective  $\sin \theta_{eff}^f$ on the $Z$-pole, printed by {\tt Dizet} is 
shown in Table~\ref{Tab:Dizet_EWeff_printout}. It is numerically slightly different than 
of Table~\ref{Tab:MC_sw2eff}, which is an average over  mass window close to $Z$-pole.
Note, that the observed very good agreement at the $Z$-pole between $A_{FB}$ predictions  of effective Born
with {\it (LEP)} or  {\it (LEP with improved norm.)} parametrisations and fully EW corrected
is not reflected for predictions of flavour dependent effective weak Weinberg angles.
Effective Born {\it (LEP)} and  {\it (LEP with improved norm.)} are parametrised with $s^2_W = 0.23152$,
while  {\tt Dizet} library predicts leptonic effective weak mixing angle  $\sin^2\theta_{eff}^{\ell}(M_Z^2)$ = 0.23176 which is about $20 \cdot 10^{-5}$ different.
Why then such a good agreement on $\Delta A_{FB}$ as seen on Fig.~\ref{Fig:Afb} bottom plots?
%{\bf sprawdzic czy dobrze cytowany rysunek}
Certainly this requires further attention.

\begin{table}
%\begin{sidewaystable}
 \vspace{2mm}
 \caption{From the {\tt Dizet} library printout: effective weak mixing angles and $\alpha (M_Z^2)$.
   For details of $ZPAR$ parameter matrix definition  see technical documentation of
   {\tt KKMC} interface and {\tt DIZET} library itself~\cite{Jadach:2013aha,Bardin:1999yd}. } 
 \label{Tab:Dizet_EWeff_printout}
  \begin{center}
    \begin{tabular}{|l|c|c|}
        \hline\hline
         Parameter  & Value & Description \\ 
         \hline\hline
         $\alpha (M_Z^2)$     & 0.00775995   & From eq.~(\ref{Eq:runningalpha}) \\
         $1/\alpha (M_Z^2)$   & 128.86674 & \\
         \hline
         $ZPAR(6)-ZPAR(8)$ & 0.23176 & $sin^{2}\theta_{eff}^{\ell}(M_Z^2)$\\
                           &         & ($\ell = e, \mu, \tau$)\\
         $ZPAR(9)$  & 0.23165 & $sin^2\theta_{eff}^{up}(M_Z^2)$  \\
         $ZPAR(10)$ & 0.23152 & $sin^2\theta_{eff}^{down}(M_Z^2)$  \\
   \hline
 \end{tabular}
  \end{center}
%\end{sidewaystable}
%\end{table}
%\begin{table}
%\begin{sidewaystable}
 \vspace{2mm}
 \caption{The effective weak mixing angles $\sin^2 \theta_{eff}^{f}$, for different mass windows
   with/without box corrections. The form-factor corrections are averaged with realistic line-shape
   and $\cos \theta$ distribution. } 
 \label{Tab:MC_sw2eff}
  \begin{center}
    \begin{tabular}{|l|c|c|c|}
        \hline\hline
         Parameter  [GeV]              & $\sin^2 \theta_{eff}^{\ell}$ &  $\sin^2 \theta_{eff}^{\ up}$ &  $\sin^2 \theta_{eff}^{ down}$ \\ 
         \hline\hline
         \cline{1-4}\multicolumn{1}{|c|}{}&\multicolumn{3}{|c|}{ EW loops without box corrections }\\
         \hline
         $ 80 < m_{ee} < 100 $      & 0.23171 & 0.23171 &  0.23146 \\
         $ 78 < m_{ee} < 82 $       & 0.23179 & 0.23172 &  0.23159 \\
         $ 89 < m_{ee} < 93 $       & 0.23170 & 0.23169 &  0.23147 \\
         $ 108 < m_{ee} < 112$     & 0.23168 & 0.23175 &  0.23137 \\
         \hline
         \cline{1-4}\multicolumn{1}{|c|}{}&\multicolumn{3}{|c|}{ EW loops with box corrections }\\
         \hline
         $ 80 <  m_{ee} < 100$     & 0.23171 & 0.23171 &  0.23146 \\
         $ 78 <  m_{ee} < 82$       & 0.23136 & 0.23167 &  0.23158 \\
         $ 89 <  m_{ee} < 93$       & 0.23168 & 0.23169 &  0.23147 \\
         $ 108 < m_{ee} < 112$     & 0.23246 & 0.23174 &  0.23130 \\
   \hline
 \end{tabular}
  \end{center}
%\end{sidewaystable}
\end{table}

\subsection{The $A_{4}$, $A_{3}$ angular coefficients}
\label{sec:EWAis}

To complete the discussion on doubly-deconvoluted observables, we  turn our attention
to angular coefficients $A_{4}$ and $A_{3}$ (proportional to product of vector and axial couplings) and to EW corrections.
 The coefficients are calculated from the event sample with the 
moments methods~\cite{Mirkes:1994dp} and
in the {\it Collins-Soper} frame. The  EW weight $wt^{EW}$ is used to introduce EW corrections
and is calculated with the help of 
 $\cos \theta^*$, $\cos \theta^{Mustraal}$ or $\cos \theta^{CS}$ angles.  

\begin{figure*}
  \begin{center}                               
{
  \includegraphics[width=7.5cm,angle=0]{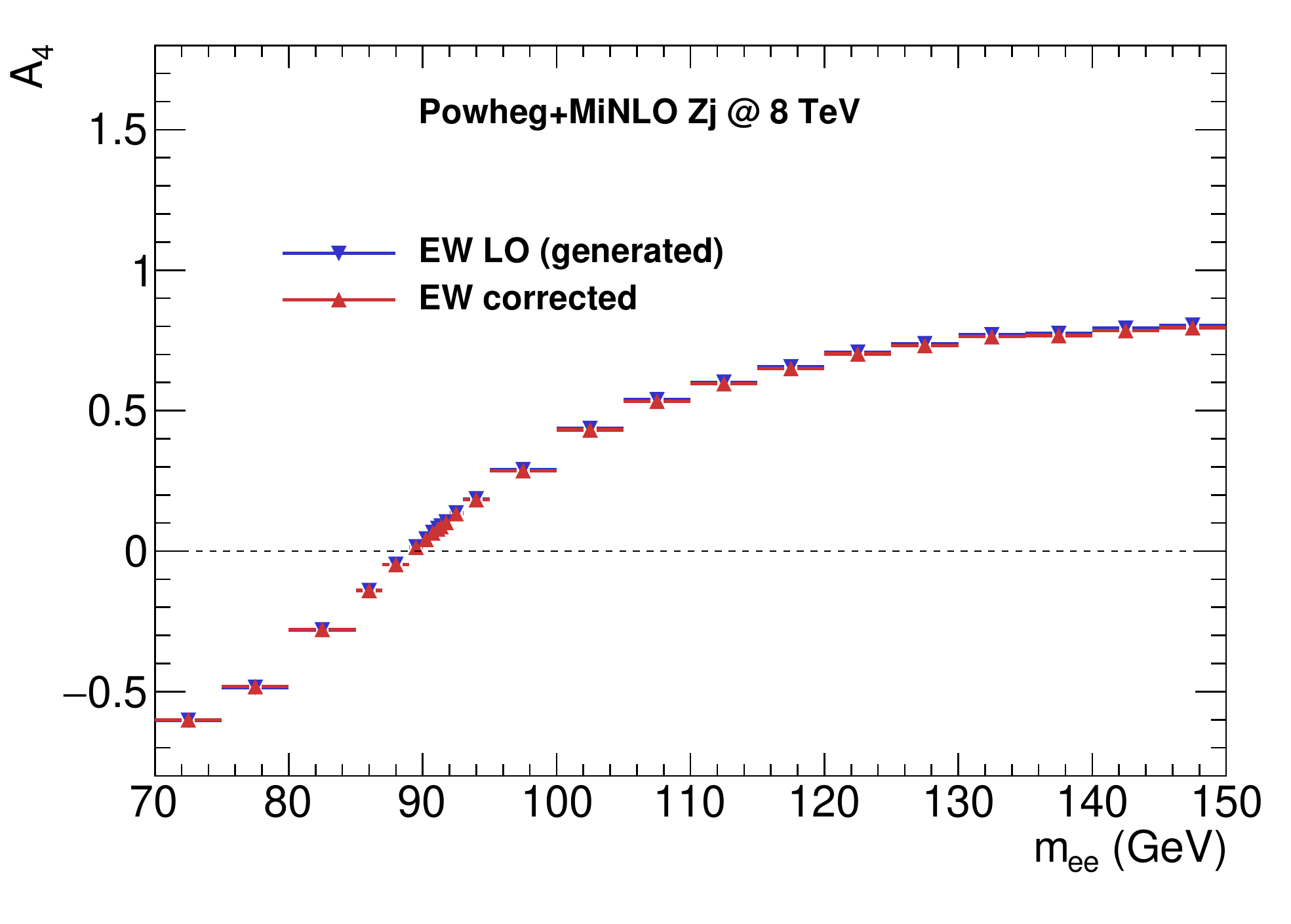}
  \includegraphics[width=7.5cm,angle=0]{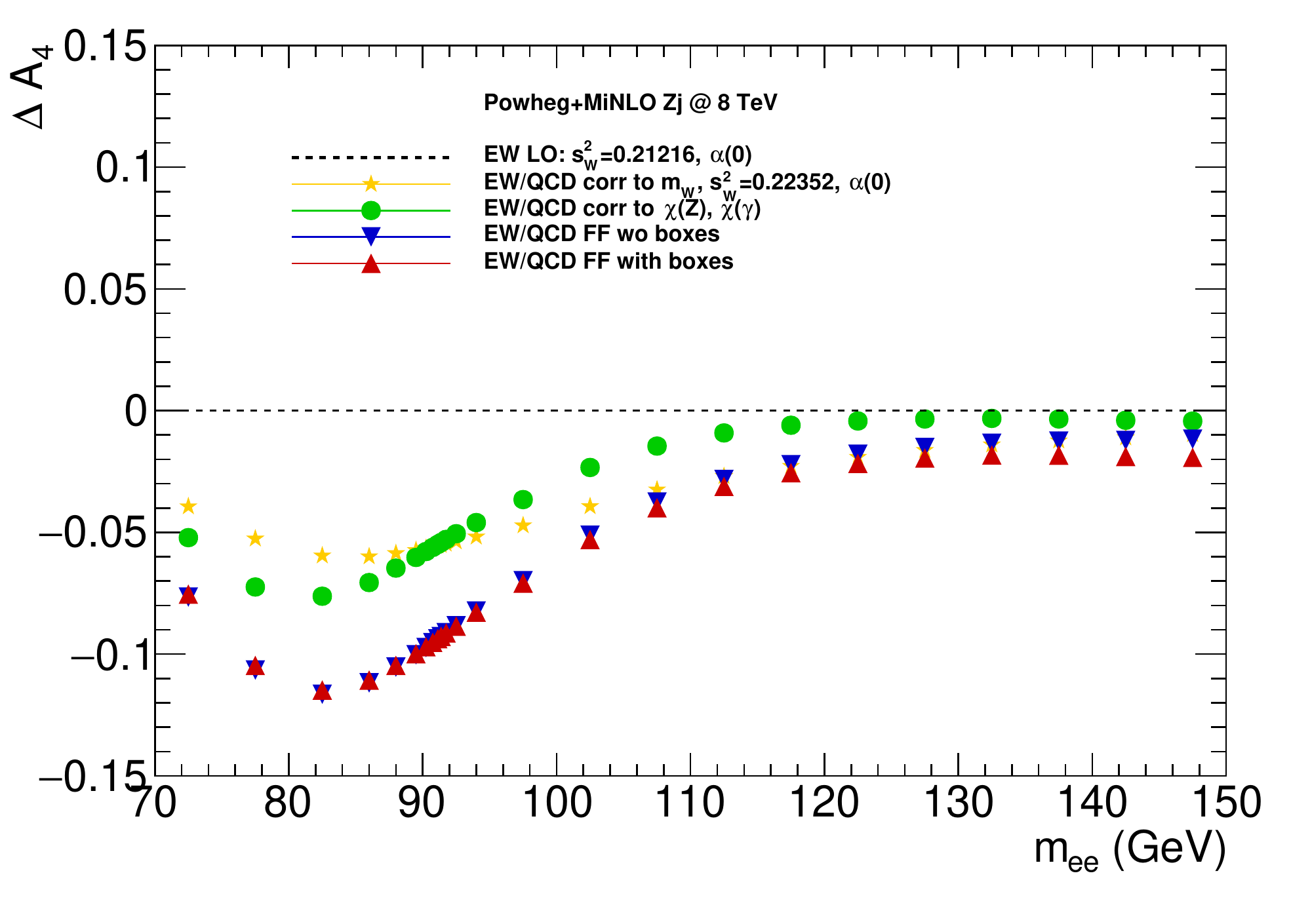}
  \includegraphics[width=7.5cm,angle=0]{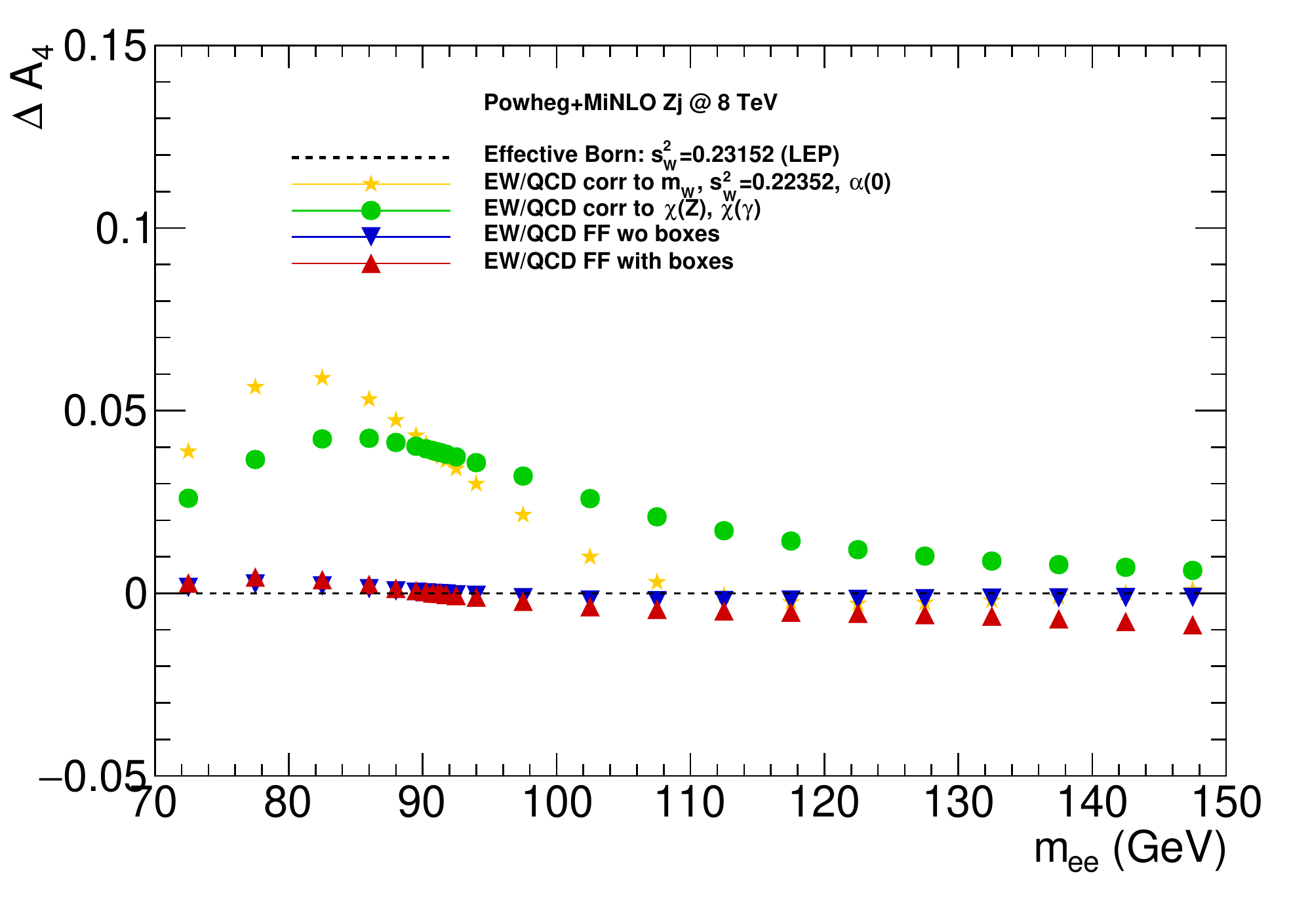}
  \includegraphics[width=7.5cm,angle=0]{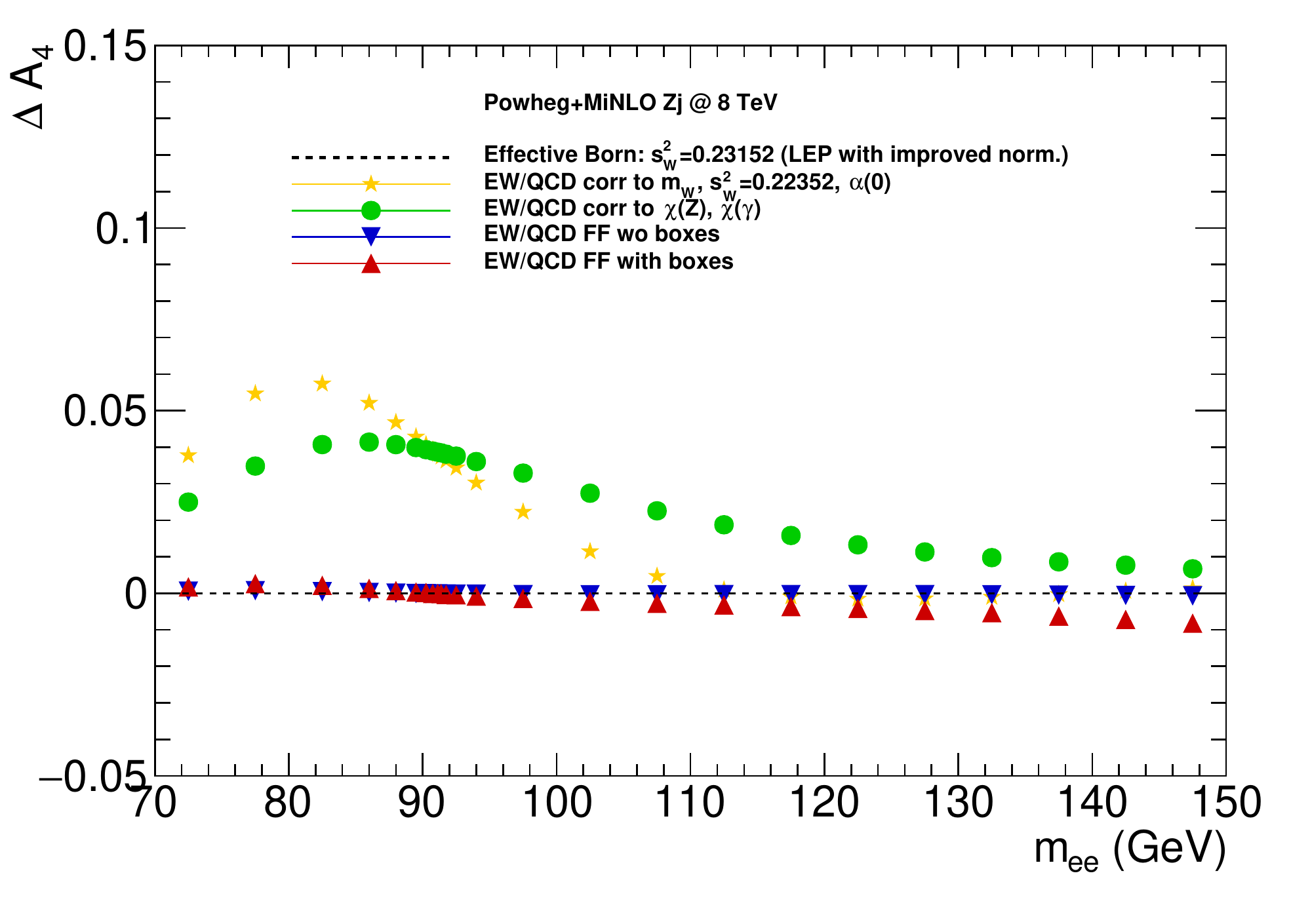}
}
\end{center}
  \caption{Top-left: the $A_4$ as function of $m_{ee}$. Overlayed are generated 
    and EW corrected  $A_4$ predictions. These results are barely distinguishable.
    The differences $\Delta A_{4} = A_4 - A_4^{ref}$ due to gradually introduced EW corrections are
    shown in consecutive plots, where  as a reference $A_4^{ref}$  (black dashed line) respectively
    (i)  EW LO  $\alpha(0)$ scheme (top-right),
    (ii)  effective Born ({\it LEP}) (bottom-left) and (iii) 
    effective Born ({\it LEP with improved norm.}) (bottom-right) was used.
\label{Fig:A4_mee} }
\end{figure*}

Similarly as for $A_{FB}$, the EW corrections change overall size and the shape of $A_4$
as a function of $m_{ee}$;
particularly around the $Z$-pole.  In Fig.~\ref{Fig:A4_mee} (top-right), the $A_4$  for generated
sample (EW LO)
and EW corrected   is shown as a function of  $m_{ee}$.
In the following plots of the figure  details are studied. The $\Delta A_4 = A_4 - A_4^{ref}$
with gradually introduced EW corrections is shown and
compared with the following  reference choices for $A_{4}^{ref}$:
%For the reference the following choices:
(i) EW LO $\alpha(0)$ scheme,
(ii) effective Born {\it (LEP)} and (iii) effective Born {\it (LEP with improved norm.)}.
Conclusions are very similar as for previous $\Delta A_{FB}$ discussion. Note that $\Delta A_4$ and $\Delta A_{FB}$
scale approximately with the  relation $A_4 = 8/3 A_{FB}$. 

\begin{figure*}
  \begin{center}                               
{
  \includegraphics[width=7.5cm,angle=0]{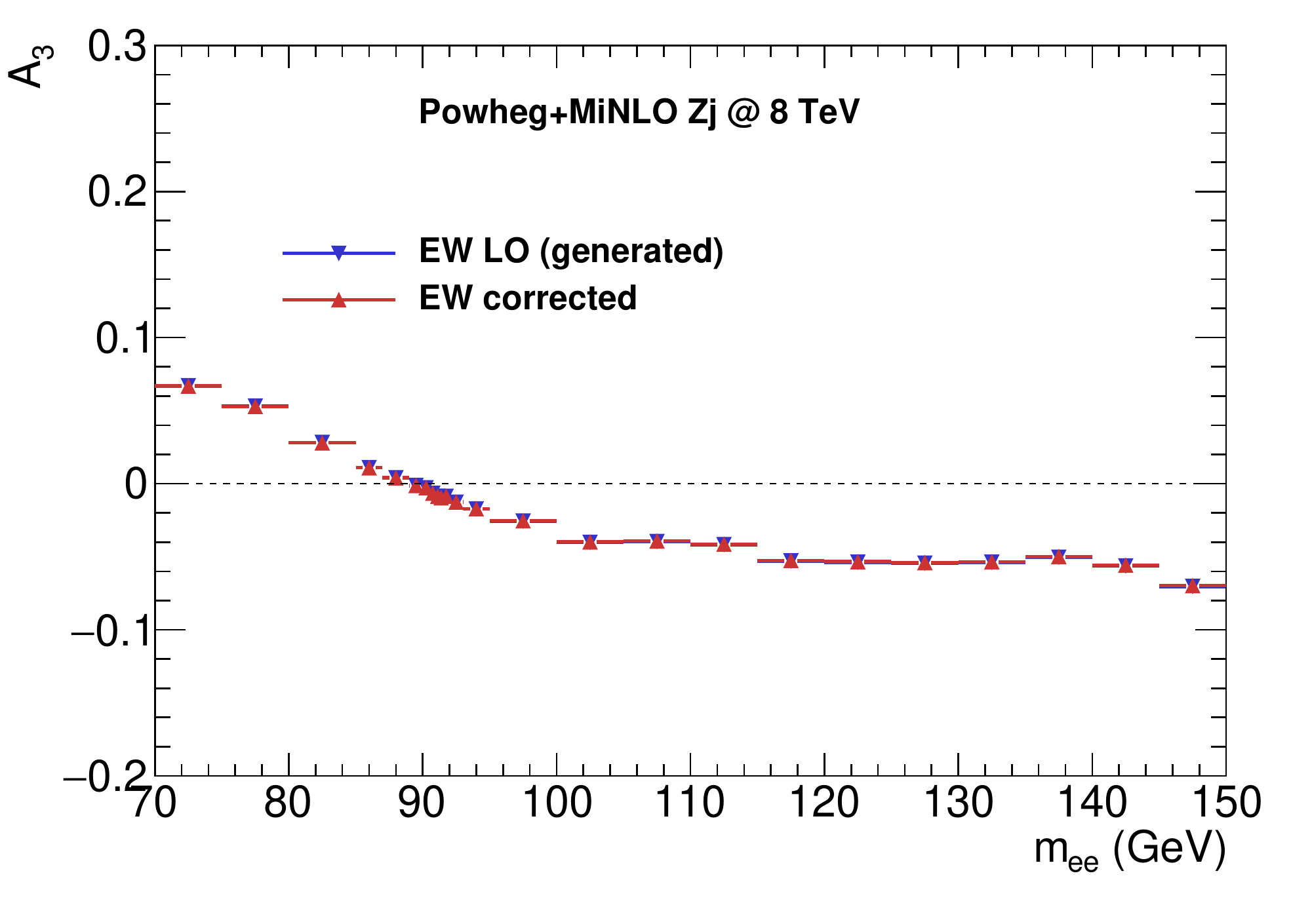}
  \includegraphics[width=7.5cm,angle=0]{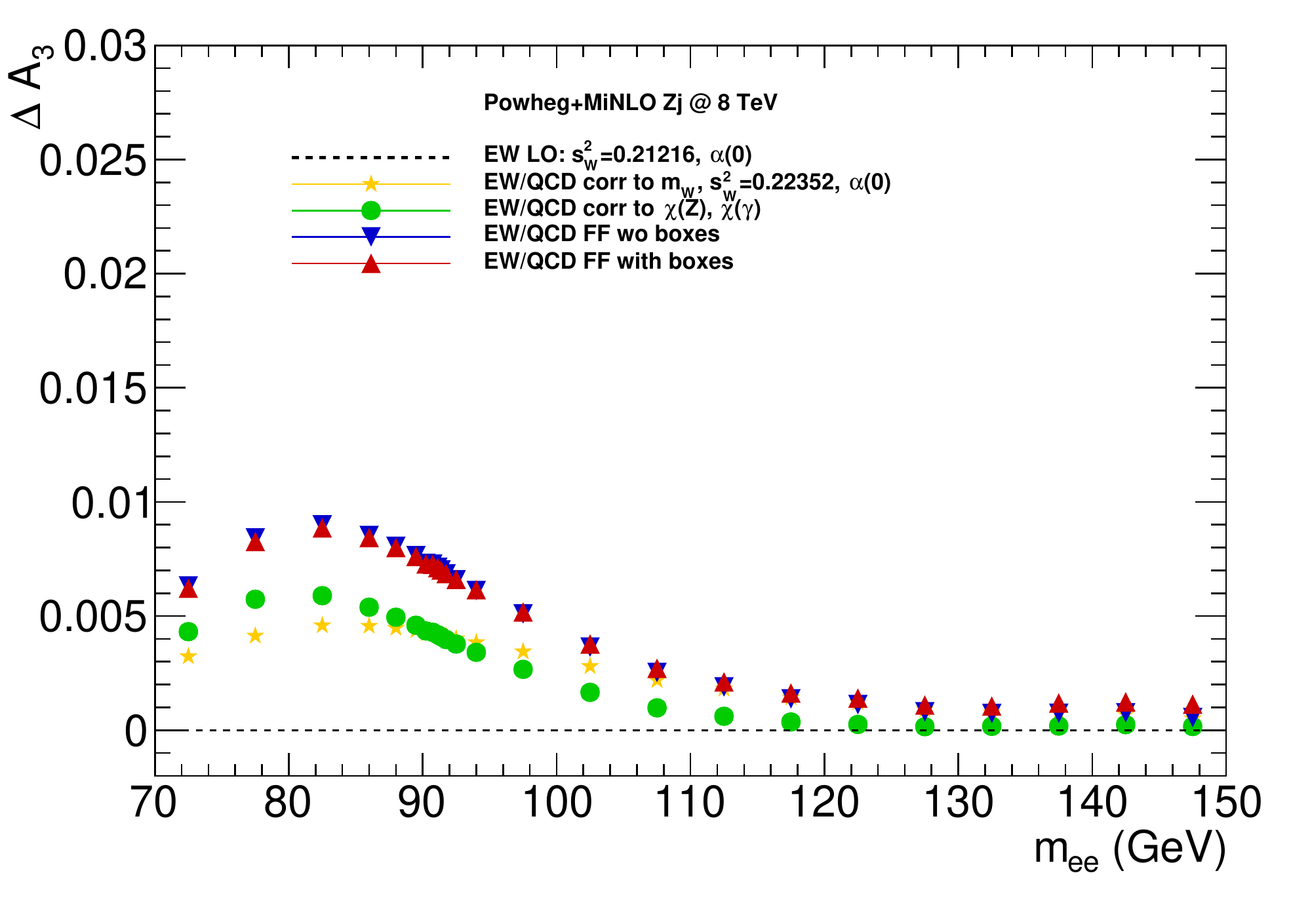}
  \includegraphics[width=7.5cm,angle=0]{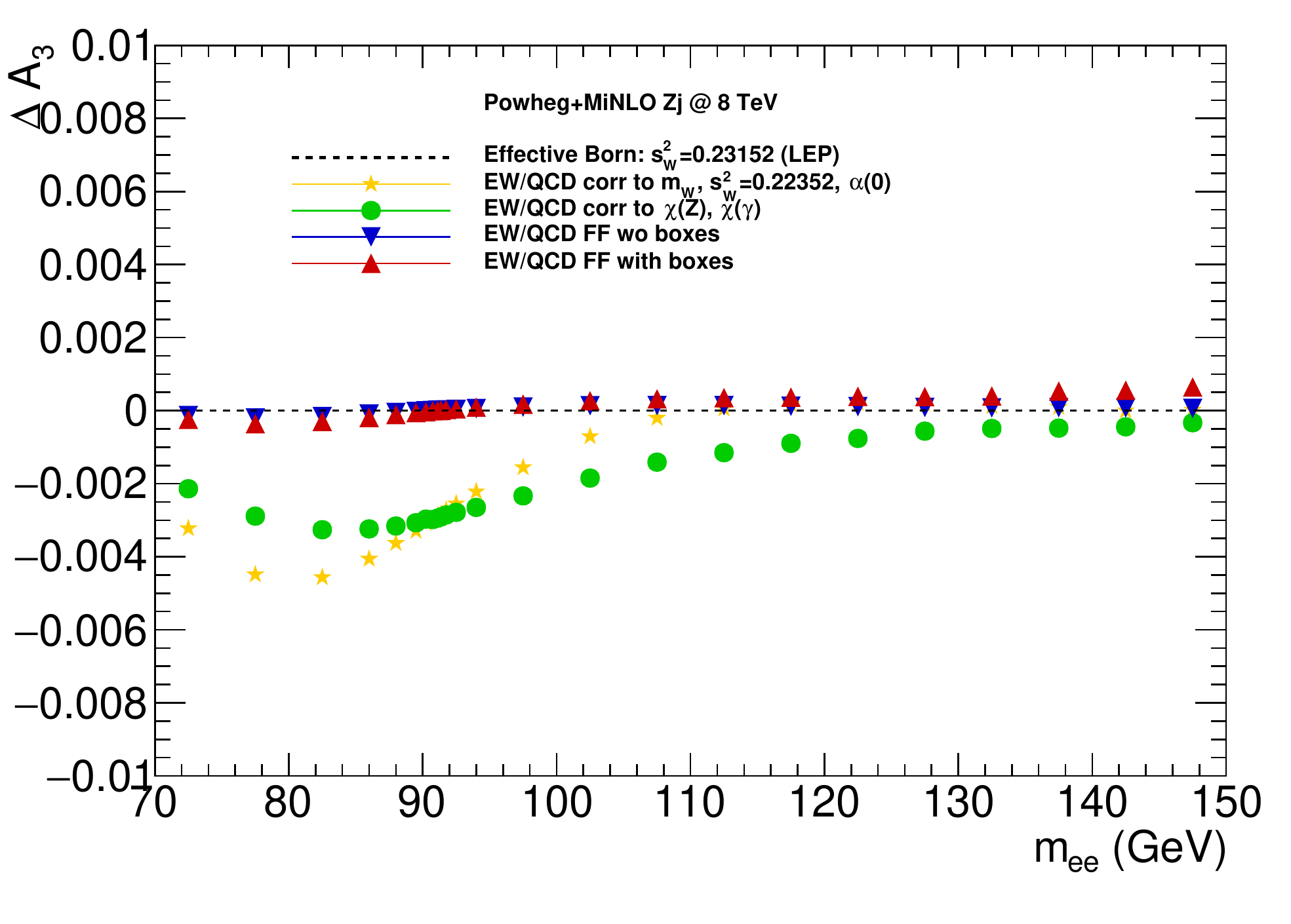}
  \includegraphics[width=7.5cm,angle=0]{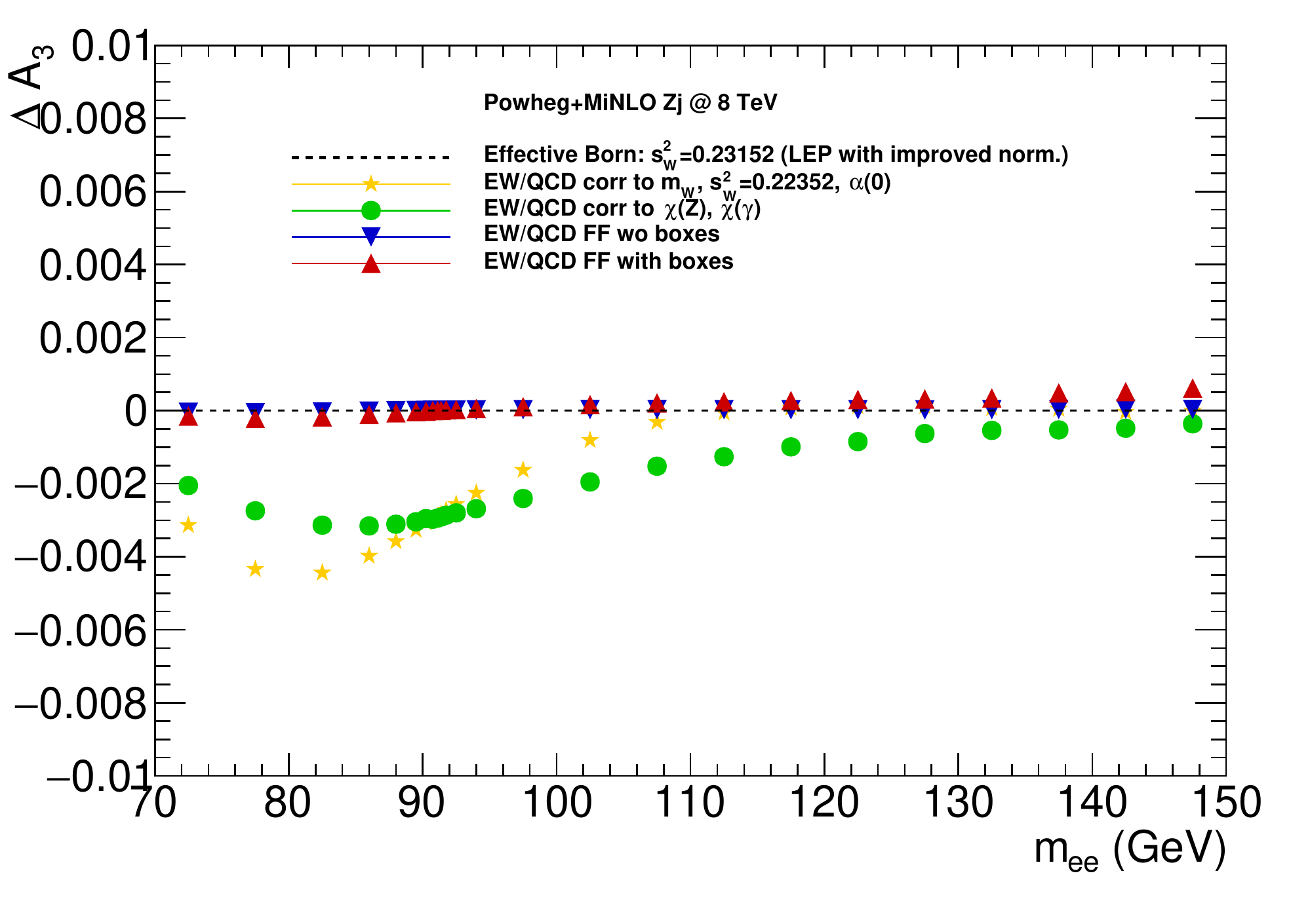}
}
\end{center}
  \caption{Top-left: the $A_3$ as function of $m_{ee}$. Overlayed are generated  %$A_3^{ref}$  (EW LO $G_{\mu}$ scheme)
    and EW corrected $A_3$ predictions. These results are barely distinguishable.
    The differences $\Delta A_3 = A_3 - A_3^{ref}$ due to gradually introduced EW corrections are
     shown in consecutive plots, where  as a reference $A_3^{ref}$ (black dashed line) respectively
    (i) EW LO  $\alpha(0)$ scheme (top-right),
    (ii) effective Born ({\it LEP}) (bottom-left) and (iii) 
    effective Born ({\it LEP with improved norm.}) (bottom-right) was used. In this case,
    the EW weight is calculated with $\cos \theta^{Mustraal}$.
\label{Fig:A3_mee_optFR1} }
\end{figure*}

The analogous set of plots, Fig.~\ref{Fig:A3_mee_optFR1}, is prepared
for $A_3$. In this case, only the {\it Mustraal} frame turned out to be adequate for
$wt^{EW}$ calculation. Both the $\cos \theta^*$ and $\cos \theta^{CS}$  were unable to fully capture
the effects of EW corrections.

%In Tables~\ref{Tab:A4EWcorr} and~\ref{Tab:A3EWcorr},
The results for
 $\Delta A_{3}$ are collected in Table~\ref{Tab:A3EWcorr}.
 The mass  window $80 < m_{ee} <100$~GeV and $p_T^{ee} < 30$~GeV 
are chosen. The estimation for   $\Delta A_4$ differ little
if  $\cos \theta^*$,  $\cos \theta^{CS}$ or $\cos \theta^{Mustraal}$ is
used for calculations of EW corrections. The  $\Delta A_3$ is non-zero,
as it should be, only
if the  $\cos \theta^{Mustraal}$ is  used in $wt^{EW}$ calculation. 
For  $A_{4}$, multiplied by $\frac{8}{3}$ entries of Table~\ref{Tab:AFBEWcorr}
are good enough.

%\begin{table}
% \vspace{2mm}
% \caption{The $\Delta A_4$ shift  of the $A_4$, due to EW corrections. It is averaged over 
%   $p_T^{ee} < $ 30 GeV and $80 < m_{ee} < 100$~GeV ranges.
%   The  $\cos \theta^{CS}$ is used for angular polynomials but for the EW weight calculation
%   $\cos \theta^*$, $\cos \theta^{Mustraal}$ or $\cos \theta^{CS}$ are respectively used.} 
% \label{Tab:A4EWcorr}
% \begin{center}
%    \begin{tabular}{|l|c|c|c|}
%        \hline\hline
%         Corrections to $A_4$  ($p_T^{ee} < $ 30 GeV) & $wt^{EW}(\cos \theta^*)$ & $wt^{EW}(\cos \theta^{Mustraal})$  & $wt^{EW}(\cos \theta^{CS})$   \\ 
%         \hline \hline
%         $A_4$({\it LEP}) - $A_4$(EW/QCD FF with boxes)                       &   0.0001 &   0.0002  &   0.0002    \\
%         \hline
%         $A_4$({\it LEP with improved norm.}) - $A_4$(EW/QCD FF with boxes)   &   0.0001 &   0.0002  &   0.0002     \\
%    \hline
% \end{tabular}
%  \end{center}
%\end{table}
\begin{table*}
 \vspace{2mm}
 \caption{The $\Delta A_3$ shift of the $A_3$, due to EW corrections, averaged over
   $p_T^{ee} < $ 30 GeV and $80 < m_{ee} < 100$ GeV ranges.
 The  $\cos \theta^{CS}$ is used for  angular polynomials but for the EW weight  calculation 
 $\cos \theta^*$, $\cos \theta^{Mustraal}$ or $\cos \theta^{CS} $ are
 used respectively.} 
 \label{Tab:A3EWcorr}
 \begin{center}
    \begin{tabular}{|l|c|c|c|}
        \hline\hline
         Corrections to $A_3$ ($p_T^{ee} < $ 30 GeV) & $wt^{EW}(\cos \theta^*)$ & $wt^{EW}(\cos \theta^{Mustraal})$  & $wt^{EW}(\cos \theta^{CS})$   \\ 
         \hline \hline
         $A_3$(EW/QCD corr. to $m_W$) - $A_3$(EW LO $\alpha(0)$)                            &  -0.00060 &  -0.00321  &  -0.00060     \\
         \hline 
         $A_3$(EW/QCD corr. to $\chi(Z),\chi(\gamma)$) - $A_3$(EW LO $\alpha(0)$)    &  -0.00061 &  -0.00322  &  -0.00061    \\
         \hline 
         $A_3$(EW/QCD FF no boxes) - $A_3$(EW LO $\alpha(0)$)                  &  -0.00103 &  -0.00546  &  -0.00102     \\
         \hline 
         $A_3$(EW/QCD FF with boxes) - $A_3$(EW LO $\alpha(0))$                &  -0.00103 &  -0.00545  &  -0.00102    \\
         \hline \hline
         $A_3$({\it LEP}) - $A_3$(EW/QCD FF with boxes)                       &   0.00000 &   0.00000  &   0.00000    \\
         \hline
         $A_3$({\it LEP with improved norm.}) - $A_3$(EW/QCD FF with boxes)   &   0.00000 &   0.00000  &   0.00000     \\
    \hline
 \end{tabular}
  \end{center}
\end{table*}

\section{Summary} \label{sec:sumary}

In this paper we have shown how the EW corrections for double-deconvoluted observables at LHC
can be evaluated using {\it Improved Born Approximation}.
We have exploited a wealth of the LEP era results encapsulated in
the {\tt Dizet} library developed at  that time. We have used that
formalism to calculate and present numerically EW corrections for
doubly-deconvoluted observables,
such as $Z$-boson line-shape, forward-backward asymmetry $A_{FB}$, effective
weak mixing angles or lepton direction angular coefficients.

We have followed  largely discussions available in {\tt Dizet} documentation.
We have introduced the notion of the effective
Born and explained how Monte Carlo events generated at NLO QCD can be transformed to reduced kinematics,
of strong interaction lowest
order, for the calculation of spin amplitudes $q \bar q \to Z/\gamma^* \to \ell \ell$.
This could be achieved thanks to  properties of spin
amplitudes discussed
in ~\cite{Richter-Was:2016mal,Richter-Was:2016avq}. We explained how per-event weight
$wt^{EW}$, can be build and used to attribute EW corrections to  already generated
events.

We have  re-visited the notion of {\it Effective Born} with {\it LEP} (or {\it with LEP of improved norm.})
parametrisations where  dominant parts of EW corrections are taken into accout
with a redefinition of coupling constants. 
We have  evaluated how well it works for observables of the paper.
The discussed approach for treating EW corrections for Drell-Yan process in pp collisions has been implemented
in the {\tt Tauola/TauSpinner} package~\cite{Davidson:2010rw,Czyczula:2012ny} to be available starting from the forthcoming release.

Once the formalism was explained, numerical results of EW corrections to the $Z$-boson line-shape,
for\-ward-backward asymmetries, lepton angular coefficients were presented.  Results were obtained
using {\tt Dizet}  for calculating EW form-factors and {\tt Tauola/TauSpinner} for calculating respective
EW weights of {\it Improved Born Approximation} or {\it Effective Born} with {\it LEP} (or with {\it LEP  improved norm.})
parametrisations.

The choice of the version of EW library was dictated by the compatibility with
the {\tt KKMC} Monte Carlo~\cite{Jadach:2013aha}, the program widely used at the LEP times.
%but in the same manner results of other versions of EW calculations
%~\cite{Andonov:2008ga, Akhundov:2013ons}
%can be interfaced as well.
It relies on a published version of {\tt Dizet}, thus suits the purposes of a reference point well. Also,
  omitted effects are rather small. In the future,
  the algorithm of {\tt TauSpinner}  can be useful to quantify the
   differences among distinct 
   implementations of the electroweak sector.
   
   The numerical studies with the updates to {\tt Dizet} version 6.42 ~\cite{Andonov:2008ga,Akhundov:2013ons} and with other, sometimes
   unpublished electroweak codes are left  for the future work.
   One should stress the  necessity of such future  numerical discussion and
   updates, in particular 
due to the photonic vacuum polarization, e.g. as provided in refs.~\cite{Burkhardt:2005se,Jegerlehner:2017zsb} but absent
in  the last published (or presently public) version of {\tt Dizet} 6.42.
This update is required already at LHC precision of $Z$-boson couplings measurements.

In many applications focused on challenges of strong interactions, electroweak corrections
are receiving rather minimal attention and in particular $Z$ boson fixed value width,
or   running only in proportion to   the energy transfer, is used. This may be inappropriate
for large $s$ as found e.g. in \cite{Seymour:1995np}. {\tt TauSpinner} can be used to evaluate numerical consequences
of such approximation.
Finally let us mention that 
presented  implementation of EW corrections as per-even weight, was already found  useful
for experimental measurements \cite{ATLAS:2018gqq} at LHC  and for
discussions during  recent workshops, see  e.g. Ref.~\cite{workshop}.

\vskip 0.2 cm
\centerline{\bf \Large Acknowledgements}
\vskip 0.2 cm
E.R-W. would like to thank  Daniel Froidevaux, Aaron Ambruster and colleagues from ATLAS Collaboration Standard Model Working Group
for numerous inspiring discussions on the applications of presented here implementation of EW corrections to the
$\sin^2\theta_{eff}^{lep}$ measurement at LHC.

This project was supported in part from funds of Polish National Science
Centre under decision UMO-2014/15/B/ST2/00049.
Majority of the numerical calculations were performed at the PLGrid Infrastructure of 
the Academic Computer Centre CYFRONET AGH in Krakow, Poland.

%\bibliography{paper-EWcorr}{}
%\bibliographystyle{utphys_spires}
%\providecommand{\href}[2]{#2}\begingroup\begin{thebibliography}{10}
%\end{thebibliography}\endgroup

%\bibliography{paper-resume}{}
\vskip 0.1 cm
%\bibliography{paper_main}{}
%\bibliographystyle{utphys_spires}
\providecommand{\href}[2]{#2}
\appendix

\section{Comment on technical details of {\tt TauSpinner} EW effects  implementation } \label{app:TauSpinnerInit}

Although the framework of {\tt Tauola/TauSpinner} package~\cite{Davidson:2010rw,Czyczula:2012ny}
has been used for numerical results presented in this paper, the code is not yet available
with the public release but only in the private distribution and only partly in development
version~\cite{wwwTauola} which updates
itself daily from our work repository.
Tests and some of the code developments need to be completed. Once we achieve confidence
the official stable version of the code will become public at~\cite{wwwTauola} .
Let us nonetheless list  main points of the implementation which was
already used to obtain numerical results:
\begin{itemize}
\item
  Pre-tabulated EW corrections: form-factors, vacuum polarization corrections in form
  of 2D root histograms or alternatively ASCII files
  of the {\tt KMMC} project~\cite{Jadach:2013aha} were used to assure modularity and to enable
  graphic tests.
\item
  Functions to calculate $\cos \theta^*$, $\cos \theta^{Mustraal}$,  $\cos \theta^{CS}$ from
  kinematics of outgoing final state (leptons and partons/jets) used for numerical results
  are already in part available in
  {\tt TAUOLA/TauSpinner/examples/} {\tt Dizet-example} directory. The {\tt README} file of that  
  directory is gradually filled with technical details.
\item
  Routine to initialize parameters of the Born function is provided. The
   {\tt SUBROUTINE INITWK} of {\tt TAUOLA/}\\
  {\tt src/tauolaFortranInterfaces/tauola\_extras.f}
  has been copied and extended. It is available under the name {\tt INITWKSWDELT }, with the following input:
  \begin{itemize}
  \item
    $G_{\mu}$, $\alpha$, $M_Z$, $s$, 
  \item
    EW form-factors and vacuum polarization corrections,
  \item
    $s^2_W$ and parameters for couplings variations $\delta_{s2W}$, $\delta_{V}$, see Section~\ref{app:SW2scan}
    for details.
  \end{itemize}
\item
  To calculate $d\sigma_{Born}$ and the   $wt^{EW}$ the  {\tt t\_bornew} function
  with flexible  options for EW scheme and $\delta_{s2W}$, $\delta_{V}$, is prepared.
  It is used by {\tt TauSpinner} library function  
  {\tt default\_nonSM\_born(ID, S, cost, H1, H2, key)}
  now.
\item
  It is premature for complete  documentation, but  comments on the software used to obtain numerical results
  are  in place.
\end{itemize}

\section{How to vary $s^2_W$ beyond the EW LO schemes.} 
\label{app:SW2scan}

In the discussed EW scheme $(\alpha(0), G_\mu, M_Z)$, the  $s^2_W$ is not directly available for fits.
It is calculated from  relation~(\ref{Eq:sw2onshell}) of the Standard Model.
One possibility to vary  $s^2_W$, but stay within Standard Model framework is to
 vary some other constants which impact $s^2_W$.
The candidates  within Standard Model, which are also inputs to the
{\tt Dizet} library, are $G_{\mu}$ or $m_{t}$.
From the simple estimates, to allow $\pm 100 \cdot 10^{-5}$ variation of  $s^2_W$,
those parameter will have to be varied far beyond their
experimental ambiguities\footnote{%
Range would be $\pm 10$ GeV for $m_t$ or $\pm 4\cdot 10^{-8} GeV^{-2}$ for $G_\mu$.}.

\begin{figure*}
  \begin{center}                               
{
  \includegraphics[width=7.5cm,angle=0]{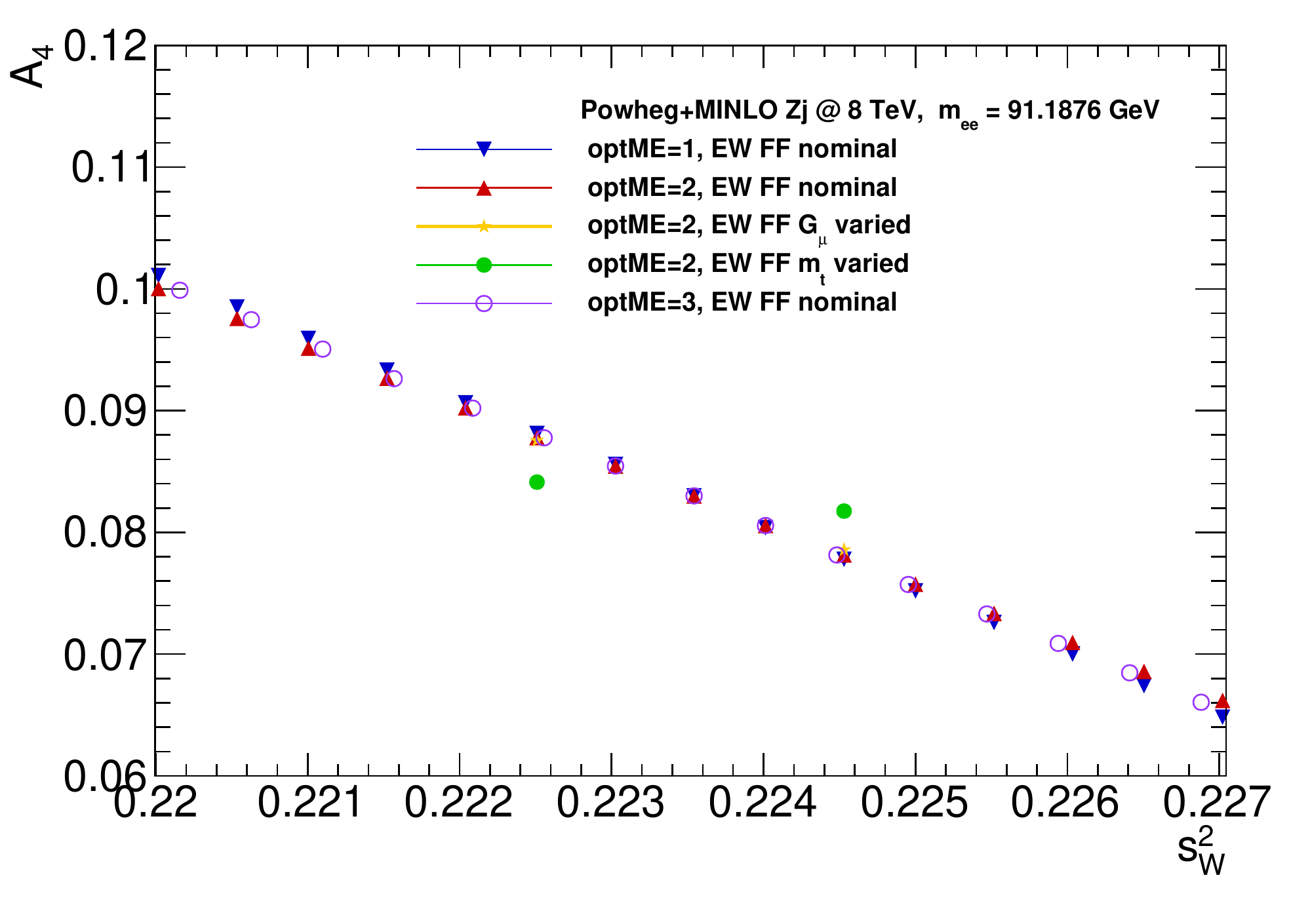}
  \includegraphics[width=7.5cm,angle=0]{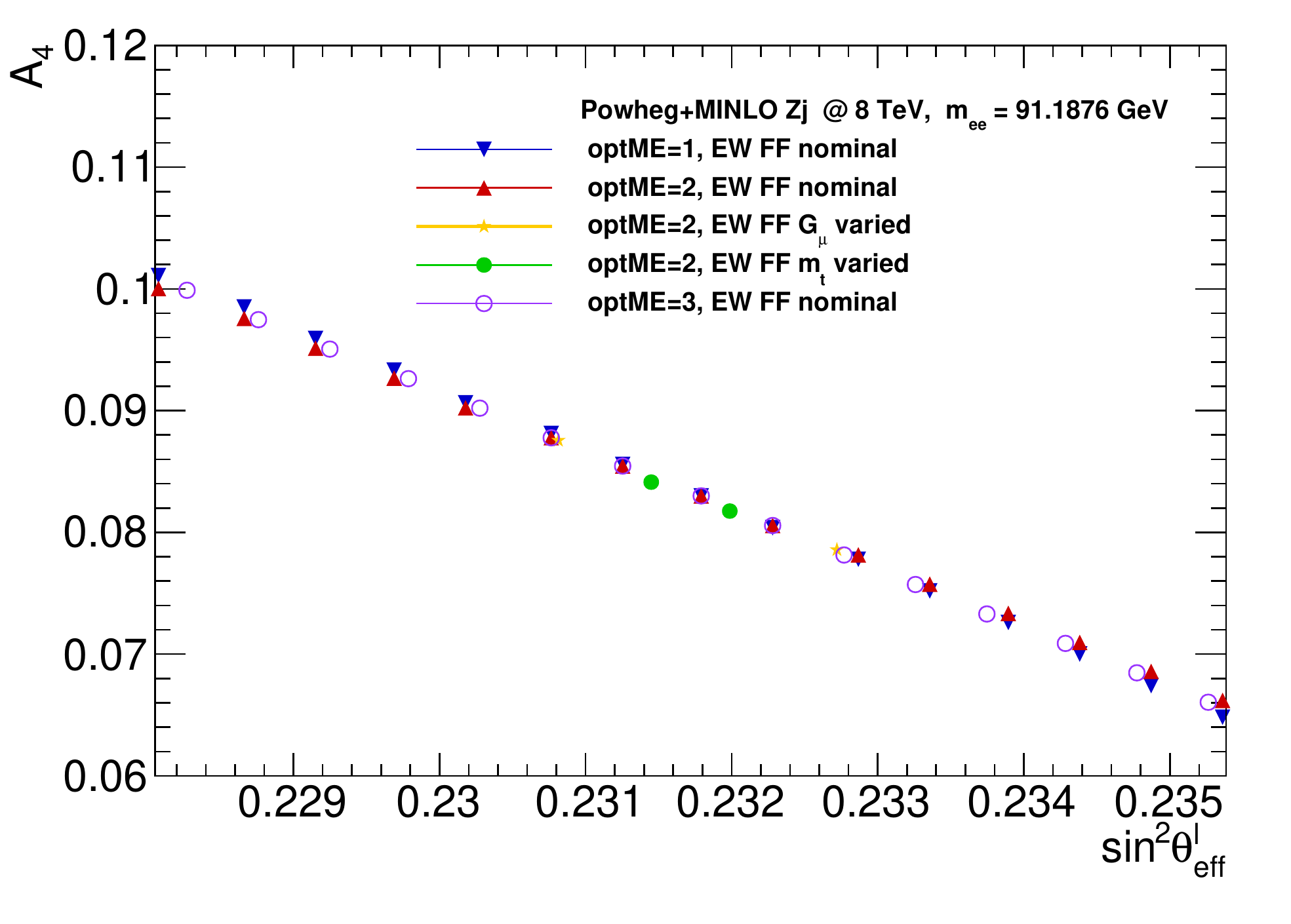}
}
\end{center}
  \caption{The  $A_4$ variation due to shifts induced with the presented
    in Appendix \ref{app:SW2scan} options;
    as a function
    of $s^2_W$ (left-hand side) and as a function of  $sin^2\theta_{eff}^l$ (right-hand side).
    The ``FF $G_\mu$ varied'', FF $m_t$ varied'' correspond to the case when
    form-factors were recalculated. Otherwise they were kept at nominal values.
\label{Fig:A4_scan} }
\end{figure*}

One can extend formulae for $\mathscr A^{Born+EW}$ (\ref{Eq:BornEW})  beyond the Standard Model too.
Additional v-like contribution  to $Z$-boson $v_{\ell},  v_f$ couplings can be introduced with $\delta_{S2W}$ or
$\delta_{V}$ as presented later.  
Below few details and options on implementation
into $\mathscr A^{Born+EW}$ amplitudes are given:

\begin{itemize}
\item
  {\tt optME = 1}: introduce unspecified heavy particle coupling to the $Z$-boson, to modify  fermions
  vector couplings
\begin{eqnarray}
   \label{Eq:avLODelt}
  v_{\ell} = && (2 \cdot T_3^{\ell} - 4 \cdot q_{\ell} \cdot (s^2_W + \delta_{S2W}) \cdot {\mathscr K}_{\ell}(s,t))/\Delta, \nonumber \\
  v_f     = && (2 \cdot T_3^f - 4 \cdot q_f \cdot (s^2_W + \delta_{S2W})  \cdot {\mathscr K}_f(s,t))/\Delta, \nonumber \\
%\end{eqnarray}
%\begin{eqnarray}  
  vv_{\ell f} = && \frac{1}{v_{\ell} \cdot v_f} [
    ( 2 \cdot T_3^{\ell}) (2 \cdot T_3^f) \nonumber\\
    && - 4 \cdot q_{\ell} \cdot (s^2_W+ \delta_{S2W}) \cdot {\mathscr K}_f(s,t)( 2 \cdot T_3^{\ell}) \\
    &&  - 4 \cdot q_f \cdot (s^2_W+ \delta_{S2W}) \cdot {\mathscr K}_{\ell}(s,t) (2 \cdot T_3^f) \nonumber \\
    && + (4 \cdot q_{\ell} \cdot s^2_W) (4 \cdot q_f \cdot s^2_W) {\mathscr K}_{\ell f}(s,t)  \nonumber \\
    && + 2 \cdot (4 \cdot q_{\ell})) (4 \cdot q_f \cdot) \cdot s^2_W \cdot \delta_{S2W} ) {\mathscr K}_{\ell f}(s,t) ]\ \frac{1}{\Delta^2} \nonumber 
   \label{Eq:vvNLODelt}
\end{eqnarray}
but do not  alter 
\begin{equation}
  \Delta = \sqrt{ 16 \cdot s^2_W \cdot (1 - s^2_W)} 
\end{equation}
or any other  $\mathscr A^{Born+EW}$ (\ref{Eq:BornEW}) couplings or  calculations of the EW form-factors.%, see Ref.~\cite{MamyCosPytanie}. 
\item
  {\tt optME = 2}:  recalculate $M_W$ for numerically modified
  $m_t$ or $G_\mu$ and modify accordingly Standard Model
  $s^2_W = 1 -M_W^2/M_Z^2$,
  for  $s^2_W $ present in  $\mathscr A^{Born+EW}$.
  The form-factors are (are not) recalculated\footnote{
The {\tt optME = 1, 2}, if form-factors are not recalculated,
formally differ by the term proportional to $\delta^2_{S2W}$ and only in the expression for  $vv_{\ell f}$.
Change of input parameters $G_{\mu}$ or $m_{t}$  as a source for $s^2_W$ variations in  {\tt optME = 2},
 implies changes of the couplings and thus for consistency, recalculation of form-factors.
All these options can be realized with the {\tt Tauola/TauSpinner}
package, of the development version.
 }.
  In total, 3 variants
  of this option were used for Fig.~\ref{Fig:A4_scan}. 
\item
  {\tt optME = 3}: similar as {\tt optME = 1} but redefine directly fermions vector couplings  with
  $\delta_{V}$.
   We keep relative normalization (charge structure) of $\delta_{V}$
    similar to $\delta_{S2W}$, to facilitate comparisons.
  Then
\begin{eqnarray}
   \label{Eq:avLODeltBis}
  v_{\ell} = && (2 \cdot T_3^{\ell} - 4 \cdot q_{\ell} \cdot (s^2_W \cdot {\mathscr K}_{\ell}(s,t) + \delta_{V} ))/\Delta, \nonumber \\
  v_f     = && (2 \cdot T_3^f - 4 \cdot q_f \cdot (s^2_W \cdot {\mathscr K}_f(s,t) + \delta_{V} ))/\Delta, \nonumber \\
%\end{eqnarray}
%\begin{eqnarray}  
  vv_{\ell f} = && \frac{1}{v_{\ell} \cdot v_f} [
    ( 2 \cdot T_3^{\ell}) (2 \cdot T_3^f) \nonumber\\
    && - 4 \cdot q_{\ell} \cdot (s^2_W \cdot {\mathscr K}_f(s,t) + \delta_{V} ) ( 2 \cdot T_3^{\ell}) \\
    && - 4 \cdot q_f \cdot (s^2_W \cdot {\mathscr K}_{\ell}(s,t) + \delta_{V} ) (2 \cdot T_3^f) \nonumber \\
    && + (4 \cdot q_{\ell} \cdot s^2_W) (4 \cdot q_f \cdot s^2_W) {\mathscr K}_{\ell f}(s,t)  \nonumber \\
    && + 2 \cdot (4 \cdot q_{\ell})) (4 \cdot q_f \cdot) \cdot s^2_W \cdot  {\mathscr K}_{\ell f}(s,t) \cdot \delta_{V} ]\ \frac{1}{\Delta^2}. \nonumber 
   \label{Eq:vvNLODeltBis}
\end{eqnarray}
The $\delta_{V}$ shift is almost equivalent to $\delta_{S2W}$  shift, but affects couplings in a $(s,t)$
independent manner.
\end{itemize}

 Even though discussion of $s^2_W$ variation necessary for fits, is generally out of scope of the present paper
  and it can not be now exhausted, let us provide some numerical results to illustrate
  stability of the method\footnote{%
    For {\tt optME=2}, the $m_t$
(or $G_{\mu}$) have been   shifted  to move  $s^2_W$ by $\pm 100 \cdot 10^{-5}$.
Then  the form-factors
were recalculated, or optionally kept at nominal values.}.
The variations for   $A_4(M_Z)$ are presented in Figure~\ref{Fig:A4_scan}:
 as a function of $s^2_W$ on the left-hand side plot and  as a function of $sin^2\theta_{eff}^l$
on the right-hand side plot.
It is very reassuring, that all presented {\tt optME} methods lead
to the same slope of the  $A_4(M_Z)$ as a function of  $sin^2\theta_{eff}$.
Very similar curve could be presented for $A_{FB}$, which would be scaled
by $\frac{3}{8}$ with respect to  $A_4$ only.
% with these new values with the impact on  $sin^2\theta_{eff}$. 
%{The technical arrangements for further studies of the phenomenological significance
%  are left for the forthcoming paper and for the work within experiments.}

\section{Initialization of the {\tt Dizet} library} \label{app:DizetInit}
There is a wealth of  initialization constants and 
options available  for  {\tt Dizet} library.
%and EW table writing program as installed in {\tt KKMC}
%(directory {\tt KKMC/KK-all/dizet}).
The documentation of that program and of its interface for {\tt KKMC},
explains options available for the {\tt TauSpinner} users as well.
Tables \ref{TabApp:GammaZCoupl_1} and \ref{TabApp:GammaZCoupl_2} recall  available
{\tt Dizet} initialization,
 Table \ref{TabApp:GammaZCoupl_3}
 lists calculated by {\tt Dizet} quantities for the use in {\tt TauSpinner}
 library.

%\begin{sidewaystable}
\begin{table}
 \vspace{2mm}
 \caption{ {\tt Dizet} initialization parameters: masses and couplings.} 
 \label{TabApp:GammaZCoupl_1}
  \begin{center}
    \begin{tabular}{|l|c|c|}
        \hline\hline
         Parameter  & Value & Description \\ 
         \hline\hline
      $M_Z$    &  91.1876 GeV  &  mass of $Z$ boson \\
      $M_h$    &  125.0  GeV  &  mass of Higgs boson \\
      $m_t$    &  173.0  GeV  &  mass of top quark \\
      $1/\alpha(0)$    &  137.0359895(61)   & $\alpha_{QED}(0)$ \\
         $G_{\mu}$    &  $1.166389(22) \cdot 10^{-5}$   & Fermi constant  \\
                   &  GeV$^{-2}$  & in $\mu$-decay \\
  \hline
  \end{tabular}
  \end{center}
  %\end{sidewaystable}
\end{table}

%The {\tt OMS} scheme uses the masses of all fundamental particles, both fermions and bosons, and coupling constants
%$\alpha_{QED}(0)$ or $\alpha_{QED}(M^2_Z)$. The ill-defined masses of light quarks are replaced by $\alpha(M_Z^2)$ making use of
%dispersion relation  relating the imaginary part of hadronic vacuum polarization
%with the total cross-section $\sigma_{tot}(e^+ e^- \to \gamma^* \to hadrons)$.

{ In the present work, we have relied on the {\tt Dizet} library version as installed in the 
  {\tt KKMC} Monte Carlo \cite{Jadach:1999vf} and used at a time of LEP 1 in detector simulations.
  Already for the  data analysis and in particular for final
  fits \cite{ALEPH:2005ab}, further effects of minor, but non-negligible numerical impact were
  taken into account. Gradually, effects such as
  improved top contributions \cite{Awramik:2004ge} or better photonic vacuum
  polarization \cite{Davier:2017zfy}, were taken into account.
  This has to be updated
  for   {\tt Dizet} library too.
  
  Such update is of importance also for the {\tt KKMC} project itself because of forthcoming
  applications for the Future Circular Collider or for LHC~\cite{Ward:2018qkh}. % The necessary work is forthcoming.
}

\begin{table*}
%\begin{sidewaystable}
 \vspace{2mm}
 \caption{{\tt Dizet} initialization flags. Unmodified  comments taken from the {\tt KKMC} code. } 
 \label{TabApp:GammaZCoupl_2}
  \begin{center}
    \begin{tabular}{|l|c|c|c|}
        \hline\hline
         Internal flag  & Default value & Optional values &  Description \\ 
  \hline\hline 
      ibox  &  1  &  0,1         & EW boxes on/off \\
  \hline
      Ihvp  &  1  &  1,2,3       & Jegerlehner/Eidelman, Jegerlehner(1988), Burkhardt et al. \\
      Iamt4 &  4  &  0,1,2,3,4   & =4 the best, Degrassi/Gambino \\
      Iqcd  &  3  &  1,2,3       &approx/fast/lep1, exact/Slow!/Bardin/, exact/fast/Kniehl \\
      Imoms &  1  &  0,1         &=1 W mass recalculated \\
      Imass &  0  &  0,1         &=1 test only, effective quark masses \\
      Iscre &  0  &  0,1,2       & Remainder terms \\
      Ialem &  3  &  1,3 or 0,2, & for 1,3 DALH5 not input \\
      Imask &  0  &  0,1         & =0: Quark masses everywhere; =1 Phys. threshold in the ph.sp. \\
      Iscal &  0  &  0,1,2,3     & Kniehl=1,2,3, Sirlin=4 \\
      Ibarb &  2  &  -1,0,1,2    & Barbieri??? \\
      Iftjr &  1  &  0,1         & FTJR corrections \\
      Ifacr &  0  &  0,1,2,3     & Expansion of $\delta_r$; =0 none; =3 fully, unrecommed. \\
      Ifact &  0  &  0,1,2,3,4,5 & Expansion of kappa; =0 none \\
      Ihigs &  0  &  0,1         & Leading Higgs contribution re-summation \\
      Iafmt &  1  &  0,1         & =0 for old ZF \\
      Iewlc &  1  &  0,1         & ??? \\
      Iczak &  1  &  0,1         & Czarnecki/Kuehn corrections \\
      Ihig2 &  1  &  0,1         & Two-loop higgs  corrections off,on \\ 
      Iale2 &  3  &  1,2,3       & Two-loop constant corrections in $\delta_{\alpha}$ \\
      Igfer &  2  &  0,1,2       & QED corrections for fermi constant \\
      Iddzz &  1  &  0,1         & ??? DD-ZZ game, internal flag\\
  \hline
  \end{tabular}
  \end{center}
%\end{sidewaystable}
%\end{table}
%\begin{table}
%\begin{sidewaystable}
 \vspace{2mm}
 \caption{{\tt Dizet} recalculated quantities available for the {\tt TauSpinner} use. For details  of the {\it ZPAR} table see Refs.~\cite{Jadach:2013aha,Bardin:1999yd} } 
 \label{TabApp:GammaZCoupl_3}
  \begin{center}
    \begin{tabular}{|l|c|c|}
        \hline\hline
         Parameter  & Value & Description \\ 
         \hline\hline
      $\alpha_{QED} (M_Z^2)$     & 0.007759 & calculated from $\Delta \alpha_h^{(5)}(M_Z)$ by {\tt Dizet} \\
      $1/\alpha_{QED} (M_Z^2)$   & 128.882588 &  \\
      $\alpha_s (M_Z^2)$  & 0.1250   & recalculated by {\tt Dizet} \\
         $\alpha_s (m_t^2)$  & 0.1134   & recalculated by {\tt Dizet} \\
         \hline
         $ZPAR(1) = \delta r$ & 0.03694272 & the loop corrections to $G_{\mu}$\\
         $ZPAR(2) = \delta r_{rem}$ & 0.01169749 & the remainder contribution $O(\alpha)$\\
         $ZPAR(3) = s_W^2$ & 0.22352 & weak mixing angle defined by weak masses\\
         $ZPAR(4) = G_{\mu}$ & $1.166370 \cdot 10^{-5}$ &  muon decay constant\\
         $ZPAR(6) - ZPAR(14)$ & 0.23176-0.23152 &  effective weak mixing angles\\
         $ZPAR(15) = \alpha_s (M_Z^2)$ & 0.12500 &   recalculated by {\tt Dizet} \\
         $ZPAR(16) - ZPAR(30)$ &  &  QCD corrections \\
   \hline
 \end{tabular}
  \end{center}
%\end{sidewaystable}
\end{table*}

\end{document}